\PassOptionsToPackage{prologue,table,xcdraw}{xcolor}

\documentclass[a4paper,UKenglish,cleveref,autoref,thm-restate]{lipics-v2021}

\usepackage[utf8]{inputenc}
\usepackage{graphicx}
\usepackage{latexsym}
\usepackage{listings}
\usepackage{tikz}
\usepackage{color,soul}
\usetikzlibrary{arrows,automata,calc,fit,positioning,shapes}
\usepackage[draft]{fixme}
\fxsetup{theme=color}
\usepackage{underscore}
\usepackage{wrapfig}
\usepackage{listings}
\usepackage{listings-rust}
\usepackage{graphics}
\usepackage{amsmath}
\usepackage{mathtools}
\usepackage{pdfpages}
\usepackage{subcaption}
\usepackage{pifont}
\usepackage{makecell}
\usepackage{tabularx}
\usepackage{etoolbox}
\usepackage{setspace}
\usepackage[inline]{enumitem}
\usepackage{pgfplots}
\usepackage{xspace}
\usepackage{scribble}
\usepackage{textcomp}
\usepackage{longtable}
\usepackage{thm-restate}
\usepackage{amsfonts}
\usepackage{cleveref}
\usepackage[page,header]{appendix}
\usepackage{titletoc}
\usepackage{setspace}
\usepackage{enumitem}
\usepackage{wrapfig}
\usepackage[framemethod=TikZ]{mdframed}
\usepackage[T1]{fontenc}
\usepackage[utf8]{inputenc} 
\usepackage{subcaption}
\usepackage{etoolbox}
\usepackage{xifthen}
\usepackage{xspace}
\usepackage{thmtools}
\usepackage{thm-restate}

\usepackage[all]{hypcap}
\usepackage{nicefrac}
\usepackage{upquote}

\usepackage{stmaryrd}
\SetSymbolFont{stmry}{bold}{U}{stmry}{m}{n}
\usepackage{anyfontsize}

\usepackage{proof}
\usepackage{mathtools} 
\usepackage{mathrsfs}
\usepackage[most]{tcolorbox}
\usepackage{xcolor}
\usepackage{graphicx,graphics}
\usepackage{url}
\usepackage{color}
\usepackage{comment}
\usepackage{rotating}
\usepackage{listings}
\usepackage{multicol}
\usepackage{multirow}
\usepackage[noend]{algpseudocode}
\usepackage{tikz}
\usepackage{fancybox}
\usetikzlibrary{positioning,calc,backgrounds,fit,shapes,arrows.meta,cd}
\usepackage{import}
\usepackage{makecell}
\usepackage{relsize}
\usepackage{listings}

\usepackage{csquotes}

\usepackage{booktabs}

\usepackage[group-minimum-digits=4,per-mode=fraction]{siunitx}

\usepackage{adjustbox} 
\usepackage{arydshln} 

\usepackage{pdfcomment}

\hypersetup{hidelinks,
  unicode,
  colorlinks=true,
  allcolors=black,
  pdfstartview=Fit,
  breaklinks=true}

\makeatletter
\g@addto@macro{\UrlBreaks}{\UrlOrds}
\makeatother

\SetExpansion
  [ context = sloppy,
    stretch = 30,
    shrink = 60,
    step = 5 ]
  { encoding = {OT1,T1,TS1} }
  { }

\newcommand{\mpstrust}{\texttt{MultiCrusty}\xspace}

\title{Stay Safe under Panic: Affine Rust Programming with Multiparty Session Types}

\makeatletter
\newcommand{\thickhline}{%
	\noalign {\ifnum 0=`}\fi \hrule height 1pt
	\futurelet \reserved@a \@xhline
}
\newcolumntype{"}{@{\hskip\tabcolsep\vrule width 1pt\hskip\tabcolsep}}
\newcolumntype{C}[1]{>{\centering\arraybackslash}m{#1}}
\makeatother

\author{Nicolas Lagaillardie}{
	Department of Computing,
	Imperial College London,
	London, SW7 2AZ,
	United Kingdom
}{n.lagaillardie19@imperial.ac.uk}{0000-0002-6431-4100}{}

\author{Rumyana Neykova}{
	Department of Computer Science,
	Brunel University London,
	London,
	UB8 3PH,
	United Kingdom
}{rumyana.neykova@brunel.ac.uk}{0000-0002-2755-7728}{}

\author{Nobuko Yoshida}{
	Department of Computing,
	Imperial College London,
	London, SW7 2AZ,
	United Kingdom
}{n.yoshida@imperial.ac.uk}{0000-0002-3925-8557}{}

\authorrunning{N. Lagaillardie, R. Neykova and N. Yoshida} 

\Copyright{Nicolas Lagaillardie, Rumyana Neykova and Nobuko Yoshida} 



\acknowledgements{We thank the ECOOP reviewers for their
insightful comments and suggestions, 
and (alphabetical order) Zak Cutner, Wen Kokke, Roland Kuhn,
Dimitris Mostrous and Martin Vassor for discussions.}

\funding{
The work is supported by EPSRC EP/T006544/1,
EP/K011715/1, EP/K034413/1, EP/L00058X/1,
EP/N027833/1, EP/N028201/1, EP/T014709/1
and EP/V000462/1, and NCSS/EPSRC
VeTSS.}

\nolinenumbers 

\EventEditors{Editors}
\EventNoEds{2}
\EventLongTitle{ECOOP 2022}
\EventShortTitle{ECOOP 2022}
\EventAcronym{ECOOP}
\EventYear{2022}
\EventDate{2022}
\EventLocation{US}
\EventLogo{}
\SeriesVolume{1}
\ArticleNo{1}

\numberwithin{theorem}{section}
\numberwithin{definition}{section}
\numberwithin{example}{section}
\numberwithin{corollary}{section}
\numberwithin{remark}{section}
\numberwithin{lemma}{section}

\newtoggle{full}
\settoggle{full}{true}
\iftoggle{full}
{
	\newcommand{\inApp}{See~\Cref{app:AMPST}.~}
}{
	\newcommand{\inApp}{See~\cite{fullVersion}.~}
}

\makeatletter
\newcommand{\iftoggleverb}[1]{%
  \ifcsdef{etb@tgl@#1}
    {\csname etb@tgl@#1\endcsname\iftrue\iffalse}
    {\etb@noglobal\etb@err@notoggle{#1}\iffalse}%
}
\makeatother

\begin{document}

\maketitle


\newcommand{\must}{\ourlibrary}
\newcommand{\Fig}{Fig.}
\newcommand{\mustL}{{\sf{MiO}}\xspace}
\newcommand{\globalcombinators}{global combinators}

\newcommand{\ruto}[1]{{\color{red} \texttt{#1}}}

\newcommand{\encc}[1]{{\ensuremath{\mathtt{#1}}}}
\newcommand{\SET}[1]{\ensuremath{\{#1\}}}
\newcommand{\lbl}[1]{\ensuremath{l}}
\definecolor{dkblue}{rgb}{0,0.1,0.5}
\definecolor{dkgreen}{rgb}{0,0.4,0.1}
\definecolor{dkred}{rgb}{0.4,0,0}

\newcommand{\highlightcolor}{orange!30}
\definecolor{linkColor}{rgb}{0,0,0.5}
\definecolor{pblue}{rgb}{0.13,0.13,1}
\definecolor{purple}{rgb}{0.5,0,0.5}
\definecolor{beige}{rgb}{0,0.5,0.5}
\definecolor{pred}{rgb}{0.9,0,0}
\definecolor{WhiteSmoke}{rgb}{0.96, 0.96, 0.96}
\definecolor{mygray}{gray}{0.6}
\definecolor{DarkRed}{rgb}{0.55, 0.0, 0.0}

\definecolor{types}{rgb}{0.282, 0.714, 0.627}
\definecolor{functions}{rgb}{0.463, 0.463, 0.373}
\definecolor{typetype}{rgb}{0.318, 0.576, 0.792}
\definecolor{let}{rgb}{0.137, 0.412, 0.639}
\definecolor{cond}{rgb}{0.169, 0.075, 0.165}
\definecolor{str}{rgb}{0.392, 0.227, 0.157}
\definecolor{var}{rgb}{0.1, 0.1, 0.5}
\definecolor{primitive}{rgb}{0, 0.5, 0}


\newcommand{\mpststack}{MPST-stack {}}

\newcommand{\CODESIZETINY}{\footnotesize}
\newcommand{\CODESIZE}{\small} 
\newcommand{\CODESIZENORMAL}{\small} 
\newcommand{\LSTCODESIZE}{\fontsize{8pt}{9.5pt}} 
\newcommand{\CODESTYLE}{\ttfamily\bfseries}
\newcommand{\CODE}[1]{\sloppy{\CODESIZE\CODESTYLE\lstinline[language=rust]{#1}}}
\newcommand{\fCODE}[1]{{\CODESIZE\CODESTYLE #1}}
\newcommand{\MCODE}[1]{\textnormal{\small\fCODE{#1}}}
\newcommand{\CODEKW}[1]{{\color{purple}\CODESIZE\CODESTYLE #1}}

\newcommand{\TICK}{\color{dkgreen}{\ding{51}}}
\newcommand{\CROSS}{\color{red}{\ding{55}}}

\newcommand{\WARNING}{\color{orange}{\ding{115}\hspace{-6pt}\color{black}{\small\textbf{!}}}}
\newcommand{\QUESTION}{{\normalsize\color{BurntOrange}\textbf{?}}}

\newcommand{\todo}[1]{{\color{purple} #1}}

\newcommand{\listing}[1]{Figure~\ref{#1}}

\newcommand{\fig}[1]{Figure~\ref{#1}}

\newcommand{\role}[1]{role \textbf{\color{teal}\emph{#1}}}
\newcommand{\norole}[1]{\textbf{\color{teal}\emph{#1}}}


\lstnewenvironment{rustlisting}%
{
  \lstset{
    language=Rust,
    style = colouredRust,
    xleftmargin= 2em,
    xrightmargin= 2em,
    framexleftmargin = 0.2em,
    numbers=left,
    backgroundcolor=\color{white},
    showstringspaces=false,
    formfeed=\newpage,
    tabsize=2,
    commentstyle=\color{dkgreen}\itshape,
    keywordstyle =\color{purple}\CODESTYLE,
    keywordstyle = \color{dkblue}\CODESTYLE,
    stringstyle = \color{DarkRed}\CODESTYLE, 
    emph={global, protocol, role, par, and, from, to, choice, at, or, rec, continue},
    morekeywords = [2]{A,B,C,D,E,S,P},
    keywordstyle = [2]\color{beige}\CODESTYLE,
    emphstyle=\color{dkblue}\CODESTYLE,
    basicstyle=\ttfamily\scriptsize,
    escapeinside={*@}{@*},
    moredelim=**[is][\btHL]{`}{`},
    breaklines=true,
  }
}{}



\lstnewenvironment{SCRIBBLELISTING}%
{
  \lstset{
    language=Scribble,
    xleftmargin= 2.5em,
    xrightmargin= 2.5em,
    framexleftmargin = 0.5em,
    showstringspaces=false,
    formfeed=\newpage,
    tabsize=2,
    mathescape=true,
    commentstyle=\color{gray}\itshape,
    backgroundcolor=\color{white},
    keywordstyle=\color{purple}\CODESTYLE,
    emph={global, protocol, role, par, and, from, to, choice, at, or, rec, continue},
    morekeywords = [2]{A,B,C,D,E, S, P},
    keywordstyle = [2]\color{beige}\CODESTYLE,
    emphstyle=\color{dkblue}\CODESTYLE,
    basicstyle=\ttfamily\scriptsize,
    escapeinside={*@}{@*},
    moredelim=**[is][\btHL]{`}{`},
  }
}{}
\newcommand{\mrg}{\mathit{mrg}}

\newlist{cdescription}{description}{1}
\mdfdefinestyle{cdescription}{%
  hidealllines=true,
  skipabove= 5pt,
  skipbelow= 5pt,
  roundcorner=5pt,
}

\surroundwithmdframed[style=cdescription]{cdescription}

\newcommand{\bCODE}{\lstinline[
   language=Rust,
	basicstyle=\scriptsize\CODESTYLE,
     keywords= {def},
     morekeywords= {func},
  	emph={func},
  	 keywordstyle=\color{blue},
  	 emphstyle = \color{purple}
  	morekeywords = {choice_at, finish},
  mathescape=true,
  breaklines=true
]}

\newcommand{\kmc}{$k$-MC}
\newcommand{\rolea}{\CODE{A} {}}
\newcommand{\rolec}{\CODE{C} {}}
\newcommand{\rumi}[1]{{\color{red}\textbf{Rumi}:  #1}}

\definecolor{darkgreen}{rgb}{0, 0.5, 0}

\newcommand{\cmark}{\ding{51}}%
\newcommand{\xmark}{\ding{55}}%

\newcommand{\cfig}[1]{Fig.~\ref{#1}}
\newcommand{\id}{\ensuremath{\mathsf{id}}}
\newcommand{\grmeq}{{::=}}
\newcommand{\grmor}{{\;\text{\Large$\mid$}\;}}

\newcommand\TableStrut{\rule{0pt}{1.25\normalbaselineskip}\rule[-1\normalbaselineskip]{0pt}{0pt}}
\newcommand\TableStrutDouble{\rule{0pt}{1.75\normalbaselineskip}\rule[-1.50\normalbaselineskip]{0pt}{0pt}}
\newcommand\TableStrutTriple{\rule{0pt}{2.25\normalbaselineskip}\rule[-2.00\normalbaselineskip]{0pt}{0pt}}

\newcommand{\oroleU}{{\color{roleColor}\roleFmt{u}}}
\newcommand{\oroleP}{{\color{roleColor}\roleFmt{p}}}
\newcommand{\oroleC}{{\color{roleColor}\roleFmt{c}}}

\newcommand{\OCaml}{OCaml\xspace}
\newcommand{\ourlibrary}{\texttt{ocaml-mpst}\xspace}
\newcommand{\Sec}{\textsection}
\newcommand{\Thm}{Thm.}
\newcommand{\channelvector}{channel vector\xspace}
\newcommand{\ourencoding}{channel vector encoding }

\newcommand{\R}{\mathtt{R}}
\newcommand{\Rptp}[1]{\ensuremath{{\color{roleColor}\roleFmt{#1}}}}
\newcommand{\ptp}{\participant}
\newcommand{\participant}[1]{{\color{roleColor}{\boldsymbol{\mathtt{#1}}}}}
\newcommand{\pp}{\ensuremath{\participant{p}}}
\newcommand{\qq}{\ensuremath{\participant{q}}}
\newcommand{\rr}{\ensuremath{\participant{r}}}

\newcommand{\glab}{\gtLab}
\newcommand{\lab}{\labFmt{m}}
\newcommand{\labHello}{{\labFmt{hello}}}
\newcommand{\labAuth}{{\labFmt{auth}}}
\newcommand{\labPass}{{\labFmt{password}}}
\newcommand{\labOk}{{\labFmt{ok}}}
\newcommand{\labCancel}{{\labFmt{cancel}}}
\newcommand{\labLeft}{{\labFmt{left}}}
\newcommand{\labRight}{{\labFmt{right}}}
\newcommand{\labMsg}{{\labFmt{msg}}}
\newcommand{\OCAMLMPST}{\mustL}
\newcommand{\myparagraph}[1]{\smallskip\noindent\emph{\bf #1\ }}

\newcommand*{\thmstart}{\setlength{\abovedisplayskip}{2pt}\setlength{\belowdisplayskip}{2pt}\rm}

\newtheorem{mycounter}{Dummy}[section]{\bfseries}{\itshape}
\newtheorem{THM}[mycounter]{Theorem}{\bfseries}{\itshape}
\newtheorem{DEF}[mycounter]{Definition}{\bfseries}{\itshape}
\newtheorem{CONV}[mycounter]{Convention}{\bfseries}{\itshape}
\newtheorem{COL}[mycounter]{Corollary}{\bfseries}{\itshape}
\newtheorem{LEM}[mycounter]{Lemma}{\bfseries}{\itshape}
\newtheorem{PROP}[mycounter]{Proposition}{\bfseries}{\itshape}
\newtheorem{REM}[mycounter]{Remark}{\bfseries}{\itshape}
\newtheorem{EX}[mycounter]{Example}{\bfseries}{\itshape}
\Crefname{THM}{Theorem}{Theorems}
\Crefname{DEF}{Definition}{Definitions}
\Crefname{CONV}{Convention}{Conventions}
\Crefname{COL}{Corollary}{Corollaries}
\Crefname{LEM}{Lemma}{Lemmas}
\Crefname{PROP}{Proposition}{Propositions}
\Crefname{REM}{Remark}{Remarks}
\Crefname{EX}{Example}{Examples}

\newcommand{\CASE}{{\bfseries Case}\xspace}

\newcommand{\checkit}[1]{{\color{black} #1}}
\newcommand{\precameraready}[1]{{#1}}
\definecolor{modify}{rgb}{0.0, 0.0, 0.0}%
\definecolor{modified}{rgb}{0.0, 0.0, 0.0}%

\makeatletter
\newsavebox{\@brx}
\newcommand{\llangle}[1][]{\savebox{\@brx}{\(\m@th{#1\langle}\)}%
  \mathopen{\copy\@brx\kern-0.5\wd\@brx\usebox{\@brx}}}
\newcommand{\rrangle}[1][]{\savebox{\@brx}{\(\m@th{#1\rangle}\)}%
  \mathclose{\copy\@brx\kern-0.5\wd\@brx\usebox{\@brx}}}
\makeatother

\newcommand{\elip}{\ifmmode\mathinner{\ldotp\kern-0.2em\ldotp\kern-0.2em\ldotp}\else.\kern-0.13em.\kern-0.13em.\fi}
\newcommand{\elipc}{\ifmmode\mathinner{\cdotp\kern-0.2em\cdotp\kern-0.2em\cdotp}\else.\kern-0.13em.\kern-0.13em.\fi}
\newcommand{\eqdeff}{\ensuremath{\stackrel{\mathrm{def}}{=}}}

\newcommand{\roleC}{{\color{roleColor}\roleFmt{c}}}
\newcommand{\roleA}{{\color{roleColor}\roleFmt{a}}}
\newcommand{\roleS}{{\color{roleColor}\roleFmt{s}}}

\newcommand{\labLoop}{\labFmt{loop}}
\newcommand{\labStop}{\labFmt{stop}}

\newcommand{\dlabel}[1]{\textcolor{gray}{\sffamily\bfseries\mathversion{bold}#1}}
\newcommand{\Rone}{\dlabel{R1}\xspace}
\newcommand{\Rtwo}{\dlabel{R2}\xspace}
\newcommand{\Rthree}{\dlabel{R3}\xspace}
\newcommand{\Mone}{\dlabel{M1}\xspace}
\newcommand{\Mtwo}{\dlabel{M2}\xspace}

\newcommand{\tts}{\mathit{ts}}
\newcommand{\uus}{\mathit{us}}
\newcommand{\lput}{{\color{black}\mathrm{put}}}
\newcommand{\lget}{{\color{black}\mathrm{get}}}

\newcommand{\NYnote}[1]{{\color{violet}{\textbf{NY:}}\ #1}}

\newcommand{\Niconote}[1]{{\color{red}{\textbf{Nico:}}\ #1}}

\newcommand{\PSet}{\!\mathscr{P}\!}
\newcommand{\ASigma}{ \mathbb{A}}

%
%

\newcommand{\ifempty}[3]{%
  \ifthenelse{\isempty{#1}}{#2}{#3}%
}%

\newcommand{\ifTR}[1]{%
  \iftoggle{techreport}{#1}{}%
}%

\newcommand{\dom}[1]{{\color{black}\operatorname{dom}\!\left({#1}\right)}}%
\newcommand{\ran}[1]{{\color{black}\operatorname{ran}\!\left({#1}\right)}}%
\newcommand{\suchthat}{\colon}%
\newcommand{\eqdef}{\colonequals}%
\newcommand{\fn}[1]{\operatorname{fn}\!\left({#1}\right)}%
\newcommand{\sbj}[1]{\operatorname{sbj}\!\left({#1}\right)}%
\newcommand{\bv}[1]{\operatorname{bv}\!\left({#1}\right)}%
\newcommand{\fv}[1]{\operatorname{fv}\!\left({#1}\right)}%
\newcommand{\fc}[1]{\operatorname{fc}\!\left({#1}\right)}%
\newcommand{\dpv}[1]{\operatorname{dpv}\!\left({#1}\right)}%
\newcommand{\fpv}[1]{\operatorname{fpv}\!\left({#1}\right)}%
\newcommand{\sem}[2][]{\mbox{\ensuremath{\llbracket#2\rrbracket_{#1}}}}%
\newcommand{\gfp}[1]{\operatorname{gfp}\!\left({#1}\right)}%
\newcommand{\lfp}[1]{\operatorname{lfp}\!\left({#1}\right)}%
\newcommand{\mfp}[1]{\operatorname{mfp}\!\left({#1}\right)}%
\newcommand{\unfold}[1]{%
  {\color{black}\operatorname{unf^{\ast}}\!\left({#1}\right)}}%
\newcommand{\unfoldOne}[1]{%
  {\color{black}\operatorname{unf}\!\left({#1}\right)}}%
\newcommand{\unfoldN}[2]{%
  {\color{black}\operatorname{unf^{#1}}\!\left({#2}\right)}}%
\newcommand{\notImplies}{\mathrel{{\kern 0.5em}{\not{\kern -0.5em}\implies}}}%
\newcommand{\notImpliedBy}{\mathrel{{\kern 1em}{\not{\kern -1em}\impliedby}}}%
\newcommand{\fix}[1]{\operatorname{fix}\, {#1}}%
\newcommand{\lfix}[1]{\operatorname{lfix}\, {#1}}%

\newcommand{\coloncolonequals}{\Coloneqq}%
\newcommand{\bnfdef}{\coloncolonequals}%
\newcommand{\bnfsep}{\mathbin{\;\big|\;}}%

\def\cf{cf.\@\xspace}%
\def\etc{\emph{etc.}\@\xspace}%
\def\eg{e.g.,\@\xspace}%
\def\Eg{E.g.,\@\xspace}%
\def\ie{i.e.,\@\xspace}%
\def\Ie{I.e.,\@\xspace}%
\def\vs{vs.\@\xspace}%
\def\wrt{w.r.t.\@\xspace}%

\definecolor{ruleColor}{rgb}{0.1, 0.3, 0.1}%
\newcommand{\inferrule}[1]{{\color{ruleColor}\text{\scriptsize [#1]}}}%
\newcommand{\inference}[3][]{\infer[\ifempty{#1}{}{\inferrule{#1}}]{#3}{#2}}%
\newcommand{\cinference}[3][]{\infer=[\ifempty{#1}{}{\inferrule{#1}}]{#3}{#2}}%
\newcommand{\inferenceSingle}[2][]{{#2}\ifempty{#1}{}{\ \inferrule{#1}}}%
\newcommand{\nrule}[1]{{\scriptsize \textsc{#1}}}%

\newcommand{\setenum}[1]{\mathord{{\color{black}\left\{#1\right\}}}}%
\newcommand{\setcomp}[2]{\mathord{%
  {\color{black}\left\{{#1} \,\middle|\, {#2}\right\}}}}%
\newcommand{\card}[1]{\mathord{\left\vert{#1}\right\vert}}

\newcommand{\setenumSmall}[1]{\mathord{{\color{black}\{#1\}}}}%
\newcommand{\setcompSmall}[2]{\mathord{{\color{black}\{{#1} \,\mid\, {#2}\}}}}%
\newcommand{\cardSmall}[1]{\mathord{\vert{#1}\vert}}%

\newcommand{\setNat}{\mathord{\mathbb{N}}}%
\newcommand{\setNatZero}{\mathord{\mathbb{N}^0}}%
\newcommand\someSetX{\mathord{\mathbb{X}}}%
\newcommand\someSetY{\mathord{\mathbb{Y}}}%

\newcommand{\setLabelTypes}{\mathbb{L}}%

\definecolor{groundColor}{rgb}{0.38, 0.25, 0.32}%

\newcommand{\groundFmt}[1]{{\color{groundColor}#1}}%
\newcommand{\setGroundTypes}{\groundFmt{\mathbb{B}}}%
\newcommand{\groundSubtype}{\mathrel{\groundFmt{\leqslant_{\setGroundTypes}}}}%

\newcommand{\someDeriv}[1][]{\ifempty{#1}{\mathcal{D}}{\mathcal{D}_{#1}}}%

\newcommand{\bind}[2]{\nicefrac{#2}{#1}}
\newcommand{\substenum}[1]{\mathord{{\color{black}\left\{{#1}\right\}}}}
\newcommand{\subst}[2]{\substenum{\bind{#1}{#2}}}
\newcommand{\mapsubst}[2]{{#1}{\mapsto}{#2}}

\definecolor{roleColor}{rgb}{0.5, 0.0, 0.0}%
\newcommand{\roleCol}[1]{{\color{roleColor}#1}}%
\newcommand{\roleUnivSet}{\roleCol{\mathfrak{R}}}%
\newcommand{\roleSet}{\roleCol{\mathbb{R}}}%
\newcommand{\roleFmt}[1]{\boldsymbol{\roleCol{\mathtt{#1}}}}%
\newcommand{\roleP}[1][]{%
  \ifempty{#1}{{\color{roleColor}\roleFmt{p}}}{{\color{roleColor}\roleFmt{p}_{{#1}}}}%
}%
\newcommand{\rolePi}[1][]{%
  \ifempty{#1}{{\color{roleColor}\roleFmt{p}'}}{{\color{roleColor}\roleFmt{p}'_{#1}}}%
}%
\newcommand{\roleQ}[1][]{%
  \ifempty{#1}{{\color{roleColor}\roleFmt{q}}}{{\color{roleColor}\roleFmt{q}_{#1}}}%
}%
\newcommand{\roleQi}[1][]{%
  \ifempty{#1}{{\color{roleColor}\roleFmt{q}'}}{{\color{roleColor}\roleFmt{q}'_{#1}}}%
}%
\newcommand{\roleR}[1][]{%
  \ifempty{#1}{{\color{roleColor}\roleFmt{r}}}{{\color{roleColor}\roleFmt{r}_{\!#1}}}%
}%
\newcommand{\roleRi}[1][]{%
  \ifempty{#1}{{\color{roleColor}\roleFmt{r}'}}{{\color{roleColor}\roleFmt{r}'_{\!#1}}}%
}%
\newcommand{\roleRii}[1][]{%
  \ifempty{#1}{{\color{roleColor}\roleFmt{r}''}}{{\color{roleColor}\roleFmt{r}''_{\!#1}}}%
}%

\newcommand{\labFmt}[2][]{\color{roleColor}{\ifempty{#1}{\mathtt{#2}}{\mathtt{#2}\textsubscript{#1}}}}%

\definecolor{mpColor}{rgb}{0, 0, 0}%
\newcommand{\mpFmt}[1]{{\color{mpColor}#1}}%

\newcommand{\mpLab}[1][]{%
  \mpFmt{\ifempty{#1}{\labFmt{m}}{{\labFmt{m}}_{\mathnormal #1}}}%
}%

\newcommand{\mpRequest}[1][]{%
  \mpFmt{\ifempty{#1}{\labFmt{req}}{{\labFmt{req}}_{\mathnormal #1}}}%
}%

\newcommand{\mpResponse}[1][]{%
\mpFmt{\ifempty{#1}{\labFmt{res}}{{\labFmt{res}}_{\mathnormal #1}}}%
}%

\newcommand{\mpVideo}[1][]{%
\mpFmt{\ifempty{#1}{\labFmt{video}}{{\labFmt{video}}_{\mathnormal #1}}}%
}%

\newcommand{\mpClose}[1][]{%
\mpFmt{\ifempty{#1}{\labFmt{close}}{{\labFmt{close}}_{\mathnormal #1}}}%
}%

\newcommand{\mpQuestion}{{\color{blue}{\textbf{?}}}\,}%
\newcommand{\mpDagger}{\roleFmt{\dagger}\,}%

\newcommand{\mpLabi}[1][]{%
  \mpFmt{\ifempty{#1}{\labFmt{m}'}{\labFmt{m}'_{\mathnormal #1}}}%
}%
\newcommand{\mpLabFmt}[1]{\mpFmt{\labFmt{#1}}}%

\newcommand{\proclit}[1]{\ensuremath{\operatorname{\text{\sffamily\bfseries{#1}}}}}

\newcommand{\mpTrue}{\mpFmt{\text{\emph{\texttt{true}}}}}%
\newcommand{\mpFalse}{\mpFmt{\text{\emph{\texttt{false}}}}}%
\newcommand{\mpString}[1]{\text{\emph{\texttt{"{#1}"}}}}%

\newcommand{\mpChanRole}[2]{\mpFmt{{#1}[{#2}]}}%
\newcommand{\mpChanRoles}[3]{\mpFmt{{#1}[{#2}][{#3}]}}%

\newcommand{\mpNil}{\mpFmt{\mathbf{0}}}%
\newcommand{\mpSeq}{\mathbin{\mpFmt{\!.\!}}}%
\newcommand{\mpExpr}{\mpFmt{e}}%
\newcommand{\EvalExpr}{\downarrow}%

\newcommand{\mpIf}[3]{%
  \mpFmt{\proclit{if}\,{#1}\,\proclit{then}}\,{#2}\,\proclit{else}\,{#3}%
}%
\newcommand{\mpChoice}[3]{%
  \mpFmt{%
    \mpLabFmt{#1}\ifempty{#2}{}{({#2})}\ifempty{#3}{}{\vphantom{x}\mpSeq {#3}}%
  }%
}%
\newcommand{\mpChoiceNoBind}[3]{%
  \mpFmt{%
    \mpLabFmt{#1}\ifempty{#2}{}{\langle{#2}\rangle}\ifempty{#3}{}{\vphantom{x}\mpSeq {#3}}%
  }%
}%
\newcommand{\mpBranchRaw}[3]{%
  \mpFmt{%
    {#1}[\roleFmt{#2}] \sum%
      {#3}%
  }%
}%
\newcommand{\mpBranch}[6]{%
\mpFmt{%
  {#1}[\roleFmt{#2}] \mathbin{\!\ifempty{\sum}{\&}{\sum_{#3}}\!}%
  \ifempty{#3}{%
    \mpChoice{#4}{#5}{#6}%
  }{%
    \mpChoice{#4}{#5}{#6}%
  }%
}%
}%
\newcommand{\mpBranchSeq}[5]{%
\mpFmt{%
  {#1}[\roleFmt{#2}]\mpChoice{#3}{#4}{#5}%
}%
}%
\newcommand{\mpBranchSolo}[4]{%
\mpFmt{%
  {#1} \mpChoice{#2}{#3}{#4}%
}%
}%
\newcommand{\mpQuestionBranch}[6]{%
\mpFmt{%
  \mpQuestion{#1}[\roleFmt{#2}] \mathbin{\!\ifempty{\sum}{\&}{\sum_{#3}}\!}%
  \ifempty{#3}{%
    \mpChoice{#4}{#5}{#6}%
  }{%
    \mpChoice{#4}{#5}{#6}%
  }%
}%
}%
\newcommand{\mpQuestionBranchSeq}[5]{%
\mpFmt{%
  \mpQuestion{#1}[\roleFmt{#2}]\mpChoice{#3}{#4}{#5}%
}%
}%
\newcommand{\mpQuestionBranchSolo}[3]{%
\mpFmt{%
  \mpQuestion{#1}[\roleFmt{#2}]\mpFmt{({#3})}%
}%
}%
\newcommand{\mpDaggerBranch}[6]{%
  \mpFmt{%
    \mpDagger{#1}[\roleFmt{#2}] \mathbin{\!\ifempty{\sum}{\&}{\sum_{#3}}\!}%
    \ifempty{#3}{%
      \mpChoice{#4}{#5}{#6}%
    }{%
      \mpChoice{#4}{#5}{#6}%
    }%
  }%
}%

\newcommand{\mpQuestionSel}[5]{%
  \mpFmt{%
    \mpQuestion{#1}[\roleFmt{#2}] \mathbin{\!\oplus\!}%
    \mpChoiceNoBind{#3}{#4}{#5}%
  }%
}%

\newcommand{\mpDaggerSel}[5]{%
  \mpFmt{%
    \mpDagger{#1}[\roleFmt{#2}] \mathbin{\!\oplus\!}%
    \mpChoiceNoBind{#3}{#4}{#5}%
  }%
}%
\newcommand{\mpSel}[5]{%
  \mpFmt{%
    {#1}[\roleFmt{#2}] \mathbin{\!\oplus\!}%
    \mpChoiceNoBind{#3}{#4}{#5}%
  }%
}%

\newcommand{\mpRaise}[2]{\chanraise(#1).{#2}}%
\newcommand{\mpPar}{\mathbin{\mpFmt{\mid}}}%
\newcommand{\mpParTall}{\mathbin{\mpFmt{\bigm|}}}%
\newcommand{\mpBigPar}[2]{\mathbin{\mpFmt{\big|_{#1}}{#2}}}%
\newcommand{\mpRes}[2]{\mpFmt{\left(\mathbf{\nu}{#1}\right){#2}}}%
\newcommand{\mpRec}[2]{\mpFmt{\mu{#1}.{#2}}}%
\newcommand{\mpJustDef}[3]{%
  \mpFmt{{#1}(#2) = {#3}}%
}%
\newcommand{\mpDef}[4]{%
  \mpFmt{\proclit{def}\;\mpJustDef{#1}{#2}{#3}\;\proclit{in}\;{#4}}%
}%
\newcommand{\mpDefAbbrev}[2]{%
  \mpFmt{\proclit{def}\;{#1}\;\proclit{in}\;{#2}}%
}%
\newcommand{\mpAngleDef}[3]{%
\mpFmt{{#1}\langle#2\rangle = {#3}}%
}%
\newcommand{\mpShortAngleDef}[3]{%
\mpFmt{\proclit{def}\;\mpAngleDef{#1}{#2}{#3}}%
}%
\newcommand{\mpCall}[2]{\mpFmt{{#1}\!\left\langle{#2}\right\rangle}}%
\newcommand{\mpErr}{\mpFmt{\proclit{err}}}%
\newcommand{\mpRaiseProc}{\mpFmt{\proclit{raise}}}%

\newcommand{\mpCtx}[1][]{\mpFmt{\ifempty{#1}{\mathbb{C}}{\mathbb{C}_{#1}}}}%
\newcommand{\mpEtx}[1][]{\mpFmt{\ifempty{#1}{\mathbb{E}}{\mathbb{E}_{#1}}}}%
\newcommand{\mpCtxi}[1][]{\mpFmt{\ifempty{#1}{\mathbb{C}'}{\mathbb{C}'_{#1}}}}%
\newcommand{\mpCtxii}[1][]{\mpFmt{\ifempty{#1}{\mathbb{C}''}{\mathbb{C}''_{#1}}}}%
\newcommand{\mpCtxHole}{[\,]}%
\newcommand{\mpEtxHole}{[\,]}%
\newcommand{\mpCtxApp}[2]{{#1}\!\left[{#2}\right]}%

\newcommand{\mpA}[1][]{\mpFmt{\ifempty{#1}{a}{a_{#1}}}}%
\newcommand{\mpAi}[1][]{\mpFmt{\ifempty{#1}{a'}{a'_{#1}}}}%
\newcommand{\mpAii}[1][]{\mpFmt{\ifempty{#1}{a''}{a''_{#1}}}}%

\newcommand{\mpC}[1][]{\mpFmt{\ifempty{#1}{c}{c_{#1}}}}%
\newcommand{\mpCi}[1][]{\mpFmt{\ifempty{#1}{c'}{c'_{#1}}}}%
\newcommand{\mpCii}[1][]{\mpFmt{\ifempty{#1}{c''}{c''_{#1}}}}%
\newcommand{\mpD}[1][]{\mpFmt{\ifempty{#1}{d}{d_{#1}}}}%
\newcommand{\mpConst}{\mathtt{d}}%

\newcommand{\mpDi}[1][]{\mpFmt{\ifempty{#1}{d'}{d'_{#1}}}}%
\newcommand{\mpDii}[1][]{\mpFmt{\ifempty{#1}{d''}{d''_{#1}}}}%

\newcommand{\mpS}[1][]{\mpFmt{\ifempty{#1}{s}{s_{#1}}}}%
\newcommand{\mpSi}[1][]{\mpFmt{\ifempty{#1}{s'}{s'_{#1}}}}%
\newcommand{\mpSii}[1][]{\mpFmt{\ifempty{#1}{s''}{s''_{#1}}}}%
\newcommand{\mpSiii}[1][]{\mpFmt{\ifempty{#1}{s'''}{s'''_{#1}}}}%

\newcommand{\mpX}[1][]{\mpFmt{\ifempty{#1}{X}{X_{#1}}}}%
\newcommand{\mpXi}[1][]{\mpFmt{\ifempty{#1}{X'}{X'_{#1}}}}%
\newcommand{\mpY}[1][]{\mpFmt{\ifempty{#1}{Y}{Y_{#1}}}}%
\newcommand{\mpZ}[1][]{\mpFmt{\ifempty{#1}{Z}{Z_{#1}}}}%
\newcommand{\mpZi}[1][]{\mpFmt{\ifempty{#1}{Z'}{Z'_{#1}}}}%
\newcommand{\mpZii}[1][]{\mpFmt{\ifempty{#1}{Z''}{Z''_{#1}}}}%

\newcommand{\mpP}[1][]{\mpFmt{\ifempty{#1}{P}{P_{#1}}}}%
\newcommand{\mpPi}[1][]{\mpFmt{\ifempty{#1}{P'}{P'_{#1}}}}%
\newcommand{\mpPii}[1][]{\mpFmt{\ifempty{#1}{P''}{P''_{#1}}}}%
\newcommand{\mpPiii}[1][]{\mpFmt{\ifempty{#1}{P'''}{P'''_{#1}}}}%
\newcommand{\mpPx}[1][]{\mpFmt{\ifempty{#1}{P^\bullet}{P^\bullet_{#1}}}}%
\newcommand{\mpQ}[1][]{\mpFmt{\ifempty{#1}{Q}{Q_{#1}}}}%
\newcommand{\mpQi}[1][]{\mpFmt{\ifempty{#1}{Q'}{Q'_{#1}}}}%
\newcommand{\mpQii}[1][]{\mpFmt{\ifempty{#1}{Q''}{Q''_{#1}}}}%
\newcommand{\mpR}[1][]{\mpFmt{\ifempty{#1}{R}{R_{#1}}}}%
\newcommand{\mpRi}[1][]{\mpFmt{\ifempty{#1}{R'}{R'_{#1}}}}%
\newcommand{\mpRii}[1][]{\mpFmt{\ifempty{#1}{R''}{R''_{#1}}}}%

\newcommand{\mpT}[1][]{\mpFmt{\ifempty{#1}{T}{T_{#1}}}}%

\newcommand{\mpDefD}[1][]{\mpFmt{\ifempty{#1}{D}{D_{#1}}}}%
\newcommand{\mpDefDi}[1][]{\mpFmt{\ifempty{#1}{D'}{D'_{#1}}}}%
\newcommand{\mpDefDii}[1][]{\mpFmt{\ifempty{#1}{D''}{D''_{#1}}}}%

\newcommand{\mpMove}{\to}%
\newcommand{\mpMoveAnnot}[4]{\xrightarrow{\mpAnnot{#1}{#2}{#3}{#4}}}%
\newcommand{\mpMoveAnnotDef}{\xrightarrow{\mpFmt{\proclit{def}}}}%
\newcommand{\mpMoveAnnotDefStar}{\mathrel{\mpMoveAnnotDef{}^{\!\!\!\!*}}}%
\newcommand{\mpAnnot}[4]{{#1}{[#2]}{[#3]}{#4}}%
\newcommand{\mpMoveStar}{\mathrel{\mpMove{}^{\!\!\!*}}}%
\newcommand{\mpMoveP}[1]{\mpFmt{#1}\!{\mpMove}}%
\newcommand{\mpNotMoveP}[1]{\mpFmt{#1}\!\not{\mpMove}}%

\definecolor{gtColor}{rgb}{0.43, 0.21, 0.1}%
\newcommand{\gtFmt}[1]{{\color{gtColor}#1}}%
\newcommand{\gtMsgFmt}[1]{\gtFmt{\labFmt{#1}}}%
\newcommand{\gtLab}[1][]{%
  \ifempty{#1}{\gtMsgFmt{m}}{{\color{gtColor}\gtMsgFmt{m}_{#1}}}%
}%
\newcommand{\gtLabi}[1][]{%
  \ifempty{#1}{\gtMsgFmt{m}'}{{\color{gtColor}\gtMsgFmt{m}'_{#1}}}%
}%

\newcommand{\gtG}[1][]{\gtFmt{\ifempty{#1}{G}{G_{#1}}}}%
\newcommand{\gtGi}[1][]{\gtFmt{\ifempty{#1}{G'}{G'_{#1}}}}%
\newcommand{\gtGii}[1][]{\gtFmt{\ifempty{#1}{G''}{G''_{#1}}}}%
\newcommand{\gtGiii}[1][]{\gtFmt{\ifempty{#1}{G'''}{G'''_{#1}}}}%

\newcommand{\gtSeq}{\mathbin{\gtFmt{.}}}%

\newcommand{\gtCommRaw}[3]{%
  \gtFmt{%
    {#1} {\to} {#2}{:}%
    \left\{%
      {#3}%
    \right\}%
  }%
}%
\newcommand{\gtComm}[6]{%
  \gtFmt{%
    \gtCommRaw{#1}{#2}{%
      \gtCommChoice{#4}{#5}{#6}%
    }_{#3}%
  }%
}%
\newcommand{\gtCommSingle}[5]{%
  \gtFmt{%
    {#1} {\to} {#2}{:}%
    \gtCommChoice{#3}{#4}{#5}%
  }%
}%
\newcommand{\gtCommChoice}[3]{%
  \gtFmt{%
    \gtMsgFmt{#1}\ifempty{#2}{}{({#2})}%
    \ifempty{#3}{}{\vphantom{x}{\gtSeq}{#3}}%
  }%
}%
\newcommand{\gtCommChoiceSmall}[3]{%
  \gtFmt{%
    \gtMsgFmt{#1}\ifempty{#2}{}{({#2})}%
    \ifempty{#3}{}{\vphantom{x} \!\gtSeq\! {#3}}%
  }%
}%

\newcommand{\gtEnd}{\gtFmt{\mathbf{end}}}%

\newcommand{\gtRec}[2]{\gtFmt{\mu{#1}.{#2}}}%
\newcommand{\gtRecVarBase}{\gtFmt{\mathbf{t}}}%
\newcommand{\gtRecVar}[1][]{\gtFmt{\ifempty{#1}{\gtRecVarBase}{\gtRecVarBase_{#1}}}}%
\newcommand{\gtRecVari}[1][]{\gtFmt{\ifempty{#1}{\gtRecVarBase'}{\gtRecVarBase'_{#1}}}}%
\newcommand{\gtRecVarii}[1][]{\gtFmt{\ifempty{#1}{\gtRecVarBase'}{\gtRecVarBase'_{#1}}}}

\newcommand{\gtRoles}[1]{{\color{black} \operatorname{roles}(\gtFmt{#1})}}%

\newcommand{\gtProj}[2]{%
  {\color{stColor}\gtFmt{#1} {\upharpoonright} #2}%
}%

\newcommand{\gtCtx}[1][]{\gtFmt{\ifempty{#1}{\mathbb{G}}{\mathbb{G}_{#1}}}}%
\newcommand{\gtCtxi}[1][]{\gtFmt{\ifempty{#1}{\mathbb{G}'}{\mathbb{G}'_{#1}}}}%
\newcommand{\gtCtxHole}[1]{\gtFmt{[\;]^{#1}}}%
\newcommand{\gtCtxApp}[4]{\gtFmt{{#1}[{#2}]^{#3}_{#4}}}%

%
%
%
%
%
\definecolor{stColor}{rgb}{0.1, 0.4, 0.1}
\newcommand{\stFmt}[1]{{\color{stColor}#1}}%

\newcommand{\stTypeBool}{\stFmt{\operatorname{Bool}}}%
\newcommand{\stTypeInt}{\stFmt{\operatorname{Int}}}%
\newcommand{\stTypeString}{\stFmt{\operatorname{Str}}}%

\newcommand{\stInTag}{?}%
\newcommand{\stOutTag}{!}%
\newcommand{\stIn}[3]{\ifempty{#1}{}{\roleFmt{#1}}\stFmt{\&{#2}\ifempty{#3}{}{({#3})}}}%
\newcommand{\stOut}[3]{\ifempty{#1}{}{\roleFmt{#1}}\stFmt{\oplus{#2}\ifempty{#3}{}{({#3})}}}%

\newcommand{\stChoice}[2]{\stFmt{#1}\ifempty{#2}{}{\stFmt{({#2})}}}%

\newcommand{\stTau}{\stFmt{\tau}}%
\newcommand{\stSeq}{\mathbin{\!\stFmt{.}\!}}%
\newcommand{\stIntC}{\mathbin{\stFmt{\oplus}}}%
\newcommand{\stIntSum}[3]{\roleFmt{#1}\stFmt{\oplus_{#2}{#3}}}%
\newcommand{\stExtC}{\mathbin{\stFmt{\&}}}%
\newcommand{\stExtSum}[3]{\roleFmt{#1}\stFmt{\&_{#2}{#3}}}%
\newcommand{\stGenSum}[3]{\roleFmt{#1}\stFmt{\maltese_{#2}{#3}}}%
\newcommand{\stGenC}{\mathbin{\stFmt{\maltese}}}%
\newcommand{\stGenSumDual}[3]{\roleFmt{#1}\,\stFmt{\bar{\maltese}_{#2}{#3}}}%
\newcommand{\stRec}[2]{\stFmt{\mu{#1}.{#2}}}%
\newcommand{\stEnd}{\stFmt{\mathbf{end}}}%

\newcommand{\stLab}[1][]{\mpFmt{\ifempty{#1}{\labFmt{m}}{\labFmt{m}_{#1}}}}%
\newcommand{\stLabi}[1][]{\mpFmt{\ifempty{#1}{\labFmt{m}'}{\labFmt{m}'_{#1}}}}%
\newcommand{\stLabFmt}[1]{\mpFmt{\labFmt{#1}}}%

\newcommand{\stS}[1][]{\stFmt{\ifempty{#1}{S}{S_{#1}}}}%
\newcommand{\stSi}[1][]{\stFmt{\ifempty{#1}{S'}{S'_{#1}}}}%
\newcommand{\stSii}[1][]{\stFmt{\ifempty{#1}{S''}{S''_{#1}}}}%
\newcommand{\stSiii}[1][]{\stFmt{\ifempty{#1}{S'''}{S'''_{#1}}}}%
\newcommand{\stSiiii}[1][]{\stFmt{\ifempty{#1}{S''''}{S''''_{#1}}}}%

\newcommand{\stT}[1][]{\stFmt{\ifempty{#1}{T}{T_{#1}}}}%
\newcommand{\stTi}[1][]{\stFmt{\ifempty{#1}{T'}{T'_{#1}}}}%
\newcommand{\stTii}[1][]{\stFmt{\ifempty{#1}{T''}{T''_{#1}}}}%
\newcommand{\stTiii}[1][]{\stFmt{\ifempty{#1}{T'''}{T'''_{#1}}}}%

\newcommand{\stU}[1][]{\stFmt{\ifempty{#1}{U}{U_{#1}}}}%
\newcommand{\stUi}[1][]{\stFmt{\ifempty{#1}{U'}{U'_{#1}}}}%
\newcommand{\stUii}[1][]{\stFmt{\ifempty{#1}{U''}{U''_{#1}}}}%
\newcommand{\stUiii}[1][]{\stFmt{\ifempty{#1}{U'''}{U'''_{#1}}}}%

\newcommand{\stLabP}[1][]{\stFmt{\stLab[#1](\stS[#1])}}%
\newcommand{\stLabPi}[1][]{\stFmt{\stLabi[#1](\stSi[#1])}}%

\newcommand{\stRecVarBase}{\stFmt{\mathbf{t}}}%
\newcommand{\stRecVar}[1][]{\stFmt{\ifempty{#1}{\stRecVarBase}{\stRecVarBase_{#1}}}}%
\newcommand{\stRecVari}[1][]{\stFmt{\ifempty{#1}{\stRecVar'}{\stRecVar'_{#1}}}}%
\newcommand{\stRecVarii}[1][]{\stFmt{\ifempty{#1}{\stRecVar''}{\stRecVar''_{#1}}}}%

\newcommand{\stRoles}[1]{\stFmt{#1}}%
\newcommand{\stRolesSet}[1]{\operatorname{roles}(\stFmt{#1})}%

\newcommand{\stCtx}[1][]{\stFmt{\ifempty{#1}{\mathbb{S}}{\mathbb{S}_{#1}}}}%
\newcommand{\stCtxi}[1][]{\stFmt{\ifempty{#1}{\mathbb{S}'}{\mathbb{S}'_{#1}}}}%
\newcommand{\stCtxHole}[1]{\stFmt{[\;]^{#1}}}%
\newcommand{\stCtxApp}[4]{\stFmt{{#1}[{#2}]^{#3}_{#4}}}%

\newcommand{\stMerge}[2]{\stFmt{\bigsqcap_{#1}{#2}}}%
\newcommand{\stBinMerge}{\mathbin{\stFmt{\sqcap}}}%

\newcommand{\stProj}[2]{%
  {\color{ptColor}\stFmt{#1} {\upharpoonright} \roleFmt{#2}}%
}%

\newcommand{\stSub}{\mathrel{\stFmt{\leqslant}}}%
\newcommand{\stNotSub}{\mathrel{\stFmt{\not\leqslant}}}%
\newcommand{\stSup}{\mathrel{\stFmt{\geqslant}}}%

\definecolor{ptColor}{rgb}{0.20, 0.29, 0.09}%
\newcommand{\ptFmt}[1]{{\color{ptColor}#1}}%

\newcommand{\ptInTag}{?}%
\newcommand{\ptOutTag}{!}%
\newcommand{\ptIn}[2]{\ptFmt{\&{#1}\ifempty{#2}{}{({#2})}}}%
\newcommand{\ptOut}[2]{\ptFmt{\oplus{#1}\ifempty{#2}{}{({#2})}}}%

\newcommand{\ptChoice}[2]{\ptFmt{#1}\ifempty{#2}{}{\ptFmt{({#2})}}}%

\newcommand{\ptSeq}{\mathbin{\!\ptFmt{.}\!}}%
\newcommand{\ptIntC}{\mathbin{\ptFmt{\oplus}}}%
\newcommand{\ptIntSum}[2]{\ptFmt{\oplus_{#1}{#2}}}%
\newcommand{\ptExtC}{\mathbin{\ptFmt{\&}}}%
\newcommand{\ptExtSum}[2]{\ptFmt{\&_{#1}{#2}}}%
\newcommand{\ptGenSum}[2]{\ptFmt{\Large{\maltese}_{#1}{#2}}}%
\newcommand{\ptRec}[2]{\ptFmt{\mu{#1}.{#2}}}%
\newcommand{\ptEnd}{\ptFmt{\mathbf{end}}}%

\newcommand{\ptLab}[1][]{\ptFmt{\ifempty{#1}{\labFmt{m}}{\labFmt{m}_{#1}}}}%
\newcommand{\ptLabFmt}[1]{\ptFmt{\labFmt{#1}}}%

\newcommand{\ptH}[1][]{\ptFmt{\ifempty{#1}{H}{H_{#1}}}}%
\newcommand{\ptHi}[1][]{\ptFmt{\ifempty{#1}{H'}{H'_{#1}}}}%
\newcommand{\ptHii}[1][]{\ptFmt{\ifempty{#1}{H''}{H''_{#1}}}}%
\newcommand{\ptHiii}[1][]{\ptFmt{\ifempty{#1}{H'''}{H'''_{#1}}}}%

\newcommand{\ptRecVar}{\ptFmt{\mathbf{t}}}%
\newcommand{\ptRecVari}[1][]{\ptFmt{\ifempty{#1}{\ptRecVar'}{\ptRecVar'_{#1}}}}%
\newcommand{\ptRecVarii}[1][]{\ptFmt{\ifempty{#1}{\ptRecVar'}{\ptRecVar'_{#1}}}}

\newcommand{\ptMerge}[2]{\ptFmt{\bigsqcap_{#1}{#2}}}%
\newcommand{\ptBinMerge}{\mathbin{\ptFmt{\sqcap}}}%

\newcommand{\ptDual}[1]{\ptFmt{\overline{#1}}}%

\newcommand{\ptSub}{\mathrel{\ptFmt{\leqslant}}}%
\newcommand{\ptNotSub}{\mathrel{\ptFmt{\not\leqslant}}}%
\newcommand{\ptSup}{\mathrel{\ptFmt{\geqslant}}}%

\newcommand{\ptQPfx}[2]{\ptFmt{\stFmt{#1}{\bullet}\ptFmt{#2}}}%

\newcommand{\stEnv}[1][]{\stFmt{\ifempty{#1}{\Gamma}{\Gamma_{#1}}}}%
\newcommand{\stEnvi}[1][]{\stFmt{\ifempty{#1}{\Gamma'}{\Gamma'_{#1}}}}%
\newcommand{\stEnvii}[1][]{\stFmt{\ifempty{#1}{\Gamma''}{\Gamma''_{#1}}}}%
\newcommand{\stEnviii}[1][]{\stFmt{\ifempty{#1}{\Gamma'''}{\Gamma'''_{#1}}}}%
\newcommand{\stEnvEmpty}{\stFmt{\emptyset}}%
\newcommand{\stEnvMap}[2]{\stFmt{\mpFmt{#1}\mathbin{\!:\!}{#2}}}%
\newcommand{\stEnvMapBig}[2]{\stFmt{\mpFmt{#1}\mathbin{:}{#2}}}%
\newcommand{\stEnvComp}{\mathpunct{\stFmt{,}}}%
\newcommand{\stEnvApp}[2]{\stFmt{#1\!\left(\mpFmt{#2}\right)}}%

\newcommand{\stEnvMove}{\mathrel{\stFmt{\to}}}%
\newcommand{\stEnvAnnotOutSym}{\stFmt{\oplus}}%
\newcommand{\stEnvAnnotInSym}{\stFmt{\&}}%
\newcommand{\stEnvAnnotTauSym}{\stFmt{,}}%
\newcommand{\stEnvAnnotGenericSym}{\stFmt{\alpha}}%
\newcommand{\stEnvMoveAnnot}[1]{\mathrel{\stFmt{\xrightarrow{#1}}}}
\newcommand{\stEnvMoveGenAnnot}{\stEnvMoveAnnot{\stEnvAnnotGenericSym}}%
\newcommand{\stEnvMoveInAnnot}[3]{%
  \stEnvMoveAnnot{\stEnvInAnnot{#1}{#2}{#3}}%
}%
\newcommand{\stEnvMoveOutAnnot}[3]{%
  \stEnvMoveAnnot{\stEnvOutAnnot{#1}{#2}{#3}}%
}%
\newcommand{\stEnvMoveCommAnnot}[3]{%
  \stEnvMoveAnnot{\stEnvCommAnnot{#1}{#2}{#3}}%
}%
\newcommand{\stEnvInAnnot}[3]{\stFmt{\mpS:{#1}{\stEnvAnnotInSym}{#2}:{#3}}}%
\newcommand{\stEnvOutAnnot}[3]{\stFmt{\mpS:{#1}{\stEnvAnnotOutSym}{#2}:{#3}}}%
\newcommand{\stEnvCommAnnot}[3]{\stFmt{\mpS:{#1}{\stEnvAnnotTauSym}{#2}:{#3}}}%
\newcommand{\stEnvInAnnotSmall}[3]{\mpS{:}\stFmt{{#1}{\stEnvAnnotInSym}{#2}{:}{#3}}}%
\newcommand{\stEnvOutAnnotSmall}[3]{\stFmt{\mpS{:}{#1}{\stEnvAnnotOutSym}{#2}{:}{#3}}}%
\newcommand{\stEnvCommAnnotSmall}[3]{\stFmt{\mpS{:}{#1}{\stEnvAnnotTauSym}{#2}{:}{#3}}}%
\newcommand{\stEnvCommAnnotSmalli}[3]{\stFmt{\mpSi{:}{#1}{\stEnvAnnotTauSym}{#2}{:}{#3}}}%
\newcommand{\stEnvMoveP}[1]{{#1}\!\!\stEnvMove}%
\newcommand{\stEnvNotMoveP}[1]{{#1}\!\!\not\stEnvMove}%
\newcommand{\stEnvMoveStar}{\mathrel{\stFmt{\stEnvMove{}^{\!\!\!*}}}}%
\newcommand{\stEnvMoveAnnotP}[2]{{#1}\!\!\stEnvMoveAnnot{#2}}%
\newcommand{\stEnvMoveGenAnnotP}[1]{{#1}\!\!\stEnvMoveGenAnnot}%
\newcommand{\stEnvNotMoveGenAnnotP}[1]{{#1}\!\!\not\stEnvMoveGenAnnot}%

\newcommand{\stEnvBeh}[1]{\operatorname{beh}(\stFmt{#1})}%
\newcommand{\stEnvQBeh}[2]{\operatorname{a-beh_{{#1}}}(\stFmt{#2})}%

\newcommand{\stEnvEndPred}{\operatorname{end}}%
\newcommand{\stEnvEndP}[1]{\stEnvEndPred(\stFmt{#1})}%

\newcommand{\stEnvTermPred}{\operatorname{term}}%
\newcommand{\stEnvTermP}[1]{\stEnvTermPred(\stFmt{#1})}%

\newcommand{\stEnvNeverTermPred}{\operatorname{nterm}}%
\newcommand{\stEnvNeverTermP}[1]{\stEnvNeverTermPred(\stFmt{#1})}%

\newcommand{\stEnvDFPred}{\operatorname{df}}%
\newcommand{\stEnvDFP}[1]{\stEnvDFPred(\stFmt{#1})}%

\newcommand{\stEnvLivePred}{\operatorname{live}}%
\newcommand{\stEnvLiveP}[1]{\stEnvLivePred(\stFmt{#1})}%
\newcommand{\stEnvNotLiveP}[1]{\neg\stEnvLivePred(\stFmt{#1})}%

\newcommand{\stEnvKLivePred}[1][k]{\operatorname{live_{\text{${#1}$}}}}%
\newcommand{\stEnvKLiveP}[2][k]{\stEnvKLivePred[{#1}](\stFmt{#2})}%
\newcommand{\stEnvNotKLiveP}[2][k]{\neg\stEnvKLivePred[{#1}](\stFmt{#2})}%

\newcommand{\stEnvLivePlusPred}{\operatorname{live^{+}}}%
\newcommand{\stEnvLivePlusP}[1]{\stEnvLivePlusPred(\stFmt{#1})}%
\newcommand{\stEnvNotLivePlusP}[1]{\neg\stEnvLivePlusPred(\stFmt{#1})}%

\newcommand{\stEnvLivePlusPlusPred}{\operatorname{live^{++}}}%
\newcommand{\stEnvLivePlusPlusP}[1]{\stEnvLivePlusPlusPred(\stFmt{#1})}%
\newcommand{\stEnvNotLivePlusPlusP}[1]{\neg\stEnvLivePlusPlusPred(\stFmt{#1})}%

\newcommand{\stEnvSafePred}{\operatorname{safe}}%
\newcommand{\stEnvSafeP}[1]{\stEnvSafePred(\stFmt{#1})}%
\newcommand{\stEnvNotSafeP}[1]{\neg\stEnvSafePred(\stFmt{#1})}%

\newcommand{\stEnvProjPred}[2]{\operatorname{fproj_{{#1},{#2}}}}%
\newcommand{\stEnvProjP}[3]{\stEnvProjPred{#1}{#2}(\stFmt{#3})}%
\newcommand{\stEnvNotProjP}[3]{\neg\stEnvProjPred{#1}{#2}(\stFmt{#3})}%

\newcommand{\stEnvProjPlainPred}[2]{\operatorname{pproj}_{{#1},{#2}}}%
\newcommand{\stEnvProjPlainP}[3]{\stEnvProjPlainPred{#1}{#2}(\stFmt{#3})}%
\newcommand{\stEnvNotProjPlainP}[3]{\neg\stEnvProjPredPlain{#1}{#2}(\stFmt{#3})}%

\newcommand{\stEnvConsistentPred}{\operatorname{consistent}}%
\newcommand{\stEnvConsistentP}[1]{\stEnvConsistentPred(\stFmt{#1})}%
\newcommand{\stEnvNotConsistentP}[1]{\neg\stEnvConsistentPred(\stFmt{#1})}%

\newcommand{\stEnvEntails}[3]{%
  \stFmt{#1} \vdash \stFmt{\mpFmt{#2} \mathbin{\!:\!} {#3}}%
}%
\newcommand{\stEnvDel}[2]{%
  \stFmt{{#1} \mathbin{\!{\backslash}\!} {\color{black}{#2}}}%
}%
\newcommand{\mpEnv}[1][]{\stFmt{\ifempty{#1}{\Theta}{\Theta_{#1}}}}%
\newcommand{\mpEnvi}[1][]{\stFmt{\ifempty{#1}{\Theta'}{\Theta'_{#1}}}}%
\newcommand{\mpEnvii}[1][]{\stFmt{\ifempty{#1}{\Theta''}{\Theta''_{#1}}}}%
\newcommand{\mpEnvEmpty}{\stFmt{\emptyset}}%
\newcommand{\mpEnvMap}[2]{\stFmt{\mpFmt{#1}{:}\stFmt{#2}}}%
\newcommand{\mpEnvMapBig}[3]{\stFmt{\stFmt{#1}:\stEnvGR{#2}{#3}}}%
\newcommand{\mpEnvComp}{\mathpunct{\stFmt{,}}}%
\newcommand{\mpEnvApp}[2]{\stFmt{#1}\!\left(\mpFmt{#2}\right)}%

\newcommand{\mpEnvEntails}[3]{%
  \stFmt{#1} \vdash \stFmt{\mpFmt{#2} \mathbin{\!:\!} \stFmt{#3}}%
}%

\newcommand{\stJudge}[3]{%
  \stFmt{\ifempty{#1}{#2}{{#1} \cdot {#2}}%
  \mathrel{\mpFmt{\vdash}} \mpFmt{#3}}%
}%

\newcommand{\stJudgeTry}[3]{%
 \stFmt{\ifempty{#1}{#2}{{#1} \cdot {#2}}%
 \mathrel{\mpFmt{\vDash}} \mpFmt{#3}}%
}%
\newcommand{\stJudgeSplit}[3]{%
  \stFmt{\ifempty{#1}{#2}{%
    \begin{array}{@{\hskip 0mm}r@{\hskip 0mm}}
    {#1}%
    \\%
    \vphantom{x} \cdot%
    {#2}%
    \end{array}
  }%
  \mathrel{\mpFmt{\vdash}} \mpFmt{#3}}%
}%
\newcommand{\stJudgeInit}[3]{%
  \stFmt{\ifempty{#1}{#2}{{#1} \cdot {#2}}%
  \mathrel{\mpFmt{\vdash_{\rm init}}} \mpFmt{#3}}%
}%

\newcommand{\mpSessionStack}[2]{%
  \mpFmt{{#1}\mathrel{\!\blacktriangleright\!}{#2}}}%
\newcommand{\mpStack}[1][]{\mpFmt{\ifempty{#1}{\sigma}{\sigma_{#1}}}}%
\newcommand{\mpStacki}[1][]{\mpFmt{\ifempty{#1}{\sigma'}{\sigma'_{#1}}}}%
\newcommand{\mpStackii}[1][]{\mpFmt{\ifempty{#1}{\sigma''}{\sigma''_{#1}}}}%
\newcommand{\mpStackiii}[1][]{\mpFmt{\ifempty{#1}{\sigma'''}{\sigma'''_{#1}}}}%
\newcommand{\mpStackEmpty}{\mpFmt{\epsilon}}%
\newcommand{\mpStackElem}[4]{%
  \mpFmt{\left({#1},{#2},{#3}\ifempty{#4}{}{\langle{#4}\rangle}\right)}}%
\newcommand{\mpStackCons}[2]{\mpFmt{{#1}\mathrel{\!\cdot\!}{#2}}}%

\newcommand{\mpStackSenders}[1]{\operatorname{senders}\!\left({#1}\right)}%

\newcommand{\stEnvQ}[1][]{\stFmt{\ifempty{#1}{\Gamma}{\Gamma_{#1}}}}%
\newcommand{\stEnvQi}[1][]{\stFmt{\ifempty{#1}{\Gamma'}{\Gamma'_{#1}}}}%
\newcommand{\stEnvQii}[1][]{\stFmt{\ifempty{#1}{\Gamma''}{\Gamma''_{#1}}}}%
\newcommand{\stEnvQiii}[1][]{\stFmt{\ifempty{#1}{\Gamma'''}{\Gamma'''_{#1}}}}%
\newcommand{\stEnvQComp}{\mathpunct{\stFmt{,}}}%
\newcommand{\stEnvQCompStack}{\stFmt{\leftsquigarrow}}%
\newcommand{\stEnvQCompStackAppend}{\stFmt{\rightsquigarrow}}%
\newcommand{\stEnvQEmpty}{\stFmt{\emptyset}}%
\newcommand{\stQJudge}[4]{%
  \stFmt{\ifempty{#1}{#2}{{#1} \cdot {#2}}%
  \mathrel{\mpFmt{\vdash}_{\mpFmt{#3}}} \mpFmt{#4}}%
}%

\newcommand{\stEnvQAnnotStackSym}{\stFmt{!}}%
\newcommand{\stEnvQAnnotRecvSym}{\stFmt{,}}%
\newcommand{\stEnvQStackAnnot}[3]{\stFmt{\mpS:{#1}{\stEnvQAnnotStackSym}{#2}:{#3}}}%
\newcommand{\stEnvQRecvAnnot}[3]{\stFmt{\mpS:{#1}{\stEnvQAnnotRecvSym}{#2}:{#3}}}%
\newcommand{\stEnvQStackAnnotSmall}[3]{\stFmt{\mpS{:}{#1}{\stEnvQAnnotStackSym}{#2}{:}{#3}}}%
\newcommand{\stEnvQRecvAnnotSmall}[3]{\stFmt{\mpS{:}{#1}{\stEnvQAnnotRecvSym}{#2}{:}{#3}}}%
\newcommand{\stEnvQMoveAnnot}[1]{\mathrel{\stFmt{\xrightarrow{#1}}}}
\newcommand{\stEnvQMoveGenAnnot}{\stEnvMoveAnnot{\stEnvAnnotGenericSym}}%
\newcommand{\stEnvQMoveStackAnnot}[3]{%
  \stEnvQMoveAnnot{\stEnvQStackAnnotSmall{#1}{#2}{#3}}%
}%
\newcommand{\stEnvQMoveRecvAnnot}[3]{%
  \stEnvQMoveAnnot{\stEnvQRecvAnnotSmall{#1}{#2}{#3}}%
}%

\newcommand{\mpSessionSet}[1][]{%
  \mpFmt{\ifempty{#1}{\mathcal{S}}{\mathcal{S}_{#1}}}}%
\newcommand{\mpSessionSeti}[1][]{%
  \mpFmt{\ifempty{#1}{\mathcal{S}'}{\mathcal{S}'_{#1}}}}%

\newcommand{\stQ}[1][]{\stFmt{\ifempty{#1}{M}{M_{#1}}}}%
\newcommand{\stQi}[1][]{\stFmt{\ifempty{#1}{M'}{M'_{#1}}}}%
\newcommand{\stQii}[1][]{\stFmt{\ifempty{#1}{M''}{M''_{#1}}}}%
\newcommand{\stQEmpty}{\stFmt{\epsilon}}%
\newcommand{\stQCons}[2]{\stFmt{{#1}\mathbin{\!\cdot\!}{#2}}}%
\newcommand{\stQMsg}[3]{\stFmt{{#1}\mathbin{\!\mathbf{!}\!}{#2}\ifempty{#3}{}{({#3})}}}%
\newcommand{\stQMsgWide}[3]{\stFmt{{#1}\mathbin{\mathbf{!}}{#2}\ifempty{#3}{}{({#3})}}}%
\newcommand{\stQProj}[2]{\stFmt{{#1}(\roleFmt{#2})}}%

\newcommand{\stQEquiv}{\mathrel{\stFmt{\equiv}}}%

\newcommand{\stM}[1][]{\stFmt{\ifempty{#1}{\tau}{\tau_{#1}}}}%
\newcommand{\stMi}[1][]{\stFmt{\ifempty{#1}{\tau'}{\tau'_{#1}}}}%
\newcommand{\stMii}[1][]{\stFmt{\ifempty{#1}{\tau''}{\tau''_{#1}}}}%
\newcommand{\stMPair}[2]{\stFmt{({#2}{\mathbin{\mathbf{;\,}}}{#1})}}%

\newcommand{\stEnvQMove}{\mathrel{\stFmt{\to}}}%
\newcommand{\stEnvQMoveStar}{\mathrel{\stEnvQMove{}^{\!\!\!*}}}%
\newcommand{\stEnvQMoveP}[1]{{#1}\!\!\stEnvQMove}%
\newcommand{\stEnvQNotMoveP}[1]{{#1}\!\!\not\stEnvQMove}%
\newcommand{\stEnvQMoveModQ}[1]{\mathrel{\stFmt{\rightarrow}_{{#1}}}}%
\newcommand{\stEnvQMoveAnnotModQ}[2]{%
  \mathrel{\stFmt{\xrightarrow{#1}}_{{#2}}}%
}%
\newcommand{\stEnvQMoveGenAnnotModQ}[1]{%
  \stEnvQMoveAnnotModQ{\stEnvAnnotGenericSym}{#1}%
}%
\newcommand{\stEnvQMoveStackAnnotModQ}[4]{%
  \stEnvQMoveAnnotModQ{\stEnvQStackAnnot{#2}{#3}{#4}}{#1}%
}%
\newcommand{\stEnvQMoveRecvAnnotModQ}[4]{%
  \stEnvQMoveAnnotModQ{\stEnvQRecvAnnot{#2}{#3}{#4}}{#1}%
}%
\newcommand{\stEnvQMoveGenAnnotModQP}[2]{{#1}\!\!\stEnvQMoveGenAnnotModQ{#2}}%
\newcommand{\stEnvQNotMoveGenAnnotModQP}[2]{{#1}\!\!\not\stEnvQMoveGenAnnotModQ{#2}}%
\newcommand{\stEnvQMoveAnnotModQP}[3]{{#1}\!\!\stEnvQMoveAnnotModQ{#2}{#3}}%

\newcommand{\stEnvQMoveModQStar}[1]{%
  \mathrel{\stEnvQMoveModQ{{#1}}^{\!\!\!*}}}%
\newcommand{\stEnvQMoveModQP}[2]{{#2}\!\!\stEnvQMoveModQ{#1}}%
\newcommand{\stEnvQNotMoveModQP}[2]{%
  {#2}/{\kern -0.5em}{\stEnvQMoveModQ{#1}}}%

\newcommand{\stEnvQBMove}[1][]{%
  \ifempty{#1}{%
    \mathrel{\to_{\!(b)}}%
  }{%
    \mathrel{\to_{\!({#1})}}%
  }%
}%

\newcommand{\stEnvQEndPred}[1]{\operatorname{a-end_{#1}}}%
\newcommand{\stEnvQEndP}[2]{\stEnvQEndPred{#1}({#2})}%
\newcommand{\stEnvQDFPred}[1]{\operatorname{a-df_{#1}}}%
\newcommand{\stEnvQDFP}[2]{\stEnvQDFPred{#1}({#2})}%
\newcommand{\stEnvQTermPred}[1]{\operatorname{a-term_{#1}}}%
\newcommand{\stEnvQTermP}[2]{\stEnvQTermPred{#1}({#2})}%
\newcommand{\stEnvQNeverTermPred}[1]{\operatorname{a-nterm_{#1}}}%
\newcommand{\stEnvQNeverTermP}[2]{\stEnvQNeverTermPred{#1}({#2})}%
\newcommand{\stEnvQSafePred}[1]{\operatorname{a-safe_{#1}}}%
\newcommand{\stEnvQSafeP}[2]{\stEnvQSafePred{#1}(\stFmt{#2})}%
\newcommand{\stEnvQNotSafeP}[2]{\neg\stEnvQSafePred{#1}(\stFmt{#2})}%
\newcommand{\stEnvQConsistentPred}[1]{\operatorname{a-consistent_{#1}}}%
\newcommand{\stEnvQConsistentP}[2]{\stEnvQConsistentPred{#1}({#2})}%
\newcommand{\stEnvQLivePred}[1]{\operatorname{a-live_{#1}}}%
\newcommand{\stEnvQLiveP}[2]{\stEnvQLivePred{#1}({#2})}%

\newcommand{\stEnvQKLivePred}[2][k]{\operatorname{a-live_{\text{${#1}-{#2}$}}}}%
\newcommand{\stEnvQKLiveP}[3][k]{\stEnvQKLivePred[{#1}]{#2}(\stFmt{#3})}%
\newcommand{\stEnvQNotKLiveP}[3][k]{\neg\stEnvQKLivePred[{#1}]{#2}(\stFmt{#3})}%

\newcommand{\stEnvQLivePlusPred}[1]{\operatorname{a-live^{+}_{#1}}}%
\newcommand{\stEnvQLivePlusP}[2]{\stEnvQLivePlusPred{#1}(\stFmt{#2})}%
\newcommand{\stEnvQNotLivePlusP}[2]{\neg\stEnvQLivePlusPred{#1}(\stFmt{#1})}%

\newcommand{\stEnvQLivePlusPlusPred}[1]{\operatorname{a-live^{++}_{#1}}}%
\newcommand{\stEnvQLivePlusPlusP}[2]{\stEnvQLivePlusPlusPred{#1}(\stFmt{#2})}%
\newcommand{\stEnvQNotLivePlusPlusP}[2]{\neg\stEnvQLivePlusPlusPred{#1}(\stFmt{#1})}%

\newcommand{\stEnvQLiveKBoundedPred}[2][k]{\operatorname{a-live_{\text{${#2}$}}-bound_{\text{${#1}$}}}}%
\newcommand{\stEnvQLiveKBoundedP}[3][k]{\stEnvQLiveKBoundedPred[{#1}]{#2}(\stFmt{#3})}%
\newcommand{\stEnvQNotLiveKBoundedP}[3][k]{\neg\stEnvQLiveKBoundedPred[{#1}]{#2}(\stFmt{#3})}%

\newcommand{\stEnvQLiveBoundedPred}[1]{\operatorname{a-live_{\text{${#1}$}}-bound}}%
\newcommand{\stEnvQLiveBoundedP}[2]{\stEnvQLiveBoundedPred{#1}(\stFmt{#2})}%
\newcommand{\stEnvQNotLiveBoundedP}[2]{\neg\stEnvQLiveBoundedPred{#1}(\stFmt{#1})}%

\newcommand{\stEnvQKBoundedPred}[2][k]{\operatorname{a-bound_{\text{${#2}$,${#1}$}}}}%
\newcommand{\stEnvQKBoundedP}[3][k]{\stEnvQKBoundedPred[{#1}]{#2}(\stFmt{#3})}%
\newcommand{\stEnvQNotKBoundedP}[3][k]{\neg\stEnvQKBoundedPred[{#1}]{#2}(\stFmt{#3})}%

\newcommand{\stEnvQBoundedPred}[1]{\operatorname{a-bound_{{#1}}}}%
\newcommand{\stEnvQBoundedP}[2]{\stEnvQBoundedPred{#1}(\stFmt{#2})}%
\newcommand{\stEnvQNotBoundedP}[2]{\neg\stEnvQBoundedPred{#1}(\stFmt{#1})}%

\newcommand{\predP}[1][]{\ifempty{#1}{\varphi}{\varphi_{#1}}}%
\newcommand{\predPi}[1][]{\ifempty{#1}{\varphi'}{\varphi'_{#1}}}%
\newcommand{\predPii}[1][]{\ifempty{#1}{\varphi''}{\varphi''_{#1}}}%
\newcommand{\predPApp}[2][]{\ifempty{#1}{\predP}{\predP[{#1}]}\!\left({#2}\right)}%

\newcommand{\relC}{\mathrel{\mathcal{C}}}%
\newcommand{\relD}{\mathrel{\mathcal{D}}}%
\newcommand{\relDi}{\mathrel{\mathcal{D}_1}}%
\newcommand{\relDii}{\mathrel{\mathcal{D}_2}}%
\newcommand{\relR}[1][]{\mathrel{\ifempty{#1}{\mathcal{R}}{\mathcal{R}_{#1}}}}%
\newcommand{\relRi}[1][]{\mathrel{\ifempty{#1}{\mathcal{R}'}{\mathcal{R}'_{#1}}}}%
\newcommand{\relRii}[1][]{\mathrel{\ifempty{#1}{\mathcal{R}''}{\mathcal{R}''_{#1}}}}%

\newcommand{\bisim}{\sim}%
\newcommand{\wbisim}{\approx}%

\newcommand{\OKmark}{{\color{Green}\text{\ding{51}}}}%
\newcommand{\KOmark}{{\color{Red}\text{\ding{55}}}}%

\newcommand{\muCol}[1]{{\color{red}#1}}%
\newcommand{\muFmt}[1]{\muCol{\mathsf{#1}}}%

\newcommand{\muJudge}[2]{{#1} \mathrel{\muCol{\models}} \muCol{#2}}%
\newcommand{\muJudgeNeg}[2]{{#1} \mathrel{\muCol{\not\models}} {#2}}%

\newcommand{\muVar}[1][]{\muCol{\ifempty{#1}{\muFmt{Z}}{\muFmt{Z}_{#1}}}}%
\newcommand{\muVari}[1][]{\muCol{\ifempty{#1}{\muFmt{Z}'}{\muFmt{Z}'_{#1}}}}%

\newcommand{\muData}[1][]{\muCol{\ifempty{#1}{\muFmt{d}}{\muFmt{d}_{#1}}}}%

\newcommand{\muAct}[1][]{\muCol{\ifempty{#1}{\alpha}{\alpha_{#1}}}}%

\newcommand{\muStar}[1]{\mathord{{#1}^{\muCol{*}}}}%
\newcommand{\muOmega}[1]{\mathord{{#1}^{\muCol{\omega}}}}%
\newcommand{\muInf}[1]{\mathord{{#1}^{\muCol{\infty}}}}%

\newcommand{\muActSet}{\muCol{\mathbb{A}\mathrm{ct}}}%
\newcommand{\muActSetStar}{\muStar{\muActSet}}%
\newcommand{\muActSetOmega}{\muOmega{\muActSet}}%
\newcommand{\muActSetInf}{\muInf{\muActSet}}%

\newcommand{\muSomeActSet}[1][]{\muCol{\ifempty{#1}{\mathbb{A}}{\mathbb{A}_{#1}}}}%

\newcommand{\muForall}[2]{\muCol{\forall{#1}\mathbin{\!.\!}{#2}}}%
\newcommand{\muExists}[2]{\muCol{\exists{#1}\mathbin{\!.\!}{#2}}}%
\newcommand{\muNotExists}[2]{\muCol{\not\exists{#1}\mathbin{\!.\!}{#2}}}%

\newcommand{\muPred}[1][]{\muCol{\ifempty{#1}{\phi}{\phi_{#1}}}}%
\newcommand{\muPredi}[1][]{\muCol{\ifempty{#1}{\phi'}{\phi'_{#1}}}}%
\newcommand{\muPredii}[1][]{\muCol{\ifempty{#1}{\phi''}{\phi''_{#1}}}}%

\newcommand{\muNeg}[1]{\muCol{\neg{#1}}}%
\newcommand{\muAnd}{\mathbin{\muCol{\land}}}%
\newcommand{\muBigAnd}[2]{\muCol{\bigwedge_{#1}{#2}}}%
\newcommand{\muPrefix}[2]{\muCol{({#1}){#2}}}%
\newcommand{\muBox}[2]{\muCol{[{#1}]{#2}}}%
\newcommand{\muDiamond}[2]{\muCol{\langle{#1}\rangle{#2}}}%
\newcommand{\muTrue}{\muCol{\top}}%
\newcommand{\muGFP}[2]{\muCol{\nu{#1}\mathbin{\!.\!}{#2}}}%

\newcommand{\muLFP}[2]{\muCol{\mu{#1}\mathbin{\!.\!}{#2}}}%
\newcommand{\muOr}{\mathbin{\muCol{\lor}}}%
\newcommand{\muBigOr}[2]{\muCol{\bigvee_{#1}{#2}}}%
\newcommand{\muFalse}{\muCol{\bot}}%
\newcommand{\muImplies}{\mathbin{\muCol{\Rightarrow}}}%
\newcommand{\muPrefixNeg}[2]{\muCol{(-{#1}){#2}}}%

\newcommand{\muAlways}[1]{\muCol{\square{#1}}}%
\newcommand{\muEventually}[1]{\muCol{\lozenge{#1}}}%
\newcommand{\muUntil}[2]{\muCol{{#1} \mathbin{\muFmt{U}} {#2}}}%

\newcommand{\muVal}[1][]{%
  \muCol{\ifempty{#1}{\mathcal{V}}{\mathcal{V}_{#1}}}
}%
\newcommand{\muVali}[1][]{%
  \muCol{\ifempty{#1}{\mathcal{V}'}{\mathcal{V}'_{#1}}}
}%
\newcommand{\muValApp}[2]{\muCol{{#1}\!\left({#2}\right)}}%
\newcommand{\muValUpdate}[3]{\muCol{{#1}\!\subst{#2}{#3}}}%
\newcommand{\muDenot}[2]{\muCol{\left\lVert{#1}\right\rVert_{#2}}}%

\newcommand{\muWord}[1][]{\muCol{\ifempty{#1}{\sigma}{\sigma_{#1}}}}%
\newcommand{\muWordEmpty}[1][]{\muCol{\epsilon}}%
\newcommand{\muWordHead}[1]{\muCol{\operatorname{\muFmt{hd}}\!\left({#1}\right)}}%
\newcommand{\muWordTail}[1]{\muCol{\operatorname{\muFmt{tl}}\!\left({#1}\right)}}%
\newcommand{\muWordSet}{\muCol{\mathbb{W}}}%

\newcommand{\iruleMPRedComm}{R-Com}%
\newcommand{\iruleMPQuestionRedComm}{R-?Com}%
\newcommand{\iruleMPPromote}{T-$\vDash$}%
\newcommand{\iruleMPRedSt}{R-Struct}%
\newcommand{\iruleMPRedPar}{R-$\mpPar$}%
\newcommand{\iruleMPRedRes}{R-$\mpFmt{\mathbf{\nu}}$}%
\newcommand{\iruleMPRedDef}{R-$\mpFmt{\mathbf{def}}$}%
\newcommand{\iruleMPRedCall}{R-Def}%
\newcommand{\iruleMPRedCongr}{R-$\equiv$}%
\newcommand{\iruleMPRedCtx}{R-Ctx}%
\newcommand{\iruleMPRedCat}{R-Cat}%
\newcommand{\iruleMPRedCan}{R-Can}%
\newcommand{\iruleMPCSel}{C-Sel}%
\newcommand{\iruleMPCQSel}{C-?Sel}%
\newcommand{\iruleMPTQSel}{T?Sel}%
\newcommand{\iruleMPTQBra}{T?Br}%
\newcommand{\iruleMPCQBra}{C-?Br}%
\newcommand{\iruleMPCBra}{C-Br}%
\newcommand{\iruleMPCCat}{C-Cat}%
\newcommand{\iruleMPTCat}{TCat}%

\newcommand{\iruleTCtxOut}{$\stEnv$-$\stFmt{\oplus}$}%
\newcommand{\iruleTCtxIn}{$\stEnv$-$\stFmt{\&}$}%
\newcommand{\iruleTCtxCom}{$\stEnv$-Comm}%
\newcommand{\iruleTCtxRec}{$\stEnv$-$\mu$}%
\newcommand{\iruleTCtxCong}{$\stEnv$-Cong}%

\newcommand{\iruleATCtxCom}{$\stEnv$-AComm}%
\newcommand{\iruleATCtxMsg}{$\stEnv$-AMsg}%

\newcommand{\iruleDTCtxCom}{$\stEnv$-DCom}%
\newcommand{\iruleDSTCtxMsg}{$\stEnv$-DMsg}%
\newcommand{\iruleDTCtxBase}{$\stEnv$-Base}%

\newcommand{\iruleMPErrLabel}{R-Err}%
\newcommand{\iruleMPTimeOutLabel}{R-TOut}%

\newcommand{\iruleMPRedOutA}{R-AOut}%
\newcommand{\iruleMPRedInA}{R-AIn}%
\newcommand{\iruleMPErrLabelA}{R-AErr}%

\newcommand{\iruleSubEnd}{Sub-$\stEnd$}%
\newcommand{\iruleSubBranch}{Sub-$\stExtC$}%
\newcommand{\iruleSubSel}{Sub-$\stIntC$}%
\newcommand{\iruleSubRecL}{Sub-$\stFmt{\mu}$L}%
\newcommand{\iruleSubRecR}{Sub-$\stFmt{\mu}$R}%

\newcommand{\irulePTSubEnd}{PTSub-$\stEnd$}%
\newcommand{\irulePTSubBranch}{PTSub-$\stExtC$}%
\newcommand{\irulePTSubSel}{PTSub-$\stIntC$}%
\newcommand{\irulePTSubRecL}{PTSub-$\stFmt{\mu}$L}%
\newcommand{\irulePTSubRecR}{PTSub-$\stFmt{\mu}$R}%

\newcommand{\iruleMPEnd}{T-$\stEnvEndPred$}%
\newcommand{\iruleMPX}{T-$\mpFmt{X}$}%
\newcommand{\iruleMPSub}{T-sub}%
\newcommand{\iruleMPRaise}{T-raise}%
\newcommand{\iruleMPNarrow}{T-Narrow}%
\newcommand{\iruleMPNil}{T-$\mpNil$}%
\newcommand{\iruleMPDef}{T-def}%
\newcommand{\iruleMPDefTry}{T-def-try}%
\newcommand{\iruleMPCall}{T-call}%
\newcommand{\iruleMPCallTry}{T-$\mpX$-Try}%
\newcommand{\iruleMPCancel}{T-cancel}%
\newcommand{\iruleMPSusp}{Susp}%
\newcommand{\iruleMPKills}{T-kill}%
\newcommand{\iruleMPTryCatch}{T-try}%
\newcommand{\iruleMPTryCatchInit}{T-try-init}%
\newcommand{\iruleMPErr}{T-err}%
\newcommand{\iruleMPOp}{T-Op}%
\newcommand{\iruleMPPar}{T-$\mpPar$}%
\newcommand{\iruleMPResProp}{T-$\mpFmt{\mathbf{\nu}}$}%
\newcommand{\iruleMPRes}{T-$\mpFmt{\mathbf{\nu}}$Classic}%
\newcommand{\iruleMPSafeRes}{T-$\mpFmt{\mathbf{\nu}}$}%
\newcommand{\iruleMPGenRes}{TGen-$\mpFmt{\mathbf{\nu}}$}%
\newcommand{\iruleMPBranch}{T-$\mpFmt{\&}$}%
\newcommand{\iruleMPQSel}{T-$\mpQuestion\mpFmt{\oplus}$}%
\newcommand{\iruleMPQBranch}{T-$\mpQuestion\mpFmt{\&}$}%
\newcommand{\iruleMPSel}{T-$\mpFmt{\oplus}$}%
\newcommand{\iruleMPInit}{T-init}%

\newcommand{\iruleSafeComm}{S-${\stIntC}{\stExtC}$}%
\newcommand{\iruleSafeRec}{S-$\stFmt{\mu}$}%
\newcommand{\iruleSafeMove}{S-$\stEnvMove$}%

\newcommand{\iruleMPLiftA}{TA-Lift}%
\newcommand{\iruleMPParA}{TA-$\mpPar$}%
\newcommand{\iruleMPResAProp}{TA-$\mpFmt{\mathbf{\nu}}$}%
\newcommand{\iruleMPResA}{TA-$\mpFmt{\mathbf{\nu}}$Classic}%
\newcommand{\iruleMPGenResA}{TAGen-$\mpFmt{\mathbf{\nu}}$}%
\newcommand{\iruleMPStackEmptyA}{TA-$\mpStackEmpty$}%
\newcommand{\iruleMPStackA}{TA-$\mpStack$}%

\newcommand{\iruleSafeQComm}{SA-${\stExtC}{\stFmt{!}}$}%
\newcommand{\iruleSafeQExtQ}{SA-${\stExtC}{\stQ}$}%
\newcommand{\iruleSafeQRec}{SA-$\stFmt{\mu}$}%
\newcommand{\iruleSafeQMove}{SA-$\stEnvMove$}%

\newcommand{\iruleLiveBranch}{L-$\stFmt{\&}$}%
\newcommand{\iruleLiveSel}{L-$\stFmt{\oplus}$}%
\newcommand{\iruleLiveRec}{L-$\stFmt{\mu}$}%
\newcommand{\iruleLiveMove}{L-$\stEnvMove$}%

\newcommand{\iruleLivePlusBranch}{L-$\stFmt{\&}^+$}%
\newcommand{\iruleLivePlusSel}{L-$\stFmt{\oplus}^+$}%
\newcommand{\iruleLivePlusRec}{L-$\stFmt{\mu}^+$}%
\newcommand{\iruleLivePlusMove}{L-$\stEnvMove^+$}%
\newcommand{\iruleLivePlusPlusBranch}{L-$\stFmt{\&}^{++}$}%
\newcommand{\iruleLivePlusPlusSel}{L-$\stFmt{\oplus}^{++}$}%

\newcommand{\iruleLiveQBranch}{LA-$\stFmt{\&}$}%
\newcommand{\iruleLiveQOut}{LA-$\stFmt{!}$}%
\newcommand{\iruleLiveQRec}{LA-$\stFmt{\mu}$}%
\newcommand{\iruleLiveQMove}{LA-$\stEnvQMoveModQ{\!}$}%

\newcommand{\iruleLivePQBranch}{LA-$\stFmt{\&}^+$}%
\newcommand{\iruleLivePQOut}{LA-$\stFmt{!}^+$}%

\newcommand{\iruleLivePPQBranch}{LA-$\stFmt{\&}^{++}$}%
\newcommand{\iruleLivePPQOut}{LA-$\stFmt{!}^{++}$}%

\newcommand{\iruleKLiveBranch}{L$k$-$\stFmt{\&}$}%
\newcommand{\iruleKLiveSel}{L$k$-$\stFmt{\oplus}$}%

\newcommand{\iruleKLiveQBranch}{LA$k$-$\stFmt{\&}$}%
\newcommand{\iruleKLiveQOut}{LA$k$-$\stFmt{!}$}%

\newcommand{\iruleTravSetTarget}{TS-Target}%
\newcommand{\iruleTravSetComm}{TS-Comm}%

\newcommand{\iruleTravSetTargetAsync}{TSA-Target}%
\newcommand{\iruleTravSetIOAsync}{TS-IO}%
\newcommand{\AMPST}{{\sf{AMPST}}\xspace}%
\newcommand{\MPST}{{\sf{MPST}}\xspace}%
\newcommand{\BST}{{\sf{BST}}\xspace}%

\newcommand{\kills}[1]{\ensuremath{{#1} \lightning}}

\newcommand{\mpCancel}[2]{\ifempty{#1}{}{\proclit{cancel}({#1})}{\ifempty{#2}{}{.{#2}}}}
\newcommand{\trycatch}[2]{\ensuremath{\proclit{try}\ {#1} \ \proclit{catch}\ {#2}}}
\newcommand{\trycatchbreakequal}[3]{
  \ensuremath{
    \begin{array}{@{}l@{}}
      {#1} = \proclit{try}\ {#2}
    } \\ \qquad \qquad \ensuremath{\proclit{catch}\ {#3}}
  \end{array}
}
\newcommand{\trycatchbreak}[2]{
  \ensuremath{
    \begin{array}{@{}l@{}}
      \proclit{try}\ {#1}
    } \\ \qquad \ensuremath{\proclit{catch}\ {#2}}
  \end{array}
}

\definecolor{ColourGlobal}{rgb}{.816,.125,.565} %
\definecolor{ColourLocal}{rgb}{0,0,.5} %
\definecolor{ColourLabel}{rgb}{.5,0,.5} %
\definecolor{ColourParticipant}{rgb}{0,.5,.5} %
\definecolor{Blue}{rgb}{0,0,1} %
\definecolor{ColourShade}{rgb}{0.94, 1.0, 0.96} %
\definecolor{Yellow}{rgb}{1,1,0}

\newcommand{\roleAlice}{{\color{roleColor}\roleFmt{A}}}%
\newcommand{\roleBob}{{\color{roleColor}\roleFmt{B}}}%
\newcommand{\roleCarol}{{\color{roleColor}\roleFmt{C}}}%

\newcommand{\chanraise}{{\color{roleColor}\proclit{err}}}%

\newcommand{\PSEND}[2]{\ptp{#1} ! \msg{#2}}
\newcommand{\PRECEIVE}[2]{\ptp{#1} ? \msg{#2}}
\newcommand{\msg}[1]{\textit{#1}}

\tikzset{
  every state/.style={minimum size=1pt,inner sep=1.2pt, initial text={}},
  mycfsm/.style={
    font=\scriptsize,
    initial where=left,
    initial distance=0.25cm,
    ->,>=stealth,auto, node distance=0.8cm and 0.8cm,
    scale=1, every node/.style={transform shape},
    baseline=(current  bounding  box.center)
  },
  line/.style = {draw,->, rounded corners=0.07cm,>=latex},
  deliv/.style = {draw, rectangle, rounded corners, fill=white,drop shadow,align=center},
  manip/.style = {draw, ellipse, fill=white,drop shadow,align=center,inner sep=1pt},
  every picture/.style={/utils/exec={\sffamily}}
}


\lstset{
	language=Rust,
	style=colouredRust,
	basicstyle=\ttfamily\small,
	columns=fullflexible,
	xleftmargin=\parindent,
	numbers=left,
	moredelim=[is][\uwave]{~}{~}
}

\begin{abstract}
Communicating systems comprise diverse software components across networks.
To ensure their robustness, modern programming
languages such as Rust provide both strongly typed channels, whose usage
is guaranteed to be \emph{affine} (\emph{at most once}),
and \emph{cancellation} operations over \emph{binary} channels.
For coordinating
components to correctly communicate and synchronise with each other,
we use the structuring mechanism from \emph{multiparty session types},
extending it with affine communication channels and
implicit/explicit cancellation mechanisms.
This new typing discipline,
\emph{affine multiparty session types} ({\textsf{AMPST}}),
ensures \emph{cancellation termination} of multiple,
independently running components
and guarantees that communication will not get stuck due to error or abrupt termination.
Guided by \textsf{AMPST},
we implemented an automated generation tool (\mpstrust)
of Rust APIs associated with cancellation termination algorithms,
by which the Rust compiler auto-detects unsafe programs.
Our evaluation shows that~\mpstrust provides an efficient
mechanism for communication, synchronisation and propagation of
the notifications of cancellation for arbitrary processes.
We have implemented several usecases, including popular application protocols (OAuth, SMTP),
and protocols with exception handling patterns (circuit breaker, distributed logging).

\end{abstract}

\iftoggle{full}{%
}{%
}%
\section{Introduction}
\label{sec:introduction}
\iftoggle{full}{%
}{%
}%
The advantage of message-passing concurrency is well-understood:
it allows cheap horizontal scalability at a time
when technology providers have to adapt and scale
their tools and applications to various devices and platforms.
In recent years, the software industry has seen a shift towards
languages with native message-passing primitives (\eg Go, Elixir and
Rust). Rust, for example, has been named the most loved
programming language in the annual Stack Overflow survey
for five consecutive years (2016-20)~\cite{web:stackOverflow:survey}.
It has been used for the implementation of large-scale concurrent
applications such as the Firefox browser, 
and Rust libraries are part of the Windows Runtime and 
Linux kernel. 
Rust's rise in popularity is due to its efficiency and memory safety.
Rust's dedication to safety, however, does not \emph{yet} extend to communication safety.
Message-passing based software
is as liable to errors as other concurrent programming
techniques~\cite{tuUnderstanding2019} and
communication programming with Rust built-in message-passing abstractions
can lead to a plethora of communication errors~\cite{kokkeRusty2019}.


Much academic research has been done to develop rigorous theoretical frameworks for the verification
of message-passing programs.
One such framework is \emph{multiparty session types}
(\MPST)~\cite{hondaMultiparty2008} --
a type-based discipline that ensures concurrent and distributed systems are \emph{safe by design}.
It guarantees that processes following a predefined
communication protocol (also called a \emph{multiparty session})
are free from communication errors and deadlocks.
Rust may seem a particularly appealing language for the practical embedding
of session types with its message-passing abstractions and affine type system.
The core theory of session types, however,
has serious shortcomings
as its safety is guaranteed under the assumption
that a session should run until its
completion without any failure.
Adapting \MPST in the presence of failure
and realising it in Rust are closely intertwined,
and raise two major challenges:
\textbf{Challenge 1: Affine multiparty session types (\AMPST).}
There is an inherent conflict between the affinity
of Rust and the linearity of session types.
The type system of \MPST guarantees a \emph{linear} usage of
channels,
\ie communication channels in a session must be used \emph{exactly once}.
As noted in~\cite{kokkeRusty2019},
in a distributed system, it is a common behaviour
that a channel or the whole session can be cancelled
\emph{prematurely} --
for example, a node can suddenly get disconnected,
and the channels associated with that node will be dropped.
A naive implementation of
\MPST cancellation, however, will lead to incorrect error notification propagation,
orphan messages, and stuck processes.
The current theory of \MPST does not capture affinity,
hence cannot guarantee
deadlock-freedom and liveness between multiple components in a
realistic distributed system.
Classic multiparty session type systems~\cite{hondaMultiparty2008} do
not prevent any errors related to session cancellation.
An affine multiparty session type system should
(1) prevent infinitely cascading errors, and
(2) ensure deadlock-freedom and liveness
in the presence of session cancellations for arbitrary processes.
Although there are a few works on affine session types, they
are either binary
\cite{mostrousAffine2018,fowlerExceptional2019}
or modelling a very limited
cancellation over a single communication action,
and a general cancellation is not supported~\cite{harveyMultiparty2021}
\iftoggle{full}{%
  (see \S~\ref{subsec:mpst-in-other}, \S~\ref{app:subsec:affine-types}).
}{%
  (see \S~\ref{subsec:mpst-in-other}, and~\cite{fullVersion}).
}%

\textbf{Challenge 2: Realising an affine multiparty channel over binary channels.}
The extension of binary session types to multiparty is usually not
trivial.
The theory assumes multiparty channels, while channels, in practice, are binary.
To preserve
the global order specified by a global protocol,
also called the order of interactions,
when implementing
a multiparty protocol over binary channels,
existing works
\cite{huHybrid2016,neykovaSession2018,scalasLinear2017,castroDistributed2019}
use code generation from a global protocol
to local APIs,
requiring type-casts \emph{at runtime} on the underlying channels,
compromising the type safety
of the host type system.
Implementing \MPST with failure
becomes especially challenging given that cancellation messages
should be correctly propagated across multiple binary channels.

In this work,
we overcome the above two challenges by presenting
a new affine multiparty session types framework for Rust (\AMPST).
We present a shallow embedding of the theory into Rust
by developing a library for safe communication,~\mpstrust.
The library utilises a new communication data structure,
\emph{affine meshed channels},
which stores multiple binary channels without compromising type safety.
A macro operation for exception handling safely propagates failure across all in-scope channels.
We leverage an existing binary session types library, Rust's macros facilities, and optional types
to ensure that communication programs written with~\mpstrust are
\emph{correct-by-construction}.

Our implementation brings three insightful contributions:
(1) multiparty communication safety can be realised
by the native Rust type system (without external validation tools);
(2) top-down and bottom-up approaches can co-exist;
(3) Rust's destructor mechanism can be utilised to propagate session cancellation.
All other works generate not only the types but also
the communication primitives for multiparty channels
which are protocol-specific.
The crucial idea underpinning the novelty of our implementation
is that one can pre-generate the possible communication
actions without having the global protocol;
and then use the types to limit the set of permitted actions.
Without this realisation neither (1), nor (2) is possible.

\myparagraph{\bf Paper Summary and Contributions: }
\begin{enumerate}
  \item[\S~\ref{sec:overview}] outlines the gains
  of programming with \emph{affine meshed channels} by
  introducing our running example, a Video streaming service, and
  its Rust implementation using~\mpstrust.
  \item[\S~\ref{sec:MPST}] establishes the metatheory of~\AMPST.
    We present a core multiparty session $\pi$-calculus
    with session delegation and recursion, together with new constructs for
    exception handling, and affine selection and branching (from Rust
    optional types).
    The calculus enjoys session-fidelity
    (Theorem~\ref{lem:session-fidelity}),
    deadlock-freedom 
    (Theorem~\ref{lem:process-df-live-from-ctx}),
    liveness
    (Theorem~\ref{lem:process-live-from-ctx}),
    and a novel
    cancellation termination property (Theorem~\ref{thm:ctermination}).
  \item[\S~\ref{sec:implementation}] describes
  the main challenges of embedding \AMPST in Rust,
  and the design and implementation of~\mpstrust,
  a library for safe multiparty communication Rust programming.
  \item[\S~\ref{sec:benchmarks}] evaluates the execution and
    compilation overhead of~\mpstrust.
    Microbenchmarks show negligible overhead when compared
    with the built-in unsafe Rust channels,
    provided by \texttt{crossbeam-channel},
    and up to two-fold runtime improvement to
    a binary session types library on protocols with
    high-degree of synchronisation.
    We have implemented, using~\mpstrust,
    examples from the literature,
    and application protocols (see~\cite{fullVersion}).
  \iftoggle{full}{%
}{%
}%
\end{enumerate}

Additionally, \S~\ref{sec:related-works} discusses
related works and \S~\ref{sec:conclusion} concludes.
\iftoggle{full}{%
Appendix includes the proofs of our theorems (\S~\ref{app:AMPST})
and additional technical details of~\mpstrust (\S~\ref{app:implementation}).
}{%
The proofs of our theorems are included in~\cite{fullVersion}.
}%
Our library is available in this public library:
\url{https://github.com/NicolasLagaillardie/mpst_rust_github/}.
An ECOOP artifact is also available.

\iftoggle{full}{%
}{%
}%
\section{Overview: affine multiparty session types (\AMPST) in Rust}
\label{sec:mpst-rust}
\iftoggle{full}{%
}{%
}%
%
\label{sec:overview}
\myparagraph{Framework overview: \AMPST in Rust}
~\Cref{fig:globalProtocol:mpstrustWorkflow:topdown}
depicts the overall design of~\mpstrust.
Our design combines the
top-down~\cite{hondaMultiparty2008} and
bottom-up~\cite{langeCommunicating2015_1}
methodologies of multiparty session types in a single framework.
\iftoggle{full}{%
}{%
Our bottom-up approach is discussed in details in~\cite{fullVersion}.
}%

\begin{figure}[t!]
	\begin{subfigure}[l]{0.48\textwidth}
		\centering
		\includegraphics[scale=0.6]{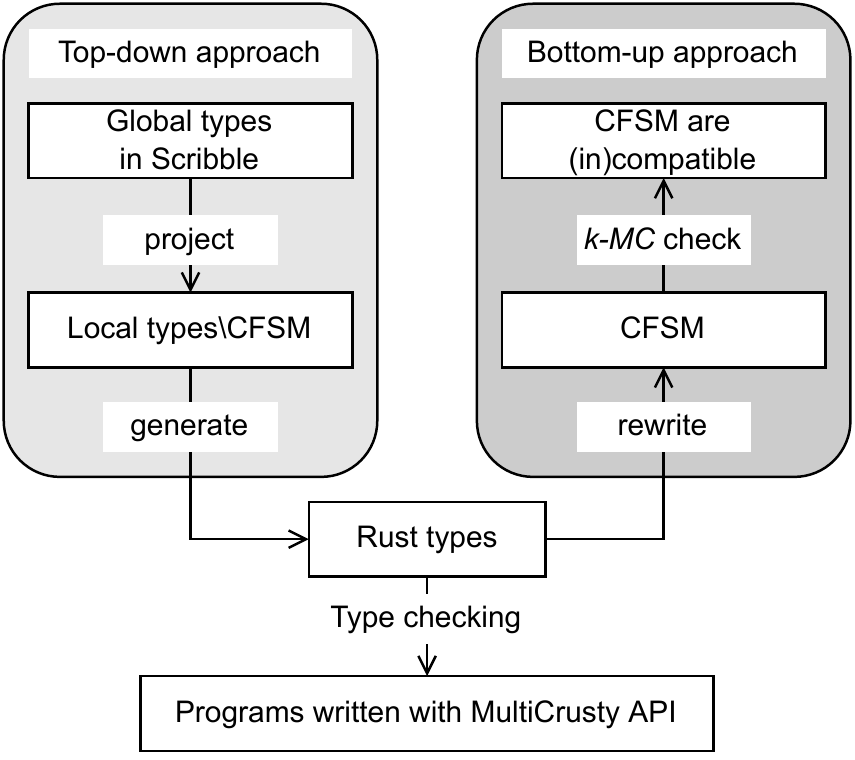}
		\caption{\mpstrust Workflow (top-down)}
		\label{fig:globalProtocol:mpstrustWorkflow:topdown}
	\end{subfigure}
	\hspace*{\fill}
	\begin{subfigure}[r]{0.48\textwidth}
		\iftoggle{full}{%
}{%
}%
		\centering
		\includegraphics[scale=0.4]{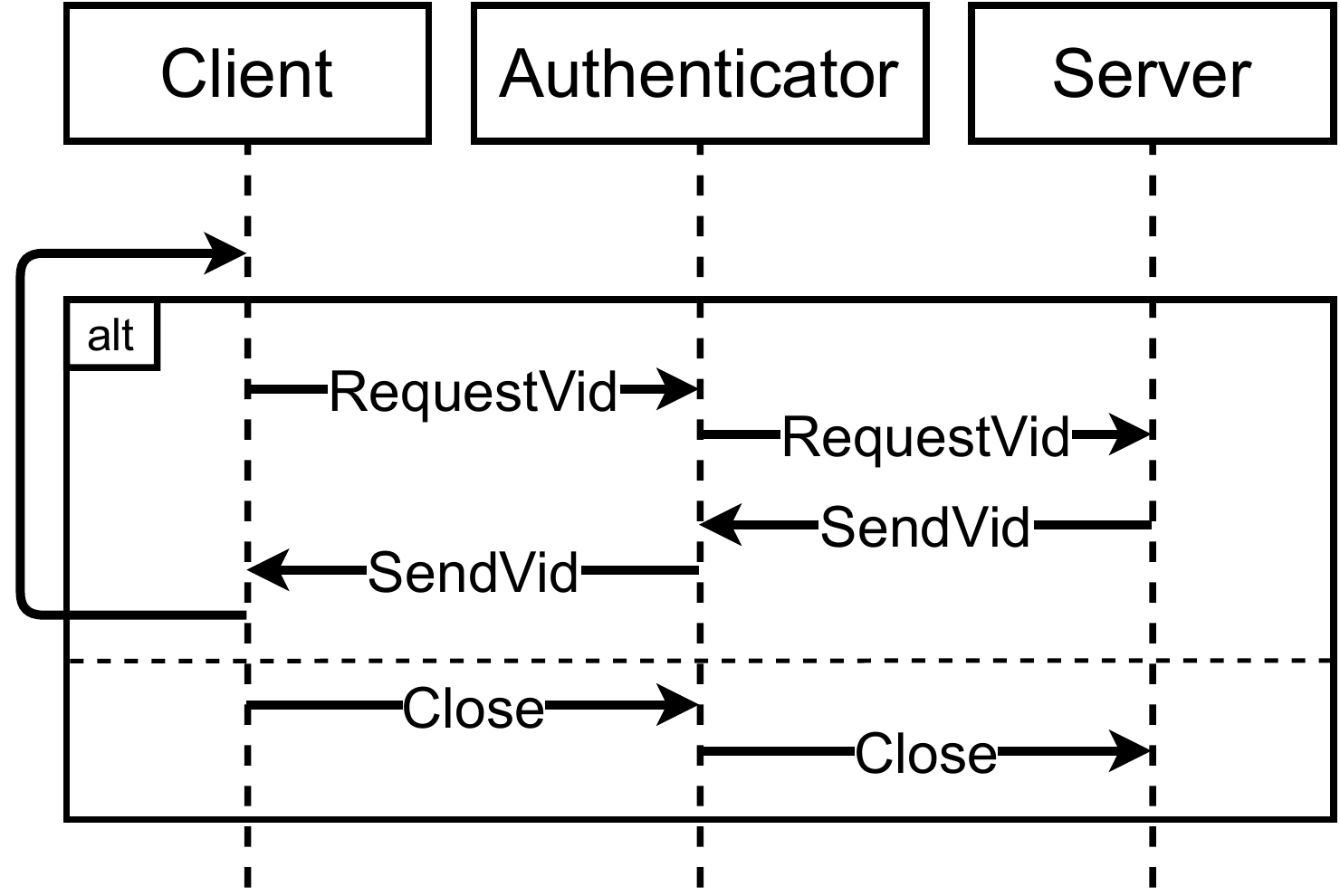}
		\iftoggle{full}{%
}{%
}%
		\caption{Video streaming service usecase}
		\label{fig:globalProtocol:amazonPrimeVideousecase}
	\end{subfigure}
	
	\iftoggle{full}{%
}{%
}%
	\caption{Programming with multiparty session types}
	\label{fig:globalProtocol}
	\iftoggle{full}{%
}{%
}%
\end{figure}

The top-down approach enables {\em correctness-by-construction}
and requires that a developer specifies a global type (hereafter a global protocol) describing the
communication behaviour of the program.
We utilise the Scribble toolchain~\cite{imperialCollegeLondonScribble2021}
for writing and verifying global protocols.
The toolchain projects local types for each role in a protocol.
We have augmented the toolchain to further generate
those local types into Rust types, 
\ie types that stipulate the behaviour of communication channels.

\iftoggle{full}{%
On the other hand, the bottom-up approach requires
a developer to only implement the concurrent programs
and their respective types, the types are then checked for compatibility.
We leverage a recent bounded model checking tool for communicating finite state machines.
In essence, we translate the Rust types written by the developer
to communicating finite state machines (CFSMs) and invoke the
\kmc~tool~\cite{langeVerifying2019} which checks if the two types
are compatible, \ie if their execution may lead to safety errors,
or in other words, no thread gets permanently stuck and
all sent messages are eventually received.
}{%
}%

\iftoggle{full}{%
Finally, our Rust API (\mpstrust API) integrates both approaches,
as illustrated
in~\Cref{fig:globalProtocol:mpstrustWorkflow:topdown}.
}{%
Our Rust API (\mpstrust API) integrates both approaches,
as illustrated
in~\Cref{fig:globalProtocol:mpstrustWorkflow:topdown}.
}%
Developers can choose to either (1) 
write the global protocol and have the Rust types generated, or
(2) write the Rust types manually and check that the types are compatible.
Note that both approaches rely on concurrent programs written with~\mpstrust API,
and both approaches rely on the Rust compiler to type check the concurrent programs against
their respective types. Overall, the framework guarantees that well-typed
concurrent programs implemented
using~\mpstrust API
with Scribble-generated types
or \kmc-compatible types,
will be free from deadlocks,
reception errors, and protocol deviations.

The main primitives of~\mpstrust API are summarised in~\Cref{fig:primitives}.
Next, we briefly explain them through an example.
A more detailed explanation is provided
in~\S~\ref{sec:implementation}.

\begin{figure}[t!]
	\begin{subfigure}{1\textwidth}
		\begin{rustlisting}
// generates at compile-time communication primitives for 3-party affine meshed channels
gen_mpst!(MeshedChannelsThree, A, C, S);*@\label{line:gen}@*\end{rustlisting}
	\end{subfigure}\vfill%
	\iftoggle{full}{%
}{%
}%
	\begin{subfigure}{0.33\textwidth}
	\begin{rustlisting}
fn client(
	s: RecC<i32>,
	i: i32
	) -> R {
	if (i<MAX) {
	let s = choose_c!(s,
	ChoiceA::Video, ChoiceS::Video) *@\label{line:c:choice:video}@*
	let n = get_video(i); *@\label{line:s:panic}@*
	let s = s.send(n)?; *@\label{line:first:send}@*
	let (_,s) = s.recv()?;*@\label{line:first:recv}@*
	client(s, i+1) *@\label{line:s:recurs}@*
	} else {
	let s = choose_c!(s,
	ChoiceA::Close, ChoiceS::Close); *@\label{line:c:choice:close}@*
	s.close()
	}
}\end{rustlisting}
	\caption{\role{C}}
	\label{fig:implementation:role:c}
	\end{subfigure}
	\begin{subfigure}{0.32\textwidth}
\begin{rustlisting}
fn auth(s: RecA<i32>)*@\label{line:start}@*
	-> R {
	offer_mpst!(*@\label{line:offer}@*
	s, {
	ChoiceA::Video(s) *@\label{line:video:beg}@*
	=> {
	let (x,s) = s.recv()?;*@\label{line:recv:c}@*
	let s = s.send(x)?; *@\label{line:send:a}@*
	let (x,s) = s.recv()?;
	let s = s.send(x)?;
	auth(s) *@\label{line:rec}@*
	}, *@\label{line:video:end}@*
	ChoiceA::Close(s) *@\label{line:close:beg}@*
	=> {
	s.close()
	} } *@\label{line:close:end}@*
	)
}\end{rustlisting}
	\caption{\role{A}}
	\label{fig:implementation:role:a}
	\end{subfigure}
	\begin{subfigure}[l]{0.336\textwidth}
\begin{rustlisting}
fn server(s: RecS<i32>)
	-> R {
	offer_mpst!(
	s, {
	ChoiceS::Video(s)
	=> {
	let v = attempt!{{ *@\label{line:attempt:start}@*
		let (x, s) = s.recv()?;
		let f = get_file(x);
		read_video_file(f)
	} catch (e) { *@\label{line:catch}@*
		cancel(s); *@\label{line:attempt:cancel}@*
		panic!("Err: {:?}", e)*@\label{line:attempt:panic}@*
	} }()?; *@\label{line:attempt:end}@*
	let s = s.send(x)?;
	server(s)} *@\label{line:attempt:server}@*
	ChoiceS::Close(s)
	=> {s.close() *@\label{line:server:close}@*
} } ) }\end{rustlisting}
	\caption{\role{S}}
	\label{fig:implementation:role:s}
	\end{subfigure}
	\iftoggle{full}{%
}{%
}%
\begin{tabular}{c@{\;}c@{\;}c}
    %
    \begin{tikzpicture}[mycfsm, node distance=1.5cm and 1.5cm]
      \node[state, initial, initial where=above] (s0) {};
      \node[state, below=of s0] (s1) {};
      \node[state, right=2.3cm of s0] (s3) {};

      \path
      (s0) edge node [below,sloped] {$\PSEND{a}{close}$} (s1)
      (s0) edge node [above] {$\PSEND{a}{ReqVideo}$} (s3)
      (s3) edge [bend left=70] node [above,sloped] {$\PRECEIVE{a}{ResVideo}$} (s0)
      ;
  \end{tikzpicture}
  & \hspace{1.7cm}
    \begin{tikzpicture}[mycfsm, node distance=1.5cm and 1.5cm]
      \node[state, initial, initial where=above] (s0) {};
      \node[state, below=of s0] (s1) {};
      \node[state, right=of s1] (s2) {};
      \node[state, right=of s0] (s3) {};
      \node[state, right=of s3] (s4) {};
      \node[state, below=of s4] (s5) {};

      \path
      (s0) edge node [above,sloped] {$\PRECEIVE{c}{close}$} (s1)
      (s1) edge node [below] {$\PSEND{s}{close}$} (s2)
      (s0) edge node [above] {$\PRECEIVE{c}{ReqVideo}$} (s3)
      (s3) edge node [above] {$\PSEND{s}{ReqVideo}$} (s4)
      (s4) edge node [above,sloped] {$\PRECEIVE{s}{ResVideo}$} (s5)
      (s5) edge node [above, sloped] {$\PSEND{c}{ResVideo}$} (s0)
      ;
    \end{tikzpicture}
    & \hspace{0.4cm}
    \begin{tikzpicture}[mycfsm, node distance=1.5cm and 1.5cm]
        \node[state, initial, initial where=above] (s0) {};
        \node[state, below=of s0] (s1) {};
        \node[state, right=2.3cm of s0] (s3) {};

        \path
        (s0) edge node [below,sloped] {$\PRECEIVE{a}{close}$} (s1)
        (s0) edge node [above] {$\PRECEIVE{a}{ReqVideo}$} (s3)
        (s3) edge [bend left=60] node [above,sloped] {$\PSEND{a}{ResVideo}$} (s0)
        ;
    \end{tikzpicture}
  \end{tabular}

	\iftoggle{full}{%
}{%
}%
	\caption{Rust implementations and respective CFSMs of \role{C} (a), \role{A} (b) and \role{S} (c)}
	\label{fig:implementation}
	\iftoggle{full}{%
}{%
}%
\end{figure}

\iftoggle{full}{%
}{%
}%
\subsection{Example: Video streaming service}
\iftoggle{full}{%
}{%
}%
The \emph{Video streaming service} is a usecase that can
take full advantage
of affine multiparty session types and demonstrate
the need for multiparty channels with cancellation.
Each streaming application connects to servers,
and possibly other devices, to access services and follows
a specific protocol.
To present our design, we use a simplified
version of the protocol,
omitting the authentication part,
illustrated in the diagram
of~\Cref{fig:globalProtocol:amazonPrimeVideousecase}.
The diagram should be read from top to bottom.
The protocol involves three services -- an
\emph{Authenticator} (\role{A}) service,
a \emph{Server} (\role{S}) and a \emph{Client} (\role{C}).
The protocol starts with a choice on the \emph{Client}
to either request a video or end the session.
The first branch is, \emph{a priori},
the main service provided,
\ie request for a video.
The \emph{Client} cannot directly request videos from the
\emph{Server} and has to go
through the \emph{Authenticator} instead.
On the diagram, the choice is denoted as the frame
\texttt{alt} and the choices are separated
by the horizontal dotted line.
The protocol is recursive, and the \emph{Client} can request new videos as many times as needed.
This recursive behaviour is represented by
the arrow going back on the \emph{Client} side
in~\Cref{fig:globalProtocol:amazonPrimeVideousecase}.
To end the session, the \emph{Client} first sends a
\texttt{Close} message to the \emph{Authenticator},
which then subsequently sends a \texttt{Close} message to the
\emph{Server}.

\myparagraph{Affine meshed channels and multiparty session programming with~\mpstrust}
The implementations in~\mpstrust of the three roles are given in~\Cref{fig:implementation}.
They closely follow the behaviour that is prescribed by the protocol.
The global protocol does not explicitly specify cancellation.
However, in a distributed setting, timeout or failure can
happen at any time: a request from a CDN network
or cloud storage to the server might be a timeout or
the result message might be lost.
Our implementation accounts for failure by providing
communication primitives for two different types of
session cancellation, called
\emph{implicit} and \emph{explicit}:
either we run a block of code and upon any error at any point,
we go to the \textbf{catch} branch,
or we explicitly test each step.
\iftoggle{full}{%
	They are displayed in~\Cref{fig:cancellation}
	and the detailed difference between both is
	explained later.
}{%
}%

The implementation of the three concurrent programs starts by generating
all communication primitives for affine channels between three roles.
This is done by the macro
\CODE{gen_mpst!(MeshedChannelsThree, A, C, S)},
see line~\ref{line:gen}
in~\Cref{fig:implementation}.
The macro \CODE{gen_mpst!} takes two kinds of arguments:
the name of the data structure for affine meshed channels,
\CODE{MeshedChannelsThree},
and the name for each role, \CODE{A}, \CODE{C} and \CODE{S}.
\CODE{MeshedChannelsThree} is a string literal
that must be supplied by the developer,
any name can be chosen.
In our case,~\CODE{gen_mpst!} will generate
a data structure called \CODE{MeshedChannelsThree} that can
be used for communication between three participants.
In our example, three roles are provided,
but the macro can handle any number of roles.
Then, using the Rust procedural macro system,
it generates communication primitives for programming
between affine meshed channels.
This generation is done at compile time.
For instance,
the primitive \CODE{s.send(p)} sends a payload \CODE{p} on an
affine meshed channel \CODE{s}.
Note that we do not have to explicitly specify the destination channel,
this is determined from the type:
the stack specifies which binary channel can be used,
regardless of the type of those binary channels.
See~\Cref{fig:UsecaseDeclarationTypes}
for an example of \CODE{MeshedChannels}.

To explain affine meshed channels and all~\mpstrust communication primitives,
we focus on the implementation for \role{A} given in~\Cref{fig:implementation:role:a}.
The implementations of the other roles are similar.
First, line~\ref{line:start} declares an \CODE{auth(s)} function
that is parametric on an affine meshed channel \CODE{s} of
type \CODE{RecA<i32>}, the result type of the function is
irrelevant to our explanation, hence we have simply denoted it by \CODE{R}.
The type \CODE{RecA<i32>},
an alias for the full type described in~\Cref{fig:UsecaseDeclarationTypes},
specifies the operations allowed on \CODE{s}.
As mentioned previously,
this type can be either written by the developer
or generated by Scribble.
We defer the explanation of the (generated) types to \S~\ref{sec:implementation},
\ie the full Rust type is given in~\Cref{fig:UsecaseDeclarationTypes}.
For clarity, here we only give a high-level view
of the behaviour for each channel
by representing its respective local session types as a communicating
finite state machine (CFSM~\cite{cfsm83}), where ! (resp. ?)
denotes sending (resp. receiving).
The CFSMs for each role (channel)
can be seen in~\Cref*{fig:implementation}.
For example,
\CODE{c!ResVideo}
means that~\role{A}
is receiving from the \role{C}
a message labelled as \CODE{Video},
while \CODE{s!ReqVideo} says that~\role{A}
sends a message to \role{S}.

The thread for~\role{A} uses an affine meshed channel \CODE{s} to implement
the given CFSM behaviour. In essence, the meshed channel is implemented as an indexed tuple of binary
channels -- one binary channel for each pair of interacting processes, \ie
a binary channel for~\role{A} and~\role{S} and a binary channel for~\role{A} and~\role{C}.


The implementation starts by realising a choice:
\role{C} broadcasts its choice,
which can either be to request a video at line~\ref{line:c:choice:video}
or to close the connection at line~\ref{line:c:choice:close}
(\Cref*{fig:implementation:role:c}).
Role \norole{C} broadcasts the choice to every other role.
\iftoggle{full}{%
	This can be seen in~\Cref*{app:fig:MPSTMeshedChannelsPrimitives},
	lines~\ref{app:line:broadcast:one} and~\ref{app:line:broadcast:two},
	where the~\CODE{choose_c!} macro encapsulates
	two sends of the new branch, one for each other participant.
}{%
}%
This choice is received by~\role{A},
which will either receive a \CODE{Video} or a \CODE{Close} label.
This behaviour is implemented by the~\mpstrust macro
\CODE{offer_mpst!} (line~\ref{line:offer}),
which is applied to a multiparty channel \CODE{s} and a sum type, either
\CODE{ChoiceA::Video} or \CODE{ChoiceA::Close} here.
The behaviour of each branch in the protocol
is implemented as an anonymous function.
For example, the code in lines~\ref{line:video:beg}
--~\ref{line:video:end} supplies such a function
that implements the behaviour
when \role{C} sends a \CODE{Video} label,
while lines~\ref{line:close:beg} --~\ref{line:close:end}
handle the \CODE{Close} request.
At each step, the channel \CODE{s} is rebound to a new meshed channel \CODE{s}
returned by the respective communication primitive.
For example, the communication primitive \CODE{recv()} at line~\ref{line:recv:c}
is for receiving a value on the binary channel
between \role{A} and \role{C}
that is stored in the meshed channel \CODE{s}.
Note that this is the only primitive available
for that type at that point.
This communication primitive, if everything goes right,
returns a tuple containing the received value and the new meshed channel
to be used for subsequent communications which are bound
resp. to the variable \CODE{x} and \CODE{s}.
An error is returned instead of the tuple
if the reception of the value fails.
Similarly, \CODE{send(x)} in line~\ref{line:send:a}
sends the value \CODE{x} to the process
that implements the \role{S} and rebinds the new meshed channel to \CODE{s}.
Finally, because the protocol is recursive, line~\ref{line:rec}
calls the recursive function \CODE{auth(s)}.

Alternatively, the anonymous function for branch \CODE{Close}
calls the primitive \CODE{close()}
which safely and cleanly closes all binary channels stored
inside the affine meshed channel \CODE{s}.
This last primitive ensures that the type of all the binary channels of \CODE{s}
is~\CODE{Close}
~\CODE{End}: the only primitive implemented
for such meshed channels is~\CODE{close()}.
Forgetting either \CODE{s.close()}
or \CODE{auth(s)} at the end of their
respective branch will throw an error
during compilation because the output type
of the \CODE{auth(s)} function
will be the wrong one.
All communication functions used in the example
(\ie \CODE{s.recv()}, \CODE{s.send()},
\CODE{s.close()}, \CODE{offer_mpst!})
are generated for all roles and all possible interactions
through the macro \CODE{gen_mpst!},
see line~\ref{line:gen}.

The types of the affine meshed channels, as well as the generic types in
the declaration of the~\mpstrust communication functions, enable
compile-time detection of protocol violations.
Examples of protocol violations include
swapping lines~\ref{line:first:recv} and~\ref{line:first:send},
using another communication primitive or using the wrong payload type.
The Rust type system, on the other hand,
ensures that all affine channels are used at most once.
For example, using channel \CODE{s} twice (without rebinding)
will be detected by the compiler.
All the errors mentioned above will be reported as compile-time errors.
In the case that an unexpected runtime error occurs,
all roles are guaranteed to terminate safely.
This is ensured by two mechanisms -- explicit session
cancellation (that can be triggered by the user)
and implicit session cancellation (that is embedded in
the~\mpstrust primitives and the channel destructors).

\begin{table}[t!]
	\centering
	\caption{Primitives provided by~\mpstrust. \CODE{s} is an affine meshed channel;
		\CODE{p} is a payload of a given type;
		$I$ is a subset of all roles in the protocol
		but the current role; $K$ is a subset of all branches;}
\iftoggle{full}{%
}{%
}%
	\resizebox{\textwidth}{!}{
		\begin{tabular}{l l}
Primitives                                                                                                                  & Description \\
\hline
\CODE{let s = s.send(p)?;}                                                                                             &
\makecell[l]{Sends a payload \CODE{p} on a channel \CODE{s}
and assigns the continuation                                                                                                               \\of the session (a new meshed channel) to a new variable \CODE{s}.} \\
\hline
\CODE{let (p, s) = s.recv()?;}                                                                                         &
\makecell[l]{Receives a payload \CODE{p} on channel \CODE{s} and assigns the continuation                                                  \\
of the session to a new variable \CODE{s}.}                                                                                               \\
\hline
\CODE{s.close()}                                                                                                       &
Closes the channel \CODE{s} and returns a unit type.                                                                                      \\
\hline
\CODE{attempt! \{\{ ... \} catch(e) \{ ... \}\}}                                                                       &
\makecell[l]{Attempts to run the first block of code and,
upon error, catches                                                                                                                       \\ the error in the variable \CODE{e} and runs the second block of code.} \\
\hline
\CODE{offer_mpst!\(s, \{} $\texttt{enum}_\texttt{i}::\texttt{variant}_\texttt{k}(\texttt{e}) => \{ ... \}_{k \in K}$ \CODE{\} \)} &
\makecell[l]{Role \norole{i} receives a message label on channel \CODE{s}, and,
\\depending on the label value which should match
\\ one of the variants $\texttt{variant}_\texttt{k}$ of $\texttt{enum}_\texttt{i}$, runs the related block of code.} \\
\hline
\CODE{choose!\(s,} $ \{ \texttt{enum}_\texttt{i}::\texttt{variant}_\texttt{k} \}_{i \in I} $ \CODE{\)}                &
\makecell[l]{Sends the chosen label, which corresponds to $\texttt{variant}_k$,
\\ to all other roles.} \\
		\end{tabular}
	}
	\label{fig:primitives}
	\iftoggle{full}{%
}{%
}%
\end{table}

\myparagraph{Implicit and explicit cancellations}
\label{paragraph:cancellation}
The processes of \role{A} and \role{S} illustrate
resp. implicit and explicit cancellations.
The primitive \CODE{cancel(s)} drops the affine meshed channel \CODE{s},
and its binary channels,
making it inaccessible to other participants.
This is convenient when an error
related to the computation aspects of the program occurs.
For example, In~\Cref{fig:implementation:role:s} the session is cancelled in
line~\ref{line:attempt:cancel} after an error occurs as a result
of reading a corrupted video.
We have used the
\CODE{attempt!\{ \{ ... \} catch(e) \{ ... \} \}} macro
(Rust version of a \CODE{try-catch} block)
in lines~\ref{line:attempt:start} to~\ref{line:attempt:end}
to catch the error message, and explicitly cancel the session.
The macro tries to go through the \CODE{attempt}-block of code,
and upon any error in this block, stops the process
and calls the \CODE{catch(e)}-block with the error message \CODE{e}.
Line~\ref{line:attempt:panic} executes a \CODE{panic!},
which allows a program to terminate immediately and
provide feedback to the caller of the program.
Forgetting to call \CODE{cancel(s)} before \CODE{panic!}
will result in the same outcome as when both \CODE{cancel(s)}
and \CODE{panic!} drop \CODE{s}.
Forgetting both will throw
an error because the output type will
not match the one of \CODE{fn server(s)},
unless replaced with an \CODE{Ok(())}.
In any cases,
an error will be thrown on other threads
linked to other roles
because \role{S}'s sessions
are inaccessible
in the \CODE{catch(e)}-block.
Alternatively, we explain implicit cancellation as
implemented by \role{A} in~\Cref{fig:implementation:role:a}.
The construct \CODE{let x = f()?},
as seen in line~\ref{line:recv:c},
is Rust’s \emph{monadic bind} notation for programs
and functions that may return errors:
their usual output type is \CODE{Result<T, Error>} where
\CODE{T} is the expected type if everything goes right
and \CODE{Error} is the error type returned.
For any program and function returning such type, the
users have to \emph{unwrap} it.
The two usual ways of doing so are by using the \CODE{?}-operator, or
by pattern matching on the result using \CODE{match}.
In our case, if \CODE{recv()} succeeds,
the \CODE{?}-operator unpacks
the result and returns the tuple containing
the received payload and the continuation.
If \CODE{recv()} fails, the \CODE{?}-operator short-circuits,
skips the rest of the statements, and returns the error.
We use this mechanism to catch any session cancellation.
In the case that a \CODE{recv()} (or \CODE{send()}) does not succeed,
the implementation of the underlying communication primitive will cancel the
channel and broadcast the cancellation to all
other binary channels that are part of the session.

Finally, we look at the implementation of \role{C}
to demonstrate the final mechanism of session cancellation.
For this purpose, we have to comment out
lines~\ref{line:first:send} --~\ref{line:s:recurs}
in~\Cref{fig:implementation:role:c} or replace them with a \CODE{panic!} as
to simulate a wrongly implemented \role{C}.
With such modification, this function will still compile despite the
protocol not being fully implemented
(since the last received action from \role{A} is missing,
the meshed channel \CODE{s} will be dropped prematurely).
Even in this case,~\mpstrust ensures that
all processes will terminate safely, \ie
all parties are notified that an affine channel has been dropped.
Prematurely dropping a channel
can happen due to
incorrectly implemented behaviour (as we demonstrated above),
or by unhandled user error,
for example, the function \CODE{get_video()} in
line~\ref{line:s:panic} can invoke a \CODE{panic!}
because there is no video associated with the index \CODE{i}.
Safe session termination is realised by customising
the native destructor \CODE{Drop} in Rust,
as proposed for binary meshed channels by~\cite{kokkeRusty2019}.
When an affine meshed channel goes out of scope,
the channel destructor is called, the session is cancelled,
dropping every channel value used in the session,
and only then is the memory deallocated.

Memory management should not pose a problem in our case.
Our library uses only the safe fragment of Rust.
This ensures that variables that are out of scope are automatically collected.
We utilise this mechanism to ensure that closed and
cancelled channels are collected in the same way.
Hence, memory leaks are ruled out.

In short, a session can be cancelled for three reasons:
(1) an error affecting the computation aspects of the program, as in~\Cref{fig:implementation:role:s};
(2) an error during communication, \eg a timeout on a channel, as in~\Cref{fig:implementation:role:a};
or (3) a premature drop of the affine meshed channel due to incorrect implementation,
as in~\Cref{fig:implementation:role:c}.
Our mechanisms for session cancellation cover all the above cases.
In this way, our framework provides \emph{affine multiparty session compliance} by
ensuring that (1) if all results are returned without failure,
the processes follow the given Scribble global protocol
(Theorem~\ref{lem:session-fidelity}) or (2) once a cancellation happens, all processes in the same
session terminate with an error (Theorem~\ref{thm:ctermination}).
We have proven the above results by formalising affine meshed channels in
an extension of a multiparty $\pi$-calculus.

\iftoggle{full}{%
}{%
}%
\section{Affine multiparty session processes for Rust programming}
\label{sec:MPST}
\iftoggle{full}{%
}{%
}%
\subsection{Affine multiparty session processes}
\label{subsec:processes}
\iftoggle{full}{%
}{%
}%
Our calculus (\AMPST) is an extension of a full multiparty session
$\pi$-calculus~\cite{scalasLess2019} which includes session delegation
(channel passing) and session recursion.
We shade additions to~\cite{scalasLess2019} in~\colorbox{ColourShade}{this colour}.

\begin{restatable}[]{definition}{defMPSTSyntax}\rm
  \label{def:mpst-syntax-terms}
  The \textbf{affine multiparty session $\pi$-calculus} (\AMPST) is defined as follows:%

  \smallskip%
  \centerline{$%
      \begin{array}{r@{\hskip 2mm}c@{\hskip 2mm}l@{\hskip 2mm}l}
        \textstyle%
        \mpC, \mpD%
         & \coloncolonequals                                                       & %
        \mpFmt{x} \bnfsep \mpChanRole{\mpS}{\roleP}%
        \quad\quad
        \colorbox{ColourShade}{$\mpDagger$}%
        \coloncolonequals
        \colorbox{ColourShade}{$\emptyset \bnfsep \mpQuestion$}
         & \mbox{\footnotesize(variable, channel with role $\roleP$, error, flag)}
        \\[1mm]
        \mpP, \mpQ
         & \coloncolonequals                                                       & %
        \mpNil \;\bnfsep\; \mpP \mpPar \mpQ \;\bnfsep\; \mpRes{\mpS}{\mpP}%
         &
        \mbox{\footnotesize(inaction, composition, restriction)}
        \\
         &
         &
        \colorbox{ColourShade}{$\mpQuestionSel{\mpC}{\roleQ}{\mpLab}{\mpD}{\mpP}$}%
        \bnfsep
        \colorbox{ColourShade}{$\mpQuestionBranch{\mpC}{\roleQ}{i \in I}{\mpLab[i]}{x_i}{\mpP[i]}$}%
         &
        \mbox{\footnotesize(affine selection, branching $I \neq \emptyset$)}
        \\
         &                                                                         &
        \;\;\;
        \mpSel{\mpC}{\roleQ}{\mpLab}{\mpD}{\mpP}%
        \;\bnfsep\;\;\;\;
        \mpBranch{\mpC}{\roleQ}{i \in I}{\mpLab[i]}{x_i}{\mpP[i]}%
         &
        \mbox{\footnotesize(selection, branching $I \neq \emptyset$)}
        \\
         &                                                                         &
        \;\;\;
        \mpDefAbbrev{\mpDefD}{\mpP}%
        \bnfsep%
        \;\;\;
        \mpCall{\mpX}{\widetilde{\mpC}}
         & \mbox{\footnotesize(process definition, process call)}
        \\
         &                                                                         &
        \colorbox{ColourShade}{$\trycatch{P}{Q}$}%
        \bnfsep%
        \colorbox{ColourShade}{$\mpCancel{\mpC}{\mpP}$}
        \bnfsep%
        \colorbox{ColourShade}{$\kills{\mpS}$}
        %
         &
        \mbox{\footnotesize(catch, cancel, kill)}
        \\[1mm]
        \mpDefD%
         & \coloncolonequals                                                       & %
        \mpJustDef{\mpX}{\widetilde{x}}{P}
         &
        \mbox{\footnotesize(declaration of process variable $\mpX$)}
        \\[1mm]
      \end{array}
    $}
\end{restatable}
\noindent
A set $\PSet$ denotes \textbf{participants}: $\PSet =\{\roleP, \roleQ, \roleR, \dots\}$,
and $\ASigma$ is a set of alphabets.
A \textbf{channel $\mpC$} can be either a variable or
a \textbf{channel with role $\mpChanRole{\mpS}{\roleP}$},
\ie a multiparty communication endpoint %
whose user plays role $\roleP$ in the session $\mpS$.
$\widetilde{\mpC}$ denotes a vector $c_1c_2\dots c_n$ ($n\geq 1$)
and similarly for $\widetilde{x}$ and $\widetilde{s}$.

The two processes with $\mpQuestion$ model the option \CODE{?}-operator in
Rust.
Process $\mpQuestionSel{\mpC}{\roleQ}{\mpLab}{\mpD}{\mpP}$ performs an %
\textbf{affine selection (internal choice)} %
towards role $\roleQ$, %
using the channel $\mpC$: %
if the \emph{message label} $\mpLab$ with the \emph{payload} channel
$\mpD$ is successfully sent, %
then the execution continues as $\mpP$; otherwise (if the receiver
has failed or timeout), it triggers an exception.
The %
\textbf{affine branching (external choice)}
$\mpQuestionBranch{\mpC}{\roleQ}{i \in I}{\mpLab[i]}{x_i}{\mpP[i]}$ %
uses channel $\mpC$ to wait for a message %
from role $\roleQ$: %
if a message label $\mpLab[k]$ with payload $\mpD$ is received %
(for some $k \!\in\! I$), %
then the execution continues as $\mpP_k$, %
with $x_k$ replaced by $\mpD$; if not received, it triggers an exception.
Note that message labels $\mpLab[i]$ are pairwise distinct and
their order is irrelevant, and %
variable $x_i$ is bound with scope $\mpP_i$.

The following two failure handling processes follow the program behaviour of
\Cref{fig:implementation:role:s}.
The \textbf{try-catch} process, $\trycatch{P}{Q}$, consists of
a \emph{try process} $P$ which is ready to communicate with
parallel composed one; and a \emph{catch process} $Q$ which
becomes active when a cancellation  or an error happens.
The \textbf{cancel} process, $\mpCancel{\mpC}{\mpP}$, cancels
other processes whose communication channel is $c$.
The \textbf{kill} $\kills{\mpS}$ kills
all processes with session $\mpS$ and \textbf{is generated only at runtime}
from affine or cancel processes.

The other syntax is from~\cite{scalasLess2019}.
The \textbf{inaction $\mpNil$} represents a terminated process %
(and is often omitted).
The \textbf{parallel composition $\mpP \mpPar \mpQ$} %
represents two processes that can execute concurrently, %
and potentially communicate.
The \textbf{session restriction $\mpRes{\mpS}{\mpP}$} %
declares a new session $\mpS$ with a scope limited to process $\mpP$.
The linear \textbf{selection} %
and the linear %
\textbf{branching}
can be understood as their affine versions
but without failure handling.
\textbf{Process definition}, $\mpDef{\mpX}{\widetilde{x}}{\mpP}{\mpQ}$ and %
\textbf{process call} $\mpCall{\mpX}{\widetilde{\mpC}}$ model
recursion: the call invokes $\mpX$ %
by expanding it into $\mpP$, %
and replacing its formal parameters with the actual ones.

Linear or affine branching and selection are denoted as either
$\mpDaggerBranch{\mpC}{\roleQ}{i \in I}{\mpLab[i]}{x_i}{\mpP[i]}$
\phantomsection
\label{def:sbj}
and $\mpDaggerSel{\mpC}{\roleQ}{\mpLab}{\mpD}{\mpP}$.
We use $\fv{P}/\fc{P}$ and $\dpv{P}/\fpv{P}$ to denote
\emph{free variables}/\emph{channels}
and \emph{bound}/\emph{free process variables} of $P$.
We call a process $P$ such that $\fv{P}=\fpv{P}=\emptyset$ \emph{closed}. 
A~\emph{set of subjects of $P$}, written $\sbj{P}$,
is defined as:
$\sbj{\mpNil}=\emptyset$;
$\sbj{\mpP \mpPar \mpQ}= \sbj{\mpP}\cup \sbj{\mpQ}$;
$\sbj{\mpRes{\mpS}{\mpP}}=\sbj{P}\setminus
\{\mpChanRole{\mpS}{\roleP[i]}\}_{i\in I}$;
$\sbj{\mpDaggerBranch{\mpC}{\roleQ}{i \in I}{\mpLab[i]}{x_i}{\mpP[i]}}
=\sbj{\mpDaggerSel{\mpC}{\roleQ}{\mpLab}{\mpD}{\mpP}}=\{\mpC\}$;
$\sbj{\mpDef{\mpX}{\widetilde{x}}{\mpP}{\mpQ}} =
\sbj{\mpQ}\cup \sbj{\mpP} \setminus \{\tilde{x}\}$ with
$\sbj{\mpCall{X}{\widetilde{c}}}= \sbj{\mpP\subst{\tilde{x}}{\tilde{c}}}$;
$\sbj{\trycatch{\mpP}{\mpQ}} = \sbj{\mpP}$;
and $\sbj{\mpCancel{\mpC}{\mpP}}=\{c\}$.


The set of subjects is the key definition which enables us to define 
the typing system for the \CODE{try-catch} process with recursive
behaviours. 
 
\begin{restatable}[Subjects of processes]{example}{exampleSubjectsOfProcesses}
  \label{ex:setOfSubjects}%
  \iftoggle{full}{%
}{%
}%
  Assume
$R_1=\mpDef{\mpX}{x}{\mpSel{x}{q}{m}{d}{\mpNil}}{\mpCall{\mpX}{\mpC}}$ which
repeats the action at $c$ and emits a message
$d$
with label
repeatedly interacting with the dual input
(but reduction with this process only happens if there
is a corresponding input at $c$, \ie on-demand).
We calculate $\sbj{R_1}$ as:
\[
    \begin{array}{l}
        \sbj{\mpDef{\mpX}{x}{\mpSel{x}{q}{m}{d}{\mpNil}}{\mpCall{\mpX}{\mpC}}}
        = \sbj{\mpCall{\mpX}{\mpC}} \cup \sbj{\mpShortAngleDef{\mpX}{x}{\mpSel{x}{q}{m}{d}{\mpNil}}} \\
        = \sbj{\mpCall{\mpX}{\mpC}}= \sbj{(\mpSel{x}{q}{m}{d}{\mpNil})\subst{x}{c}} = \{c\}
    \end{array}
\]
Another example is: 
        $\sbj{
            \trycatch{
                \mpSel{x}{\roleQ}{m}{d}{\mpNil}
            }{
                \mpCancel{\mpChanRole{x}{\roleQ}}{\mpNil}
            }
        }
        =
        \sbj{
            \mpSel{x}{\roleQ}{m}{d}{\mpNil}
        } 
        =
        \{
        x
        \}$.
  \iftoggle{full}{%
}{%
}%
\end{restatable}

\begin{restatable}[Syntax and semantics]{remark}{remarkSyntaxAndSemantics}
  \iftoggle{full}{%
}{%
}%
  \AMPST extends~\MPST incorporating some design choices
  from~\cite{mostrousAffine2018},
  aiming to distil the implementation essence of~\mpstrust.
  The design of our \CODE{try-catch} process
  follows the binary affine session types in~\cite{mostrousAffine2018},
  but models more cancellations for arbitrary processes
  with affine branchings/selections and cancel processes
  non-deterministically
  (whose semantics follow the implementation
  behaviours, see \S~\ref{sec:exceptions}).
  %
  We list the essential differences from~\cite{mostrousAffine2018}.
  \begin{itemize*}
    \item[(1)] (Nondeterministic failures)
    The kill process is a runtime syntax and generated only
    during reductions unlike~\cite{mostrousAffine2018}.
    Our calculus also allows \emph{nondeterministic failures} caused by
    either (1) affine selection/branching or (2) \CODE{try-catch} processes.
    \iftoggle{full}{%
      See~\Cref{app:ex:reduction}.
    }{%
      See~\cite{fullVersion} for examples.
    }%

    \item[(2)] (Recursion parameterised by linear names) \
    One of the novelties of our formalism which is not found in
    \cite{mostrousAffine2018} is
    a combination of session recursions, affinity,
    and interleaved sessions,
    \ie the \proclit{def} agents
    (linearly parameterised recursions),
    which are the most technical part when
    designing the typing system with \CODE{try-catch} processes.
    The combination of all features is absent
    from~\cite{mostrousAffine2018,fowlerExceptional2019,harveyMultiparty2021}:
    see \S~\ref{sec:related-works} for more detailed comparisons.
  \end{itemize*}
  \iftoggle{full}{%
}{%
}%
\end{restatable}

\begin{figure*}[t]
\small
$
    \begin{array}{rl}
      \colorbox{ColourShade}{\inferrule{\iruleMPRedComm}} & %
      \colorbox{ColourShade}{$\mpEtx_{1}[\mpDaggerBranch{\mpChanRole{\mpS}{\roleP}}{\roleQ}{i \in I}{%
              \mpLab[i]}{x_i}{\mpP[i]}]%
          \,\mpPar\,%
          \mpEtx_{2}[\mpDaggerSel{\mpChanRole{\mpS}{\roleQ}}{\roleP}{\mpLab[k]}{%
              \mpChanRole{\mpSi}{\roleR}%
            }{\mpQ}]%
          \;\;\mpMove\;\;%
          \mpP[k]\subst{\mpFmt{x_k}}{\mpChanRole{\mpSi}{\roleR}}%
          \,\mpPar\,%
          \mpQ%
          \quad%
          \text{if\, $k \!\in\! I$}$}%
      \\[2mm]%
      \colorbox{ColourShade}{\inferrule{\iruleMPCQSel}}   & %
      \colorbox{ColourShade}{$
          \mpQuestionSel{\mpChanRole{\mpS}{\roleP}}{\roleQ}{\mpLab}{
            \mpChanRole{\mpS'}{\roleR}}{\mpP}
          \mpMove
          \mpSel{\mpChanRole{\mpS}{\roleP}}{\roleQ}{\mpLab}{
            \mpChanRole{\mpS'}{\roleR}}{\mpP}
          \mpPar \kills{\mpS}$}
\\[2mm]
    \colorbox{ColourShade}{$\inferrule{\iruleMPTQSel}$} &
    \colorbox{ColourShade}{$\trycatch{\mpQuestionSel{\mpChanRole{\mpS}{\roleP}}{\roleQ}{\mpLab}{
            \mpChanRole{\mpS'}{\roleR}}{\mpP}
          }{Q}
          \mpMove Q \mpPar \kills{\mpS}$}
      \\[2mm]%
      \colorbox{ColourShade}{$\inferrule{\iruleMPCSel}$}  &
      \colorbox{ColourShade}{$\mpSel{\mpChanRole{\mpS}{\roleP}}{\roleQ}{\mpLab}{
            \mpChanRole{\mpS'}{\roleR}}{\mpP}
          \mpPar \kills{\mpS}
          \mpMove
          \mpP \mpPar \kills{\mpS} \mpPar \kills{\mpS'}$}
      \\[2mm]%
      \colorbox{ColourShade}{$\inferrule{\iruleMPCQBra}$} & %
      \colorbox{ColourShade}{$\mpQuestionBranch{\mpChanRole{\mpS}{\roleP}}{\roleQ}{i \in
            I}{\mpLab[i]}{x_i}{\mpP[i]}%
          \mpMove
          \mpBranch{\mpChanRole{\mpS}{\roleP}}{\roleQ}{i \in
            I}{\mpLab[i]}{x_i}{\mpP[i]}
          \mpPar \kills{\mpS}$}
      \\[2mm]%
    \colorbox{ColourShade}{$\inferrule{\iruleMPTQBra}$} &
    \colorbox{ColourShade}{$\trycatch{
\mpQuestionBranch{\mpChanRole{\mpS}{\roleP}}{\roleQ}{i \in
            I}{\mpLab[i]}{x_i}{\mpP[i]}
          }{Q}
          \mpMove Q \mpPar \kills{\mpS}$}
      \\[2mm]%
      \colorbox{ColourShade}{$\inferrule{\iruleMPCBra}$}  & %
      \colorbox{ColourShade}{$\mpBranch{\mpChanRole{\mpS}{\roleP}}{\roleQ}{i \in
            I}{\mpLab[i]}{x_i}{\mpP[i]} \mpPar \kills{\mpS}%
          \mpMove
          \mpRes{\mpS'}(\mpP[k]\subst{\mpFmt{x_k}}{\mpChanRole{\mpS'}{\roleR}}
          \mpPar \kills{\mpS'})
          \mpPar \kills{\mpS}
          \quad \text{$\mpS'\not\in \fc{P_k}, k\in I$}$}
      \\[2mm]%
      \colorbox{ColourShade}{\inferrule{\iruleMPRedCan}} &
      \colorbox{ColourShade}{$\mpEtx{}[\mpCancel{\mpChanRole{\mpS}{\roleP}}{Q}]
          \mpMove \kills{s}\mpPar Q$}
\quad      \colorbox{ColourShade}{$\inferrule{\iruleMPCCat}$}
      \colorbox{ColourShade}{$\trycatch{P}{Q} \mpPar \kills{\mpS}
          \mpMove Q \mpPar \kills{\mpS}
          \quad \text{$\exists \roleR.\ \mpChanRole{\mpS}{\roleR}=\sbj{P}$}$}
      \\[2mm]%
      \inferrule{\iruleMPRedCall}                         & %
      \mpDef{\mpX}{x_1,\dots,x_n}{\mpP}{(%
        \mpCall{\mpX}{%
          \mpChanRole{\mpS[1]}{\roleP[1]}, \dots,%
          \mpChanRole{\mpS[n]}{\roleP[n]}%
        }%
        \mpPar%
        \mpQ%
        )%
      }%
      \\[2mm]%
                                                          & %
      \qquad%
      \mpMove %
      \mpDef{\mpX}{x_1,..,x_n}{\mpP}{(%
        \mpP\subst{\mpFmt{x_1}}{\mpChanRole{\mpS[1]}{\roleP[1]}}%
        \cdot\cdot
        \subst{\mpFmt{x_n}}{\mpChanRole{\mpS[n]}{\roleP[n]}}%
        \mpPar%
        \mpQ%
        )%
      }%
      \\[2mm]%
      \inferrule{\iruleMPRedCtx}                          & %
      \mpP \mpMove \mpPi%
      \;\;\text{implies}\;\;%
      \mpCtx{}[\mpP] \mpMove \mpCtx{}[\mpPi]
      \quad\quad
      \inferrule{\iruleMPRedSt}\quad
      \mpP \equiv \mpPi \mpMove \mpQi \equiv \mpQ%
      \;\;\text{implies}\;\;%
      \mpP \mpMove \mpQ
      %
    \end{array}
$

  \caption{%
    \AMPST $\pi$-calculus reduction between closed processes 
    (we highlight~\colorbox{ColourShade}{the new rules}~from~\cite{scalasLess2019})
      \iftoggle{full}{%
        Additional rules can be found in our Appendix~\Cref{app:fig:structure}.
      }{%
      }%
    }%
\label{fig:mpst-pi-semantics}%
\label{fig:structure}%
\iftoggle{full}{%
}{%
}%
\end{figure*}

\begin{restatable}[Semantics]{definition}{definitionOperationalSemantics}
  \label{def:mpst-proc-context}%
  \label{def:mpst-pi-reduction-ctx}%
  \label{def:mpst-pi-semantics}%
  \label{def:mpst-pi-error}%
  A \textbf{try-catch context} $\mpEtx$ is: %
  \;
  \(%
  \mpEtx \,\coloncolonequals\,%
  \trycatch{\mpEtx}{P}%
  \bnfsep%
  \mpEtxHole%
  \)
  and a \textbf{reduction context} $\mpCtx$ is:
  \(\mpCtx \,\coloncolonequals\,
  \mpRes{\mpS}{\mpCtx}%
  \bnfsep%
  \mpDefAbbrev{\mpDefD}{\mpCtx}%
  \bnfsep%
  \mpCtx \mpPar P%
  \bnfsep%
  P \mpPar \mpCtx%
  \bnfsep%
  \mpEtxHole%
  \).
  %
  %
  \ \textbf{Reduction $\mpMove$} %
  is inductively defined %
  in \Cref{fig:mpst-pi-semantics}, which uses the \textbf{structural
    congruence $\equiv$} which is defined by 
\colorbox{ColourShade}{$\kills{\mpS} \mpPar
          \kills{\mpS}
          \;\equiv\;%
          \kills{\mpS}$} and 
      \colorbox{ColourShade}{$\mpRes{\mpS}{\kills{\mpS}}%
          \;\equiv\;%
          \mpNil$}
together with other rules in~\cite{scalasLess2019}.
\end{restatable}

\begin{restatable}[Nested try-catches and {$\mpEtx[\ ]$}]{remark}{remarkTryCatches}
The context $\mpEtx[\ ]$ is only used for
  defining the reductions at the top parallel composed processes, \emph{not} 
used nested exception handling like  
\cite{fowlerExceptional2019,harveyMultiparty2021}. 
Our (typable) try-catch
processes allow any form of processes such as 
recursions, parallel, 
session delegations, and restriction/scope opened processes 
under a guarded process: \\[1mm]
$
\small
\trycatchbreakequal{R}
{
        \mpSel{
            \mpChanRole{\mpS}{\roleP}
        }{\roleR}{\stLab[1]}{}{
            \mpRes{\mpSi}{
                (
                \mpSel{
                    \mpChanRole{\mpS}{\roleP}
                }{\roleR}{\stLab[3]}{
                    \mpChanRole{\mpSi}{\roleR}
                }{
                    \mpNil
                }
                \mpPar
                \trycatch{
                    \mpSel{
                        \mpChanRole{\mpSi}{\roleQ}
                    }{\roleR}{\stLab[2]}{}{
                        \mpNil
                    }
                }{
                    \mpCancel{\mpChanRole{\mpSi}{\roleQ}}{
                        \mpNil
                    }
                }
                )
            }
        }
    }{
        \mpCancel{\mpChanRole{\mpS}{\roleP}}{
            \mpNil
        }
    } 
$

\iftoggle{full}{%
  See~\Cref{app:ex:nestedtrycatch} for
  typable processes with nested \CODE{try-catch} blocks.
}{%
  See~\cite{fullVersion} for more 
  typable processes with nested \CODE{try-catch} blocks.
}%
\end{restatable}

\iftoggle{full}{%
  Numerous examples about \emph{nested try-catch processes}
  and their type derivation can be found in our Appendix\S~\ref{subsec:app:additional:examples}.
}{%
}%

We explain each rule highlighting the new rules.
\iftoggle{full}{%
}{%
}%
\begin{description}
  \item[Communication] Rule \inferrule{\iruleMPRedComm} is the main
    communication rule between an affine/linear selection and an
    affine/linear branching.
    Linear selections/branching are
    placed in the \emph{try} position but can interact with affine counterparts.
    Once they interact, processes are
    spawned from \emph{try-block}s (notice that $\mpEtx{}_1$,
    $\mpEtx{}_2$ are erased after the communication), and start
    communicating on parallel
    with other parallel composed processes.
    Note that the context $\mathbb{E}$ is discarded after the successful
    communication.
    \iftoggle{full}{%
      See~\Cref{app:ex:reduction}.
    }{%
    }%

  \item[Error-Cancellation]
    Rules \inferrule{\iruleMPCQSel} and
    \inferrule{\iruleMPCQBra} model the situations that
    an error handling occurs at the affine selection/branching.
    This might be the case if its counterpart has failed (hence
    \inferrule{\iruleMPRedComm} does not happen) or timeout.
    It then triggers the kill process at $s$.
    Rules \inferrule{\iruleMPTQSel} and
    \inferrule{\iruleMPTQBra} model
    the case that the affine selection/branching are placed inside the
    try-block and triggered by the error. In this case,
    it will go to the \emph{catch-block}, generating a kill process.

  \item[Cancelling Processes]
    Rule \inferrule{\iruleMPCSel} cancels the selection prefix $s$,
    additionally generating
    the kill process at the delegated channel for all the
    session processes at $s'$ to be cancelled.
    Rule \inferrule{\iruleMPCBra} cancels only one of the branches -- this
    is sufficient since all branches contain the same channels
    except $x_i$ (ensured by rule \inferrule{\iruleMPBranch}
    in \Cref{fig:mpst-rules}). After the cancellation, it additionally
    instantiates a fresh name $\mpChanRole{\mpS'}{\roleR}$ to
    $x_k$ into $\mpP[k]$. The generated kill process at $\mpS'$
    kills prefixes at
    $\mpChanRole{\mpS'}{\roleR}$
    in
    $\mpP[k]\subst{\mpFmt{x_k}}{\mpChanRole{\mpS'}{\roleR}}$.

  \item[Cancellation from Other Parties]
    Rule \inferrule{\iruleMPRedCan} is a cancellation and generates
    a kill process. Note that the \CODE{try-catch} context ${\Bbb E}$ is thrown
    away.
    Rule \inferrule{\iruleMPCCat} is prompted to move to $\mpQ$
    by kill $\kills{s}$. The side condition
    $\sbj{P}$ ensures that $P$ is a prefix at $s$ (up to $\equiv$ for a
    recursive process).
    All mimic the behaviour of the programs in
    \Cref{fig:implementation:role:s}.

  \item[Other Rules]
    Rules $\inferrule{\iruleMPRedCall}$,
    $\inferrule{\iruleMPRedCtx}$, and $\inferrule{\iruleMPRedSt}$ are standard from~\cite{scalasLess2019}. In \Cref{fig:structure},
    the two new rules are for garbage collections of
    kill processes.
  \iftoggle{full}{%
}{%
}%
\end{description}

\begin{restatable}[Syntax and reductions]{example}{exampleSyntaxAndReductions}
  \label{app:ex:reduction}
  A process might be completed, or cancelled in many ways, and also
  interacts non-deterministically.
  We demonstrate the reduction rules
  using the running example with a minor modification.
  We use a nested \CODE{try-catch} block, and
  for simplicity we use shorter label names, and we use a
  constant, \ie $\mpD$, as a message payload.
  
Assume the process for \role{S} is
$
  \mpP
  =
  \mpQuestionBranchSolo{
    \mpChanRole{\mpS}{\roleP}
  }{
    \roleQ
  }{
    \mpQ
    \mpFmt{+}
    \mpClose{(x)}.\mpNil
  }
$
where\\[1mm]
{\small
$
  \mpQ
  =
  \mpVideo{(x)}.
  \trycatch{
    \mpQuestionBranchSeq{
      \mpChanRole{\mpS}{\roleP}
    }{
      \roleQ
    }{
      \mpRequest
    }{
      x
    }
    {
      \trycatch{
        \mpQuestionSel{
          \mpChanRole{\mpS}{\roleP}
        }{
          \roleR
        }{
          \mpResponse
        }{
          \mpConst
        }{
          \mpNil
        }}{
        \mpCancel{
          \mpChanRole{\mpS}{\roleP}
        }{
          \mpNil
        }
      }
    }
  }{
    \mpCancel{
      \mpChanRole{\mpS}{\roleP}
    }{
      \mpNil
    }
  }
$
}\\[1mm]
The following shows a possible reduction.
\begin{small}
  \begin{align}
                                        &
    \mpP
    \mpPar
    \mpSel{
      \mpChanRole{\mpS}{\roleQ}
    }{
      \roleP
    }{
      video
    }{
      \mpConst
    }{
      \mpSel{
        \mpChanRole{\mpS}{\roleQ}
      }{
        \roleP
      }{
        \mpRequest
      }{
        \mpConst
      }{
        \mpBranchSeq{
          \mpChanRole{\mpS}{\roleQ}
        }{
          \roleP
        }{
          \mpResponse
        }{
          x
        }{
          \mpNil
        }
      }
    }
    \label{ex:reduction:solution:two}
    \\
    \inferrule{\iruleMPRedComm} \mpMove &
    \begin{array}{l}
      \trycatchbreak{
        (
          \mpQuestionBranchSeq{
            \mpChanRole{\mpS}{\roleP}
          }{
            \roleQ
          }{
            \mpRequest
          }{
            x
          }
          {
            \trycatch{
              (
                \mpQuestionSel{
                  \mpChanRole{\mpS}{\roleP}
                }{
                  \roleR
                }{
                  \mpResponse
                }{
                  \mpConst
                }{
                  \mpNil
                }
              )
            }{
              \mpCancel{
                \mpChanRole{\mpS}{\roleP}
              }{
                \mpNil
              }
            }
          }
        )
      }{
        \mpCancel{
          \mpChanRole{\mpS}{\roleP}
        }{
          \mpNil
        }
      }
    \end{array}
    \label{ex:reduction:solution:three}
    \\
                                        &
    \mpPar
    \mpSel{
      \mpChanRole{\mpS}{\roleQ}
    }{
      \roleP
    }{
      \mpRequest
    }{
      \mpConst
    }{
      \mpBranchSeq{
        \mpChanRole{\mpS}{\roleQ}
      }{
        \roleP
      }{
        \mpResponse
      }{
        x
      }{
        \mpNil
      }
    }
    \label{ex:reduction:solution:four}
    \\
    \inferrule{\iruleMPRedComm} \mpMove &
    \trycatch{
      \mpQuestionSel{
        \mpChanRole{\mpS}{\roleP}
      }{
        \roleR
      }{
        \mpResponse
      }{
        \mpConst
      }{
        \mpNil
      }
    }{
      \mpCancel{
        \mpChanRole{\mpS}{\roleP}
      }{
        \mpNil
      }
    }
    \mpPar
    \mpBranchSeq{
      \mpChanRole{\mpS}{\roleQ}
    }{
      \roleP
    }{
      \mpResponse
    }{
      x
    }{
      \mpNil
    }
    \label{ex:reduction:solution:five}
    \\
    \inferrule{\iruleMPTQSel} \mpMove   &
    \mpCancel{
      \mpChanRole{\mpS}{\roleP}
    }{
      \mpNil
    }
    \mpPar  \kills{s}
    \mpPar
    \mpBranchSeq{
      \mpChanRole{\mpS}{\roleQ}
    }{
      \roleP
    }{
      \mpResponse
    }{
      x
    }{
      \mpNil
    }
    \label{ex:reduction:solution:six}
    \\
    \inferrule{\iruleMPRedCan} \mpMove  &
    \kills{s}  \mpPar \kills{s}  \mpPar  \mpNil
    \mpPar
    \mpBranchSeq{
      \mpChanRole{\mpS}{\roleQ}
    }{
      \roleP
    }{
      \mpResponse
    }{
      x
    }{
      \mpNil
    }
    \label{ex:reduction:solution:seven}
    \\
    \inferrule{\iruleMPCBra} \mpMove    &
    \kills{s}  \mpPar \kills{s} \mpPar \mpNil
    \mpPar
    \mpNil \equiv  \kills{s}
    \label{ex:reduction:solution:eight}
  \end{align}
\end{small}

$\mpEtx_{6}[P_6]$ for~\Cref{ex:reduction:solution:two} is
$P_6=
  \mpQuestionBranchSeq{
    \mpChanRole{\mpS}{\roleP}
  }{
    \roleQ
  }{
    \mpRequest
  }{
    x
  }
  {
    \trycatch{
...
       }{
      \mpCancel{
      \mpChanRole{\mpS}{\roleP}
      }{
        \mpNil
      }
    }
  }
$
and 
$\mpEtx_{9}[P_9]$ for~\Cref{ex:reduction:solution:five} is
$P_9=
  \mpQuestionSel{
    \mpChanRole{\mpS}{\roleP}
  }{
    \roleR
  }{
    \mpResponse
  }{
    \mpConst
  }{
    \mpNil
  }
$
both because of rule
$\inferrule{\iruleMPCCat}$.

Initially we reduce using the communication rule for
the branching and selection.
Next, we apply $\inferrule{\iruleMPRedComm}$ demonstrating how
the affine branching reduces under \proclit{try}.
Then we apply $\inferrule{\iruleMPTQSel}$ assuming an error
(or a timeout) occurs during the selection of $\mpResponse$.
This generates a kill process $\kills{s}$ and spawns the
process in the \proclit{catch}-block.
\proclit{Cancel} spawns a kill process $\kills{s}$ and hence reduces to  $\kills{s} \mpPar \mpNil$,
following rule $\inferrule{\iruleMPRedCan}$ with $\mpEtx = \mpEtxHole$.
Finally, applying $\inferrule{\iruleMPRedCan}$ cancels the linear selection.
To conclude, we garbage collect all kill processes.
Given that our initial parallel composition
has name restrictions $\mpRes{\mpS}{}$ at the top level, 
\iftoggle{full}{(omitted from the reduction for convenience), we have}{} 
$\mpRes{\mpS}{\kills{s}} \equiv \mpNil$.

  \end{restatable}

\iftoggle{full}{%
  \begin{restatable}[
      Rules
        {$\inferrule{\iruleMPCQSel}/\inferrule{\iruleMPCSel}$}
      and
        {$\inferrule{\iruleMPCQBra}/\inferrule{\iruleMPCBra}$}
    ]{example}{exampleRulesSelAndBra}
    \label{ex:rules:sel:and:bra}
    \label{example:reduce}%
    We can obtain the reduction
    $\mpQuestionSel{\mpChanRole{\mpS}{\roleP}}{\roleQ}{m}{\mpChanRole{\mpSi}{r}}{\mpP}
      \mpMoveStar \mpP \mpPar \kills{s} \mpPar \kills{s'}$, by first
    deriving $\mpQuestionSel{\mpChanRole{\mpS}{\roleP}}{\roleQ}{m}{\mpChanRole{\mpSi}{r}}{\mpP}
      \mpMove
      \mpSel{\mpChanRole{\mpS}{\roleP}}{\roleQ}{m}{\mpChanRole{\mpSi}{r}}{\mpP}
      \mpPar \kills{s}=R$ by rule $\inferrule{\iruleMPCQSel}$,
    then
    $\mpSel{\mpChanRole{\mpS}{\roleP}}{\roleQ}{m}{\mpChanRole{\mpSi}{r}}{\mpP} \mpPar \kills{s}
      \mpMove \mpP \mpPar \kills{s} \mpPar \kills{s'}$ by rule $\inferrule{\iruleMPCSel}$.
    Suppose $\mpR \mpPar \mpQ$ and $\mpQ$ contains some prefix at s.
    Then $\mpQ$ can receive a signal from $\kills{s}$ and can fail
    before an action from
    \ref{example:r:two} to \ref{example:r:three}.
    However, this decomposition into two rules allows
    $\mpCtx{}{[R]}$ to interact with other processes in the context
    $\mathbb{C}$ before generating $\kills{s'}$,
    capturing a time difference (nondeterministic failures/asynchrony).
  \end{restatable}
}{%
}%





\iftoggle{full}{%
}{%
}%
\subsection{Affine multiparty session typing system}\label{sec:AMPST}
\label{subsec:types}
\iftoggle{full}{%
}{%
}%

\myparagraph{Global and local types}
The advantage of affine session frameworks is that no change
of the syntax of types
from the original system is required.
We follow~\cite{scalasLess2019} which is the most widely used syntax in the
literature.
A \emph{global type}, written $\gtG,\gtG',\dots$,
describes the whole
conversation scenario of a multiparty session as a type signature, and
a {\em local type}, written by $\stS,\stS', \dots$, 
represents a local protocol for each participant. 
The syntax of types is given as:

\begin{restatable}[Global types]{definition}{defMPSTGlobalTypes}\rm
  The syntax of a \textbf{global type $\gtG$} is:%

  \smallskip%
  \centerline{%
    \(%
    \gtG%
    \,\coloncolonequals\,%
    \gtComm{\roleP}{\roleQ}{i \in I}{\gtLab[i]}{\stS[i]}{\gtG[i]}%
    \bnfsep%
    \gtRec{\gtRecVar}{\gtG}%
    \bnfsep%
    \gtRecVar%
    \bnfsep%
    \gtEnd%
    \)%
    \qquad%
    \text{%
      with %
      $\roleP \!\neq\! \roleQ$, %
      \;$I \!\neq\! \emptyset$, %
      \;and\; $\forall i \!\in\! I: \fv{\stS[i]} = \emptyset$%
    }%
  }%
  The syntax of \textbf{local types} %
  is:

  \smallskip%
  \centerline{\(%
    \textstyle%
    \stS, \stT%
    \,\coloncolonequals\,%
    \stExtSum{\roleP}{i \in I}{\stChoice{\stLab[i]}{\stS[i]} \stSeq \stSi[i]}%
    \bnfsep%
    \stIntSum{\roleP}{i \in I}{\stChoice{\stLab[i]}{\stS[i]} \stSeq \stSi[i]}%
    \bnfsep%
    \stEnd%
    \bnfsep%
    \stRec{\stRecVar}{\stS}%
    \bnfsep%
    \stRecVar
    \quad%
    \text{with $I \!\neq\! \emptyset$,
      and $\stLab[i]$ pairwise distinct.}%
    \)}%
  \smallskip%

  \noindent%
  Types must be closed, %
  and recursion variables to be guarded.
\end{restatable}

$\stLab\in \ASigma$ corresponds to the usual message labels in the session type
theory.
Global branching type
$\gtComm{\roleP}{\roleQ}{i \in I}{\gtLab[i]}{\stS[i]}{\gtG[i]}$
states that participant $\roleP$ can
send a message with one of the $\gtLab[i]$ labels and a {\em message payload type} $\stS[i]$
to the participant $\roleQ$ and that interaction
described in $\gtG[i]$ follows.
We require $\roleP\neq\roleQ$ to prevent self-sent messages and
$\gtLab[i]\not= \gtLab[k]$ for all $i\not =k\in J$. Recursive types
$\gtRec{\gtRecVar}{\gtG}$ are for recursive protocols,
assuming those type variables ($\gtRecVar, \gtRecVar', \dots$) are
guarded in the
standard way, \ie they only occur under branching.
Type $\gtEnd$ represents session termination (often omitted).
We write\; $\roleP \in \gtRoles{\gtG}$ %
\;(or simply $\roleP \!\in\! \gtG$) \;iff, for some $\roleQ$, %
either $\gtFmt{\roleP {\to} \roleQ}$ %
or $\gtFmt{\roleQ {\to} \roleP}$ %
occurs in $\gtG$.
The function $\id(\gtG)$ gives the participants of $\gtG$.

For local types, the {\em branching type}
$\stExtSum{\roleP}{i \in I}{\stChoice{\stLab[i]}{\stS[i]} \stSeq \stSi[i]}$
specifies the reception of a message from
$\roleP$ with a label among the $\stLab[i]$ and a payload $\stS[i]$. The {\em selection type}
$\stIntSum{\roleP}{i \in I}{\stChoice{\stLab[i]}{\stS[i]} \stSeq \stSi[i]}$
is its \emph{dual} -- its opposite operation.
The remaining type constructors are as for global types.
We say a type is {\em guarded} if it is neither a recursive type nor a type variable.

\iftoggle{full}{%
  \myparagraph{Projection}
  The relation between global and local types is formalised by
  projection~\cite{coppoGlobal2016,hondaMultiparty2008}.
  
\begin{restatable}[Projection]{definition}{defMPSTProjection}\rm
    \label{app:def:projection}
    The {\em projection of $\gtG$ onto $\roleP$}
    (written $\gtProj{\gtG}{\roleP}$) is defined as:
  
    \smallskip%
    \centerline{\(%
      \small%
      \begin{array}{c}
        \gtProj{\left(%
          \gtComm{\roleQ}{\roleR}{i \in I}{\gtLab[i]}{\stS[i]}{\gtG[i]}%
          \right)}{\roleP}%
        \;=\;%
        \left\{%
        \begin{array}{@{\hskip 0.5mm}l@{\hskip 5mm}l@{}}
          \stIntSum{\roleR}{i \in I}{%
            \stChoice{\stLab[i]}{\stS[i]} \stSeq (\gtProj{\gtG[i]}{\roleP})%
          }%
           & \text{\footnotesize %
            if\, $\roleP = \roleQ$}%
          \\%
          \stExtSum{\roleQ}{i \in I}{%
            \stChoice{\stLab[i]}{\stS[i]} \stSeq (\gtProj{\gtG[i]}{\roleP})%
          }%
           & \text{\footnotesize %
            if\, $\roleP = \roleR$}%
          \\%
          \stMerge{i \in I}{\gtProj{\gtG[i]}{\roleP}}%
           & \text{\footnotesize %
            if\, $\roleQ \neq \roleP \neq \roleR$}%
        \end{array}
        \right.
        \\[6mm]%
        \begin{array}{rcl}
          \gtProj{(\gtRec{\gtRecVar}{\gtG})}{\roleP}%
           & = & %
          \left\{%
          \begin{array}{@{\hskip 0.5mm}l@{\hskip 5mm}l@{}}
            \stRec{\stRecVar}{(\gtProj{\gtG}{\roleP})}%
             & %
            \text{\footnotesize%
              if\, $\roleP \in \gtG$ or $\fv{\gtRec{\gtRecVar}{\gtG}}\not=\emptyset$%
            }%
  \\
            \stEnd%
             & %
            \text{\footnotesize%
              otherwise}
          \end{array}
          \right.%
  \qquad
          \gtProj{\gtRecVar}{\roleP}%
            = %
          \stRecVar%
  \qquad
          \gtProj{\gtEnd}{\roleP}%
            =  %
          \stEnd%
        \end{array}
      \end{array}
      \)}%
    \smallskip%
  
    \noindent%
    For projection of branching, we introduce a merge
    operator along the lines of~\cite{denielouDynamic2011},
    written $\stS \stBinMerge \stS'$, ensuring
    that if the locally observable behaviour of the local type is
    dependent of the chosen branch then it is identifiable via a unique
    choice/branching label.
  
    The \emph{merging operation} $\stBinMerge{}{}$ is defined as a partial commutative
    operator over two types such that:
  
    \smallskip%
    \centerline{\(%
      \small%
      \begin{array}{c}%
        \textstyle%
        \stExtSum{\roleP}{i \in I}{\stChoice{\stLab[i]}{\stS[i]} \stSeq \stSi[i]}%
        \,\stBinMerge\,%
        \stExtSum{\roleP}{\!j \in J}{\stChoice{\stLab[j]}{\stS[\!j]} \stSeq \stTi[j]}%
        \;=\;%
        \stExtSum{\roleP}{k \in I \cap J}{\stChoice{\stLab[k]}{\stS[k]} \stSeq%
          (\stSi[k] \!\stBinMerge\! \stTi[k])%
        }%
        \stExtC%
        \stExtSum{\roleP}{i \in I \setminus J}{\stChoice{\stLab[i]}{\stS[i]} \stSeq \stSi[i]}%
        \stExtC%
        \stExtSum{\roleP}{\!j \in J \setminus I}{\stChoice{\stLab[j]}{\stS[\!j]} \stSeq \stTi[j]}%
        \\[1mm]%
        \stIntSum{\roleP}{i \in I}{\stChoice{\stLab[i]}{\stS[i]} \stSeq \stSi[i]}%
        \,\stBinMerge\,%
        \stIntSum{\roleP}{i \in I}{\stChoice{\stLab[i]}{\stS[i]} \stSeq \stSi[i]}%
        \;=\;%
        \stIntSum{\roleP}{i \in I}{\stChoice{\stLab[i]}{\stS[i]} \stSeq \stSi[i]}%
        \qquad%
        \stRec{\stRecVar}{\stS} \,\stBinMerge\, \stRec{\stRecVar}{\stT}%
        \;=\;%
        \stRec{\stRecVar}{(\stS \stBinMerge \stT)}%
        \qquad%
        \stT \,\stBinMerge\, \stT%
        \;=\;%
        \stT%
      \end{array}
      \)}%
  \end{restatable}
  \noindent As an example of projection, consider
  $\gtRec{\gtRecVar}{\gtCommRaw{\roleAlice}{\roleBob}{
            \gtCommChoiceSmall{a}{}{
              \gtCommRaw{\roleBob}{\roleCarol}{
                \gtCommChoiceSmall{b}{}{
                  \gtCommRaw{\roleAlice}{\roleCarol}{\gtCommChoiceSmall{c}{}{
                      \gtRecVar}}}}},
            \gtCommChoiceSmall{d}{}{
              \gtCommRaw{\roleBob}{\roleCarol}{
                \gtCommChoiceSmall{e}{}{
                  \gtCommRaw{\roleAlice}{\roleCarol}{
                    \gtCommChoiceSmall{f}{}{\gtEnd}}}}}}}$.
  Then $\roleCarol$'s local type is given by:
  $\stRec{\stRecVar}{
      \stExtSum{\roleBob}{}{\{
        \stChoice{\stLabFmt{b}}{} \stSeq
        \stExtSum{\roleAlice}{}{\stLabFmt{c}} \stSeq \stRecVar,
        \stChoice{\stLabFmt{e}}{} \stSeq
        \stExtSum{\roleAlice}{}{\stLabFmt{f}} \stSeq \stEnd
        \}}
    }$.
  We say that $G$ is {\em well-formed} if for all $\roleP\in \PSet$,
  $\gtProj{\gtG}{\roleP}$ is defined.
  We also use the standard \emph{multiparty session subtyping relation} from
 ~\cite{ghilezanPrecise2019,dezaniCiancagliniPrecise2015}.
  \begin{restatable}[Subtyping]{definition}{definitionSubtyping}
    \label{app:def:session:subtyping}
    The subtyping relation $\stSub$ is \emph{co}inductively defined as:
    \smallskip%

    \centerline{\small\(%
        \begin{array}{c}%
            \cinference[\iruleSubBranch]{
                \forall i \in I%
             & %
                \stS[i] \stSub \stT[i]%
             & %
                \stSi[i] \stSub \stTi[i]%
            }{%
                \stExtSum{\roleP}{i \in I}{\stChoice{\stLab[i]}{\stS[i]} \stSeq \stSi[i]}%
                \stSub%
                \stExtSum{\roleP}{i \in I \cup J}{\stChoice{\stLab[i]}{\stT[i]} \stSeq \stTi[i]}%
            }%
            \qquad%
            \cinference[\iruleSubSel]{
                \forall i \in I%
             & %
                \stT[i] \stSub \stS[i]%
             & %
                \stSi[i] \stSub \stTi[i]%
            }{%
                \stIntSum{\roleP}{i \in I \cup J}{\stChoice{\stLab[i]}{\stS[i]} \stSeq \stSi[i]}%
                \stSub%
                \stIntSum{\roleP}{i \in I}{\stChoice{\stLab[i]}{\stT[i]} \stSeq \stTi[i]}%
            }%
            \\[2mm]%
            \cinference[\iruleSubEnd]{%
                \phantom{X}%
            }{%
                \stEnd \stSub \stEnd%
            }%
            \qquad%
            \cinference[\iruleSubRecL]{%
                \stS\subst{\stRecVar}{\stRec{\stRecVar}{\stS}} \stSub \stT%
            }{%
                \stRec{\stRecVar}{\stS} \stSub \stT%
            }%
            \qquad%
            \cinference[\iruleSubRecR]{%
                \stS \stSub \stT\subst{\stRecVar}{\stRec{\stRecVar}{\stT}}%
            }{%
                \stS \stSub \stRec{\stRecVar}{\stT}
            }%
        \end{array}
        \)}%
\end{restatable}
}{%
  The relation between global and local types is formalised by
  projection~\cite{coppoGlobal2016,hondaMultiparty2008}.
  The {\em projection of $\gtG$ onto $\roleP$} is
  written $\gtProj{\gtG}{\roleP}$ and
  the standard subtyping relation, $\stSub$.
  See~\cite{fullVersion}.
}%

%
%
We define typing contexts which are used to define properties of
type-level behaviours.

\begin{restatable}[Typing contexts]{definition}{definitionTypingContexts}
 \label{def:mpst-env}%
  \label{def:mpst-env-closed}%
  \label{def:mpst-env-comp}%
  \label{def:mpst-env-subtype}%
  $\mpEnv$ denotes a partial mapping %
  from process variables to $n$-tuples of types, %
  and $\stEnv$ denotes a partial mapping %
  from channels to types, %
  defined as:%

  \smallskip%
  \centerline{\(%
    \mpEnv%
    \;\;\coloncolonequals\;\;%
    \mpEnvEmpty%
    \bnfsep%
    \mpEnv \mpEnvComp\, \mpEnvMap{\mpX}{\stS[1],\dots,\stS[n]}%
    \quad\quad
    \stEnv%
    \,\coloncolonequals\,%
    \stEnvEmpty%
    \bnfsep%
    \stEnv \stEnvComp \stEnvMap{c}{\stS}%
    \)}%
  \smallskip%

  \noindent%
  The \,\emph{composition} $\stEnv[1] \stEnvComp \stEnv[2]$\,
  is defined iff $\dom{\stEnv[1]} \cap \dom{\stEnv[2]} = \emptyset$.
  We write\; %
  $\mpS \!\not\in\! \stEnv$ %
  \;iff\; %
  $\forall \roleP: \mpChanRole{\mpS}{\roleP} \!\not\in\! \dom{\stEnv}$ %
  (\ie session $\mpS$ does not occur in $\stEnv$).%
  We write\; %
  $\dom{\stEnv} \!=\! \setenum{\mpS}$ %
  \;iff\; %
  $\forall \mpC \!\in\! \dom{\stEnv}$ there is $\roleP$ such that %
  $\mpC \!=\! \mpChanRole{\mpS}{\roleP}$ %
  (\ie $\stEnv$ only contains session $\mpS$); and
  $\stEnv \!\stSub\! \stEnvi$ %
  iff %
  $\dom{\stEnv} \!=\! \dom{\stEnvi}$ %
  and %
  $\forall \mpC \!\in\! \dom{\stEnv}{:}\,%
    \stEnvApp{\stEnv}{\mpC} \!\stSub\! \stEnvApp{\stEnvi}{\mpC}$.%
\iftoggle{full}{}
{
We write $\stEnv \stEnvMove \stEnvi$ with
            $\stEnv=\stEnv[0] \stEnvComp\,%
              \stEnvMap{%
                \mpChanRole{\mpS}{\roleP}%
              }{%
                \stIntSum{\roleQ}{i \in I}{\stChoice{\stLab[i]}{\stS[i]} \stSeq \stSi[i]}%
              }%
              \stEnvComp\,%
              \stEnvMap{%
                \mpChanRole{\mpS}{\roleQ}%
              }{%
                \stExtSum{\roleP}{j \in J}{\stChoice{\stLab[j]}{\stT[j]} \stSeq \stTi[j]}}%
            $ and
            $\stEnv' =
              \stEnv[0]    \stEnvComp\,%
              \stEnvMap{%
                \mpChanRole{\mpS}{\roleP}%
              }{%
                {\stSi[i]}%
              }%
              \stEnvComp\,%
              \stEnvMap{%
                \mpChanRole{\mpS}{\roleQ}%
              }{%
                \stTi[j]}%
            $ where types are defined modulo unfolding recursive types.
            We write $\stEnv \stEnvMove^\ast \stEnvi$ for a transitive and
            reflexive closure of $\stEnvMove$; and  $\stEnv \stEnvMove$ if
            there exists $\stEnvi$ such that
            $\stEnv \stEnvMove \stEnvi$.
}
\end{restatable}

Next, we define typing context properties defined by its reduction. 

\iftoggle{full}
{

\begin{restatable}[
        Typing context reduction,
        safety,
        deadlock-freedom and liveness
    ]{definition}{definitionTypingContextReduction}
    \label{def:mpst-env-safe}%
    \-
    \begin{itemize}
        \item We write $\stEnv \stEnvMove \stEnvi$ with
              $\stEnv=\stEnv[0] \stEnvComp\,%
                  \stEnvMap{%
                      \mpChanRole{\mpS}{\roleP}%
                  }{%
                      \stIntSum{\roleQ}{i \in I}{\stChoice{\stLab[i]}{\stS[i]} \stSeq \stSi[i]}%
                  }%
                  \stEnvComp\,%
                  \stEnvMap{%
                      \mpChanRole{\mpS}{\roleQ}%
                  }{%
                      \stExtSum{\roleP}{j \in J}{\stChoice{\stLab[j]}{\stT[j]} \stSeq \stTi[j]}}%
              $ and
              $\stEnv' =
                  \stEnv[0]    \stEnvComp\,%
                  \stEnvMap{%
                      \mpChanRole{\mpS}{\roleP}%
                  }{%
                      {\stSi[i]}%
                  }%
                  \stEnvComp\,%
                  \stEnvMap{%
                      \mpChanRole{\mpS}{\roleQ}%
                  }{%
                      \stTi[j]}%
              $ where types are defined modulo unfolding recursive types.
              We write $\stEnv \stEnvMove^\ast \stEnvi$ for a transitive and
              reflexive closure of $\stEnvMove$; and  $\stEnv \stEnvMove$ if
              there exists $\stEnvi$ such that
              $\stEnv \stEnvMove \stEnvi$.
        \item
              $\predP$ is a \emph{safety property} of typing contexts %
              iff: %

              \begin{tabular}{r@{\hskip 2mm}l}
                  \inferrule{\iruleSafeComm}%
                   & %
                  $\predPApp{%
                          \stEnv \stEnvComp\,%
                          \stEnvMap{%
                              \mpChanRole{\mpS}{\roleP}%
                          }{%
                              \stIntSum{\roleQ}{i \in I}{\stChoice{\stLab[i]}{\stS[i]} \stSeq \stSi[i]}%
                          }%
                          \stEnvComp\,%
                          \stEnvMap{%
                              \mpChanRole{\mpS}{\roleQ}%
                          }{%
                              \stExtSum{\roleP}{j \in J}{\stChoice{\stLab[j]}{\stT[j]} \stSeq \stTi[j]}%
                          }%
                      }$%
                  \;\;implies\;\; %
                  $I \!\subseteq\! J$\,, %
                  \,and\, %
                  $\forall i \!\in\! I: \stS[i] \!\stSub\! \stT[i]$;
                  \\%
                  \inferrule{\iruleSafeRec}%
                   & %
                  $\predPApp{%
                          \stEnv \stEnvComp\, \stEnvMap{%
                              \mpChanRole{\mpS}{\roleP}%
                          }{%
                              \stRec{\stRecVar}{\stS}%
                          } %
                      }$ %
                  \;\;implies\;\; %
                  $\predPApp{%
                          \stEnv \stEnvComp\, \stEnvMap{%
                              \mpChanRole{\mpS}{\roleP}%
                          }{%
                              \stS\subst{\stRecVar}{\stRec{\stRecVar}{\stS}}%
                          }%
                      }$;
                  \\%
                  \inferrule{\iruleSafeMove}%
                   & %
                  $\predPApp{\stEnv}$ %
                  \;and\; $\stEnv \stEnvMove \stEnvi$ %
                  \;\;implies\;\; %
                  $\predPApp{\stEnvi}$.
              \end{tabular}
              \noindent%
              We say \emph{$\stEnv$ is safe},
              written $\stEnvSafeP{\stEnv}$, %
              if $\predPApp{\stEnv}$ %
              for some safety property $\predP$.
        \item \emph{$\stEnv$ is deadlock-free},
              written $\stEnvDFP{\stEnv}$, %
              if $\stEnv \stEnvMove^\ast \stEnvi \not\stEnvMove$ %
              \;\;implies\;\; %
              $\stEnvEndP{\stEnvi}$.
        \item \emph{$\stEnv$ is live},
              written $\stEnvLiveP{\stEnv}$, %
              iff:%
              $\predPApp{\stEnv}$, %
              for some $\predP$ such that %
              \begin{description}[labelsep=2mm,leftmargin=7mm]
                  \item[\inferrule{\iruleLiveBranch}]%
                  whenever %
                  $\predPApp{%
                          \stEnvi \stEnvComp \stEnvMap{%
                              \mpChanRole{\mpS}{\roleP}%
                          }{%
                              \stS%
                          }%
                      }$ %
                  with %
                  $\stS \!=\!%
                      \stExtSum{\roleQ}{i \in I}{%
                          \stChoice{\stLab[i]}{\stS[i]} \stSeq \stSi[i]%
                      }$, %
                  $\exists i \!\in\! I{:}$ %
                  $\exists \stEnvii{:}$ %
                  $\stEnvi \stEnvComp \stEnvMap{%
                          \mpChanRole{\mpS}{\roleP}%
                      }{%
                          \stS%
                      }%
                      \!\stEnvMoveStar\!%
                      \stEnvii \stEnvComp \stEnvMap{%
                          \mpChanRole{\mpS}{\roleP}%
                      }{%
                          \stSi[i]%
                      }$
                  \item[\inferrule{\iruleLiveSel}]%
                  whenever %
                  $\predPApp{%
                          \stEnvi \stEnvComp \stEnvMap{%
                              \mpChanRole{\mpS}{\roleP}%
                          }{%
                              \stS%
                          }%
                      }$ %
                  with %
                  $\stS \!=\!%
                      \stIntSum{\roleQ}{i \in I}{%
                          \stChoice{\stLab[i]}{\stS[i]} \stSeq \stSi[i]%
                      }$, %
                  $\forall i \!\in\! I{:}$ %
                  $\exists \stEnvii{:}$\; %
                  $\stEnvi \stEnvComp \stEnvMap{%
                          \mpChanRole{\mpS}{\roleP}%
                      }{%
                          \stS%
                      }%
                      \!\stEnvMoveStar\!%
                      \stEnvii \stEnvComp \stEnvMap{%
                          \mpChanRole{\mpS}{\roleP}%
                      }{%
                          \stSi[i]%
                      }$
              \end{description}
              plus clauses %
              \inferrule{\iruleSafeRec} and
              \inferrule{\iruleSafeMove}.
              \item\label{item:mpst-env-liveplus}%
              \emph{$\stEnv$ is live\textsuperscript{+}}, %
              written $\stEnvLivePlusP{\stEnv}$, %
              iff:%
              $\predPApp{\stEnv}$, %
              for $\predP$ such that %
              \begin{description}[labelsep=2mm,leftmargin=7mm]
                  \item[\inferrule{\iruleLivePlusBranch}]%
                  clause \inferrule{\iruleLiveBranch} above; %
                  \emph{moreover}, %
                  $\stEnvi \stEnvComp \stEnvMap{%
                          \mpChanRole{\mpS}{\roleP}%
                      }{%
                          \stS%
                      }$ %
                  belongs to some fair traversal set $\someSetX$ %
                  with targets $\someSetY$ %
                  (\cite[Definition 5.5]{scalasLess2019}) %
                  such that, %
                  $\forall \stEnv[t] \!\in\! \someSetY$, %
                  we have %
                  $\stEnv[t] \!=\!%
                      \stEnvii \stEnvComp \stEnvMap{%
                          \mpChanRole{\mpS}{\roleP}%
                      }{%
                          \stSi[i]%
                      }$ %
                  (for some $\stEnvii, i \!\in\! I$)%
                  \item[\inferrule{\iruleLivePlusSel}]%
                  clause \inferrule{\iruleLiveSel} above, %
                  \emph{plus} the \emph{``moreover\ldots''} part of %
                  \inferrule{\iruleLivePlusBranch}
              \end{description}
              plus clauses %
              \inferrule{\iruleSafeRec} and
              \inferrule{\iruleSafeMove}.
    \end{itemize}
\end{restatable}
}
{}
\iftoggle{full}
{We omit the definition of fair traversal set from
  \cite[Definition 5.5]{scalasLess2019} (it needs to be defined by
  the transition system of the typing context).
}{
We say \emph{$\stEnv$ is safe},
written $\stEnvSafeP{\stEnv}$, %
if $\predPApp{\stEnv}$ %
for some safety property $\predP$.
Similarly, for \emph{deadlock-freedom} ($\stEnvDFP{\stEnv}$)
and \emph{liveness plus} ($\stEnvLivePlusP{\stEnv}$).
See~\cite{fullVersion} for the definitions.
}%
The reader can refer to~\cite{scalasLess2019}
for more explanations of the typing context properties.

\begin{restatable}[Typing judgement]{definition}{defMPSTTyping}\rm
  The typing judgement for processes has the form:
  \iftoggle{full}{%
}{%
}%
  \begin{equation}
    \label{eq:typing-judgement}%
    \stJudge{\mpEnv}{\stEnv}{\mpP}%
    \
    \text{%
      (with $\mpEnv$/$\stEnv$ omitted when empty)%
    }%
  \end{equation}
  and are defined by the typing rules in \Cref{fig:mpst-rules} with
  the judgements for process variables and channels.
  For convenience, we type-annotate %
  channels bound by process definitions and restrictions.
  Note that $\stEnvEndP{\stEnv}$ denotes that $\stEnv$ only contains
  type $\stEnd$.
\end{restatable}

\begin{figure}[t]
\small
\[
\begin{array}{c}
          \inference[\iruleMPX]{%
            \mpEnvApp{\mpEnv}{X} = \stFmt{\stS[1],\dots,\stS[n]}%
          }{%
            \mpEnvEntails{\mpEnv}{X}{\stS[1],\dots,\stS[n]}%
          }%
\quad
          \inference[\iruleMPSub]{%
            \stS \stSub \stSi 
          }{%
            \stEnvEntails{\stEnvMap{\mpC}{\stS}}{\mpC}{\stSi}%
          }%
\quad
          \inference[\iruleMPEnd]{%
            \forall i \in 1..n%
           &                             %
            \stEnvEntails{\stEnvMap{\mpC[i]}{\stS[i]}}{%
              \mpC[i]%
            }{%
              \stEnd%
            }%
          }{%
            \stEnvEndP{%
              \stEnvMap{\mpC[1]}{\stS[1]}%
              \stEnvComp \dots \stEnvComp%
              \stEnvMap{\mpC[n]}{\stS[n]}%
            }%
          }%
          \quad
          \inference[\iruleMPNil]{%
            \stEnvEndP{\stEnv}%
          }{%
            \stJudge{\mpEnv}{\stEnv}{\mpNil}%
          }%
\\[3mm]
     \inference[\iruleMPBranch]{%
     \stEnvEntails{\stEnv[1]}{\mpC}{%
     \stExtSum{\roleQ}{i \in I}{\stChoice{\stLab[i]}{\stS[i]} \stSeq \stSi[i]}%
            }%
           &                             %
            \forall i \!\in\! I%
           &                             %
            \stJudge{\mpEnv}{%
              \stEnv \stEnvComp%
              \stEnvMap{y_i}{\stS[i]} \stEnvComp%
              \stEnvMap{\mpC}{\stSi[i]}%
            }{%
              \mpP[i]%
            }%
          }{%
            \colorbox{ColourShade}{$\stJudge{\mpEnv}{%
                  \stEnv \stEnvComp \stEnv[1]%
                }{%
                  \mpDaggerBranch{\mpC}{\roleQ}{i \in I}{\mpLab[i]}{y_i}{\mpP[i]}%
                }%
            $}
          }%
\quad
          \inference[\iruleMPPar]{%
            \stJudge{\mpEnv}{%
              \stEnv[1]%
            }{%
              \mpP[1]%
            }%
            \qquad%
            \stJudge{\mpEnv}{%
              \stEnv[2]%
            }{%
              \mpP[2]%
            }%
          }{%
            \stJudge{\mpEnv}{%
              \stEnv[1] \stEnvComp \stEnv[2]%
            }{%
              \mpP[1] \mpPar \mpP[2]%
            }%
          }%
\\[3mm]
          \inference[\iruleMPSel]{%
            \stEnvEntails{\stEnv[1]}{\mpC}{%
              \stIntSum{\roleQ}{}{\stChoice{\stLab}{\stS} \stSeq \stSi}%
            }%
           &                             %
            \stEnvEntails{\stEnv[2]}{\mpCi}{\stS}%
           &                             %
            \stJudge{\mpEnv}{%
              \stEnv \stEnvComp \stEnvMap{\mpC}{\stSi}%
            }{%
              \mpP%
            }%
          }{%
            \colorbox{ColourShade}{$\stJudge{\mpEnv}{%
                  \stEnv \stEnvComp \stEnv[1] \stEnvComp \stEnv[2]%
                }{%
                  \mpDaggerSel{\mpC}{\roleQ}{\mpLab}{\mpCi}{\mpP}%
                }$%
          }}%
\quad
         \inference[\colorbox{ColourShade}{\iruleMPTryCatch}]{%
         \colorbox{ColourShade}{$\stJudge{\mpEnv}{%
                \stEnv%
                }{%
                  \mpP%
                }$}%
            \qquad%
            \colorbox{ColourShade}{$\sbj{P}=\{c\}$}
            \qquad%
            \colorbox{ColourShade}{$\stJudge{\mpEnv}{%
                  \stEnv%
                }{%
                  {\mpQ}%
                }$}%
          }{%
            \colorbox{ColourShade}{$\stJudge{\mpEnv}{\stEnv}
                {%
                  \trycatch{P}{Q}
                }%
              $}}%
          \\[3mm]
          \inference[\colorbox{ColourShade}{\iruleMPKills}]{%
            \colorbox{ColourShade}{$\stEnvEndP{\stEnv}$}%
            \quad \colorbox{ColourShade}{$0\leq n$}
          }{%
            \colorbox{ColourShade}{$\stJudge{\mpEnv}{\stEnv \stEnvComp
                  \stEnvMap{\mpChanRole{\mpS}{\roleP_1}}{\stS[1]}\stEnvComp
                  \dots
                  \stEnvComp
                  \stEnvMap{\mpChanRole{\mpS}{\roleP_n}}{\stS[n]}
                }
                {
                  \kills{s}
                }$}%
          }%
          \qquad
          \inference[\colorbox{ColourShade}{\iruleMPCancel}]{%
            \colorbox{ColourShade}
            {$\stJudge{\mpEnv}{\stEnv}{%
                  {\mpQ}%
                }$}%
            %
            %
            %
          }{%
            \colorbox{ColourShade}{$\stJudge{\mpEnv}{\stEnv\stEnvComp \stEnvMap{c}{S}}
                {
                  \mpCancel{c}{Q}
                }$}%
          }%

\\[3mm]
          \inference[\iruleMPDef]{%
            \stJudge{%
              \mpEnv \mpEnvComp%
              \mpEnvMap{\mpX}{\stS[1],\dots,\stS[n]}%
            }{%
              \stEnvMap{x_1}{\stS[1]}%
              \stEnvComp \dots \stEnvComp%
              \stEnvMap{x_n}{\stS[n]}%
            }{%
              \mpP%
            }%
            \qquad%
            \stJudge{%
              \mpEnv \mpEnvComp%
              \mpEnvMap{\mpX}{\stS[1],\dots,\stS[n]}%
            }{%
              \stEnv%
            }{%
              \mpQ%
            }%
          }{%
            \stJudge{\mpEnv}{%
              \stEnv%
            }{%
              \mpDef{\mpX}{%
                \stEnvMap{x_1}{\stS[1]},%
                \dots,%
                \stEnvMap{x_n}{\stS[n]}%
              }{\mpP}{\mpQ}%
            }%
          }%
          \\[3mm]%
          \inference[\iruleMPCall]{%
            \mpEnvEntails{\mpEnv}{X}{%
              \stS[1],\dots,\stS[n]%
            }%
           &                             %
            \stEnvEndP{\stEnv[0]}%
           &                             %
            \forall i \in 1..n%
           &                             %
            \stEnvEntails{\stEnv[i]}{\mpC[i]}{\stS[i]}%
          }{%
            \stJudge{\mpEnv}{%
              \stEnv[0] \stEnvComp%
              \stEnv[1] \stEnvComp \dots \stEnvComp \stEnv[n]%
            }{%
              \mpCall{\mpX}{\mpC[1],\dots,\mpC[n]}%
            }%
          }%
          \\[3mm]
          \begin{array}{@{}c@{}}
            \inference[\iruleMPSafeRes]{%
              \stEnvi = \setenum{%
                \stEnvMap{\mpChanRole{\mpS}{\roleP}}{\stS[\roleP]}%
              }_{\roleP \in I}%
              \quad%
              \mpS \not\in \stEnv%
              \quad%
              \stEnvSafeP{\stEnvi}%
              \quad%
              \stJudge{\mpEnv}{%
                \stEnv \stEnvComp \stEnvi%
              }{%
                \mpP%
              }%
            }{%
              \stJudge{\mpEnv}{%
                \stEnv%
              }{%
                \mpRes{\stEnvMap{\mpS}{\stEnvi}}\mpP%
              }%
            }%
\\[3mm]
            \inference[\colorbox{ColourShade}{\iruleMPInit}]{%
              \colorbox{ColourShade}{$\stEnvi = \setenum{%
                \stEnvMap{\mpChanRole{\mpS}{\roleP}}{\gtProj{\gtG}{\roleP}}%
              }_{\roleP \in \gtRoles{\gtG}}%
              \text{ or } \stEnvEndP{\stEnvi}$}
              \quad
                    \mpS \not\in \stEnv%
\quad
              \stJudge{\Theta}{\stEnv \stEnvComp \stEnvi}{%
              }{%
                \mpP%
              }%
            }{%
              \stJudge{\Theta}{\stEnv}{%
                \mpRes{\stEnvMap{\mpS}{\stEnvi}}
                {\mpP}%
              }%
            }%
          \end{array}
\end{array}
\]
  %
  \iftoggle{full}{%
}{%
}%
  \caption{%
    Multiparty session typing rules.
    We highlight \colorbox{ColourShade}{the new rules} from~\cite{scalasLess2019}.
  }%
  \label{fig:mpst-rules}%
  \iftoggle{full}{%
}{%
}%
\end{figure}

We explain each rule highlighting the new rules from~\cite{scalasLess2019}.
\iftoggle{full}{%
}{%
}%
\begin{description}
  \item[(Affine) Branching/Selection]
\inferrule{\iruleMPBranch} and \inferrule{\iruleMPSel} are the
 standard rules for branching and selection, which can also type
affine branching and selection.
Note that
    the premise
    $\stEnv$ in $\stJudge{\mpEnv}{%
        \stEnv \stEnvComp%
        \stEnvMap{y_i}{\stS[i]} \stEnvComp%
        \stEnvMap{\mpC}{\stSi[i]}%
      }{%
        \mpP[i]%
      }%
    $ in \inferrule{\iruleMPBranch} ensures that
    selecting one branch in the reduction rule defined by
$\inferrule{\iruleMPCBra}$ is
    sufficient for ensuring type soundness.


  \item[Try-Catch and Cancellation]
    \inferrule{\iruleMPTryCatch} is typing a try process: we ensure
    $P$ has a unique subject and catch block process $Q$ has
    the same session typing (similar with branching).
    $\inferrule{\iruleMPCancel}$ generates a kill process at its
    declared session.

  \item[Kill process]
  \inferrule{\iruleMPKills} types
  a kill process that appears
      during reductions: the cancellation of
      $\mpChanRole{\mpS}{\roleP}$ is broadcasting the cancellation
      to all processes which belong to session $\mpS$.

  \item[Recursions]
    \inferrule{\iruleMPDef} and \inferrule{\iruleMPCall}
    are identical to those of~\cite{scalasLess2019}.

  \item[Restriction]
    Processes are initially typed projecting a global type by $\inferrule{\iruleMPInit}$,
    while running processes are typed by \inferrule{\iruleMPSafeRes}
    (see the proof of Theorem~\ref{thm:subjectreduction}).
\end{description}

\begin{restatable}[Typing \AMPST processes]{example}{exampleTypingAMPSTProcesses}
  \label{ex:typing}
To demonstrate the typing rules we type the inner \emph{try} process from the reduction example.
Let $\mpQ
  =
  \trycatch{
    \mpR
  }{
    \mpCancel{
      \mpChanRole{\mpS}{\roleP}
    }{
      \mpNil
    }
  }
$
where $\mpR =  \mpQuestionSel{
    \mpChanRole{\mpS}{\roleP}
  }{
    \roleR
  }{
    \mpResponse
  }{
    \mpConst
  }{
    \mpNil
  }$ and
$\mpConst$ is of type $S_1=\stEnd$.
We show that $\stJudge{}{\stEnv}{\mpQ}$
where $
  \stEnv=\stEnvMap{\mpConst}{S_1},
  \stEnvMap{\mpChanRole{\mpS}{\roleP}}{S_2}$
with $S_2 = \stIntSum{\roleR}{}{\stChoice{\mpResponse}{S_1}\stSeq\stEnd}$.

\label{ex:derivation:solution}
{\footnotesize
  \[
    \begin{array}{l}
      \inference[\iruleMPTryCatch]{
        \inference[\iruleMPSel]{
          {
            \stJudge{}{
              \stEnvMap{\mpChanRole{\mpS}{\roleP}}{S_2}
            }{
              \stEnvMap{\mpChanRole{\mpS}{\roleP}}{S_2}
            }
            \quad
            \stJudge{}{\stEnvMap{\mpConst}{S_1}}{\stEnvMap{\mpConst}{S_1}}
            \quad
            \inference[\iruleMPNil]{
              \dots
            }{
              \stJudge{}{\stEnvMap{\mpChanRole{\mpS}{\roleP}}{\stEnd}}{\mpNil}
            }
          }
        }{
          \stJudge{}{
            \stEnv
          }{
            \mpQuestionSel{
              \mpChanRole{\mpS}{\roleP}
            }{
              \roleR
            }{
              \mpResponse
            }{
              \mpConst
            }{
              \mpNil
            }
          }
        }
        \sbj{\mpR}=
        \{
        \mpChanRole{\mpS}{\roleP}
        \}
        \quad
        \inference[\iruleMPCancel]{
          \inference[\iruleMPNil]{
            \dots
          }{
            \stJudge{}{\stEnvMap{\mpConst}{S_1}}{\mpNil}
          }
        }{
          \stJudge{}{
            \stEnv
          }{
            \mpCancel{\mpChanRole{\mpS}{\roleP}}{\mpNil}
          }
        }
      }
      {
        \stJudge{}{
          \stEnv
        }{
          \mpQ
        }
      }
    \end{array}
  \]}

  \iftoggle{full}{%
}{%
}%
\end{restatable}

\subsection{Properties of affine multiparty session types}
\label{sec:properties}
This subsection proves the main properties of \AMPST processes.
We first prove basic properties such as Subject Congruence and 
Reduction Theorems, then prove important properties, 
session fidelity, deadlock-freedom and liveness. 
The highlight is cancellation termination, which guarantees 
that once an exceptional behaviour is triggered, 
all parties in a single session can terminate as nil processes. 



Unlike linear-logic based typing systems
\cite{mostrousAffine2018}, we do \emph{not}
assume that the typing system is closed modulo
$\equiv$. Instead, we prove closedness of $\equiv$ for 
tricky cases, e.g., kill and try-catches.

\begin{restatable}[Subject Congruence]{theorem}{thmMPSTSubectCongruence}\rm
  \label{thm:subjectcong}
  If $\stJudge{\mpEnv}{\stEnv}{\mpQ}$ and
  $\mpQ\!\equiv\! \mpP$, then
  we have
  $\stJudge{\mpEnv}{%
      \stEnv%
    }{%
      \mpP
    }%
  $.
\end{restatable}



By 
Theorem~\ref{thm:subjectcong},
\AMPST processes satisfy \emph{type soundness}.

\begin{restatable}[Subject Reduction]{theorem}{thmMPSTSubectReduction}\rm
  \label{thm:subjectreduction}
  Suppose
$\stJudge{\mpEnv}{\stEnv}{\mpP}$ and $\stEnv$ safe.
Then, $\mpP \mpMove \mpPi$
  implies there exists $\stEnvi$ such that $\stEnvi$ is safe and
$\stEnv \stEnvMove^\ast \stEnvi$ and
$\stJudge{\mpEnv}{\stEnvi}{\mpPi}$.%
\end{restatable}

\iftoggle{full}
{%
  \begin{proof}
    See~\Cref{app:AMPST}.
    While the syntax of types stays the
    same as~\cite{scalasLess2019},
    the proof needs extra care
    due to reduction semantics not presented in~\cite{scalasLess2019}.
  \end{proof}
}{%
}%

A single agent in a multiparty session $s$ is a
participant playing a single role $\roleP$ in $s$.
We use the definition from~\cite{scalasLess2019} except the
highlighted part, which now includes affine processes.

\begin{restatable}[A unique role process]{definition}{definitionAUniqueRoleProcess}
  \label{def:unique-role-proc}%
  Assume\; $\stJudge{\mpEnvEmpty}{\stEnv}{\mpP}$.
  \;We say that $\mpP$:
  \begin{enumerate}[leftmargin=4mm]%
    \item\label{item:guarded-definitions:stmt}%
    \textbf{has guarded definitions} %
    \;iff\; %
    in each subterm of the form\, %
    $\mpDef{\mpX}{\stEnvMap{x_1}{\stS[1]},\dots,\stEnvMap{x_n}{\stS[n]}}{
        \mpQ}{\mpPi}$, %
    \,for all $i \in 1..n$,\, %
    $\stS[i] \!\stNotSub\! \stEnd$ %
    \,implies\, %
    that a call $\mpCall{\mpY}{\dots,x_i,\dots}$ %
    can only occur in $\mpQ$ %
    as subterm of\; %
    \colorbox{ColourShade}{$\mpDagger\mpBranch{x_i}{\roleQ}{j \in J}{\mpLab[j]}{y_j}{\mpP[j]}$} %
    \,or\, %
    \colorbox{ColourShade}{$\mpDagger\mpSel{x_i}{\roleQ}{\mpLab}{\mpC}{\mpPii}$} %
    (\ie after using $x_i$ for selection/branching);%
    \item\label{item:unique-role-proc:stmt}
    \textbf{only plays role $\roleP$ in $\mpS$, by $\stEnv$\!,} %
    \,iff:\; %
    \begin{itemize*}
      \item[i)]%
      $\mpP$ has guarded definitions;\; 
      \item[ii)]%
      $\fv{\mpP} \!=\! \emptyset$;\; %
      \item[iii)]\label{item:unique-role-proc:stenv}%
      $\stEnv \!=\!%
        \stEnv[0] \stEnvComp \stEnvMap{\mpChanRole{\mpS}{\roleP}}{\stS}$
      with %
      $\stS \!\stNotSub\! \stEnd$ %
      and %
      $\stEnvEndP{\stEnv[0]}$;\; %
      \item[iv)]\label{item:unique-role-proc:res-end}%
      in all subterms %
      $\mpRes{\stEnvMap{\mpSi}{\stEnvi}}{\mpPi}$ %
      of $\mpP$, %
      we have $\stEnvi = \stEnvMap{\mpChanRole{\mpSi}{\rolePi}}{\stEnd}$ (for some $\rolePi$).
    \end{itemize*}
  \end{enumerate}
  We say ``\textbf{$\mpP$ only plays role $\roleP$ in $\mpS$}'' %
  \,iff\, %
  $\exists\stEnv: \stJudge{\mpEnvEmpty\!}{\!\stEnv\!}{\!\mpP}$, %
  and item~\ref{item:unique-role-proc:stmt} holds.
\end{restatable}
Note that by definition, a unique role process in $\mpS$ includes
$\kills{s}$.

\emph{Session fidelity} is an important property to ensure
liveness and deadlock-freedom, as well as termination.
We extend that in~\cite{scalasLess2019} by taking a kill process into account.
A set of unique role processes of a single multiparty session,  
together with kill processes always make progress if 
a typing context has progress,
satisfying a protocol compliance.  

Below we write $\kills{Q}$ if $Q$ contains only a parallel composition
of kill processes.

\begin{restatable}[Session Fidelity]{theorem}{lemSessionFidelity}%
  \label{lem:session-fidelity}%
  Assume\, $\stJudge{\mpEnvEmpty\!}{\!\stEnv}{\!\mpP}$, %
  where %
  $\stEnv$ is safe, %
  \,$\mpP \equiv \mpBigPar{\roleP \in I}{\mpP[\roleP]}\mpPar \kills{Q}$, %
  \,and\, $\stEnv = \bigcup_{\roleP \in I}\stEnv[\roleP]\cup \stEnv[0]$
  such that, for each $\mpP[\roleP]$, %
  we have\, $\stJudge{\mpEnvEmpty\!}{\stEnv[\roleP]}{\!\mpP[\roleP]}$;
  and $\stJudge{\mpEnvEmpty\!}{\stEnv_0}{\kills{Q}}$.
  Assume that each $\mpP[\roleP]$
  is either\, $\mpP[\roleP]\equiv\mpNil$, %
  or only plays $\roleP$ in $\mpS$, by $\stEnv[\roleP]$.
  Then, %
  $\stEnvMoveP{\stEnv}$ %
  \,implies\; %
  $\exists \stEnvi,\mpPi$ %
  such that\, %
  $\stEnv \!\stEnvMove\! \stEnvi$, %
  $\mpP \!\mpMoveStar\! \mpPi$ %
  \,and\, %
  $\stJudge{\mpEnvEmpty\!}{\!\stEnvi}{\mpPi}$, %
  \;with\; %
  $\stEnvi$ safe, %
  \,$\mpPi \equiv \mpBigPar{\roleP \in I}{\mpPi[\roleP]}\mpPar \kills{\mpQi}$, %
  \,and\, $\stEnvi = \bigcup_{\roleP \in I}\stEnvi[\roleP]\cup
  \stEnvi[0]$
such that, for each $\mpPi[\roleP]$, %
we have\, $\stJudge{\mpEnvEmpty\!}{\stEnvi[\roleP]}{\!\mpPi[\roleP]}$,
\,and each $\mpPi[\roleP]$
is either\, $\mpNil$, %
or only plays $\roleP$ in $\mpS$, by $\stEnvi[\roleP]$;
and $\stJudge{\mpEnvEmpty\!}{\stEnvi_0}{\kills{\mpQi}}$.
\end{restatable}


By the above theorem, we can prove deadlock-freedom and liveness for
a single session multiparty session in the presence of
affine processes.

\begin{restatable}[Deadlock-freedom and liveness]{definition}{defMPSTProcessProperties}\rm
  \label{def:process-properties}%
  \label{def:process-term}%
  \label{def:process-nterm}%
  \label{def:process-df}%
\quad
\begin{enumerate}
\item 
  $\mpP$ is \textbf{deadlock-free} %
  iff %
  $\mpP \!\mpMoveStar\! \mpNotMoveP{\mpPi}$ %
  implies %
  $\mpPi \!\equiv\! \mpNil$.
  %
\item \;$\mpP$ is \textbf{live} %
  iff %
  $\mpP \!\mpMoveStar\! \mpPi \!\equiv\! \mpCtxApp{\mpCtx}{\mpQ}$ \,implies:
  \begin{itemize*}
    \item[i)]\label{item:process-liveness:sel}%
    if\; %
    $\mpQ = \mpSel{\mpC}{\roleQ}{\mpLab}{\mpChanRole{\mpSi}{\roleR}}{\mpQi}$ %
    (for some $\mpLab, \mpSi, \roleR, \mpQi$), %
    \;then\; %
    $\exists \mpCtxi$: %
    $\mpPi \!\mpMoveStar\! \mpCtxApp{\mpCtxi}{\mpQi}$; %
    \;and
    \item[ii)]\label{item:process-liveness:branch}%
    if\; %
    $\mpQ = \mpBranch{\mpC}{\roleQ}{i \in I}{\mpLab[i]}{x_i}{\mpQi[i]}$ %
    (for some $\mpLab[i], \mpFmt{x_i}, \mpQi[i]$), 
    \;then\; %
    $\exists \mpCtxi, k \!\in\! I, \mpSi, \roleR$:\, %
    $\mpPi \!\mpMoveStar\! \mpCtxApp{\mpCtxi}{\mpQi[k]\subst{x_k}{\mpChanRole{\mpSi}{\roleR}}}$.%
\end{itemize*}
\end{enumerate}
\end{restatable}
Note that liveness is defined for 
\emph{linear} selection or \emph{linear} branching 
processes
which appear at the top level, \ie under the
reduction context $\mpCtx{}{\ }$, 
not under \CODE{try-catch} construct, cancel nor affine branching and selection 
processes. 

\begin{restatable}[Deadlock-freedom]{theorem}{thmProcessDeadfree}
\label{lem:process-df-live-from-ctx}%
  Assume\; $\stJudge{\mpEnvEmpty}{\stEnv}{\mpP}$,
  \;with $\stEnv$ safe,\;
  $\mpP \equiv \mpBigPar{\roleP \in I}{\mpP[\roleP]}$,
  each $\mpP[\roleP]$ %
  either $\mpP[\roleP]\equiv\mpNil$, 
  or only playing role $\roleP$ in $\mpS$.
  \;Then,\; %
    \,$\stEnvDFP{\stEnv}$
    implies that %
    $\mpRes{\stEnvMap{\tilde{\mpS}}{\stEnv}}
                {\mpP}$ with $\{\tilde{\mpS}\}=\dom{\stEnv}$ 
is deadlock-free.
\end{restatable}

\iftoggle{full}{%
  \begin{proof}
    We use Theorem~\ref{lem:session-fidelity}.
    \inApp
  \end{proof}
}{%
}%

As discussed in \cite[Definition 5.11]{scalasLess2019}, 
we require $\stEnvLivePlusP{\stEnv}$ for proving 
liveness. 

\begin{restatable}[Liveness]{theorem}{thmProcessLiveness}
\label{lem:process-live-from-ctx}%
  Assume\; $\stJudge{\mpEnvEmpty}{\stEnv}{\mpP}$,
  \;with $\stEnv$ safe,\;
  $\mpP \equiv \mpBigPar{\roleP \in I}{\mpP[\roleP]}$,
  each $\mpP[\roleP]$ %
  either $\mpP[\roleP]\equiv\mpNil$,  
  or only playing role $\roleP$ in $\mpS$.
  \;Then,\; %
    \,$\stEnvLivePlusP{\stEnv}$ %
    implies that %
    $\mpP$ is live. 
\end{restatable}

\iftoggle{full}{%
  \begin{proof}
    We use Theorem~\ref{lem:session-fidelity}. 
    \inApp
  \end{proof}
}{%
}%

Now we consider a user-written Rust program with one session as 
an \emph{initial program}.

\begin{restatable}[Initial program]{definition}{definitionInitialProgram}
  \label{def:initialsingle}
  \label{def:imps}%
  \rm
  We say
  $\stJudge{}{}{\mpQ}$ is an \emph{initial program}
  if
\begin{enumerate}
\item  $\mpQ \equiv \mpRes{\stEnvMap{\tilde{\mpS}}{\stEnv}}{\mpBigPar{\roleP
      \in \gtG}{\mpP[\roleP]}}$ with $\{\tilde{\mpS}\}=\dom{\stEnv}$;
\item $\mpP[\roleP]$ only plays $\roleP$ in $\mpS$;
\item in each subterm of the form, \ %
    $\mpDef{\mpX}{\widetilde{x}}{\mpQ}{\mpPi}$,
    (1) $\mpQ$ is of the form $\trycatch{\mpQi}{\mpPii}$;
and (2) $\mpPii$ does not contain any (free or bound) process call.
\item $\stEnv = \setenum{\stEnvMap{\mpChanRole{\mpS}{\roleP}}{\gtProj{\gtG}{\roleP}}}_{\roleP \in \gtG},\stEnv'$
  for some $\gtG$ and $\stEnvEndP{\stEnv'}$;
\item $\stJudge{}{}{\mpQ}$ is derived
using \inferrule{\iruleMPInit}
instead of
$\inferrule{\iruleMPSafeRes}$; and
without $\inferrule{\iruleMPKills}$.
\end{enumerate}
\end{restatable}

Condition (3) ensures that once a process moves to the
\emph{catch-block}, then it ensures  finite computation; (4,5) state
that the initial program starts conforming to a global protocol.

\begin{restatable}[Initial processes]{remark}{remarkInitialProcesses}
Condition (3) does not limit the expressiveness since
the \emph{try-block} can include  infinite computations; and 
conditions (4,5) imply that
an initial program typed by
condition (1) has started. 
Notice that running (runtime) processes generated from the initial program
are typed using
\inferrule{\iruleMPSafeRes} and \inferrule{\iruleMPKills}; 
hence the proof of the subject reduction holds with
\Cref{lem:globalsafe} below.
\end{restatable}

Before proving the main theorems, we state that a set of local types
projected from a well-formed global type satisfy the safety property.

\begin{restatable}[{\cite[Lemma 5.9]{scalasLess2019}}]{lemma}{lemmaScalas}
\label{lem:globalsafe}
\label{lem:globalsafety}
Let $\stEnv =
\{\stEnvMap{\mpChanRole{\mpS}{\roleP}}{\gtProj{\gtG}{\roleP}\}_{\roleP \in \gtRoles{\gtG}}}$.
Then $\stEnvSafeP{\stEnv}$, $\stEnvDFP{\stEnv}$ and 
$\stEnvLivePlusP{\stEnv}$.
\end{restatable}

Now we state the two main theorems of this paper:
deadlock-freedom, liveness and
cancellation termination.
The cancellation termination theorem states that
once a kill signal is produced by
cancellation or affine processes
(due to a timeout or an
error), then all processes are enabled to terminate.
We start from deadlock-freedom.

\iftoggle{full}{%
  \begin{restatable}[Deadlock-freedom for an initial program]{corollary}{colDeadlockFree}\rm
    \label{cor:liveness}
    Suppose $\stJudge{}{}{\mpQ}$ is an initial program.
    Then for all $P$ such that $\mpQ\!\mpMoveStar\! \mpP$, $\mpP$
    is deadlock-free.
  \end{restatable}

  \begin{proof}
    By Theorem~\ref{lem:process-df-live-from-ctx}
    and~\Cref{lem:globalsafe}.
  \end{proof}%

  \begin{restatable}[Liveness for an initial program]{corollary}{colLiveness}\rm
    \label{cor:livenessplus}
    Suppose $\stJudge{}{}{\mpQ}$ is an initial program.
    Then for all $P$ such that $\mpQ\!\mpMoveStar\! \mpP$, $\mpP$
    is live.
  \end{restatable}

  \begin{proof}
    By Theorem~\ref{lem:process-live-from-ctx}
    and~\Cref{lem:globalsafe}.
  \end{proof}
}{%
  \begin{restatable}[Deadlock-freedom and liveness for an initial program]{corollary}{colDeadlockFreeLiveness}\rm
    \label{cor:liveness}
    \label{cor:livenessplus}
    Suppose $\stJudge{}{}{\mpQ}$ is an initial program.
    Then for all $P$ such that $\mpQ\!\mpMoveStar\! \mpP$, $\mpP$
    is deadlock-free and live.
  \end{restatable}
}%

\begin{restatable}[Cancellation Termination]{theorem}{thmCancellationTermination}\
  \label{thm:ctermination}%
  Suppose $\stJudge{}{}{\mpQ}$ is an initial program.
         If $\mpQ\!\mpMoveStar\!
           \mpCtx{}{[\kills{s}]}=\mpP'$,
          then we have $\mpP'\mpMoveStar\mpNil$.
\end{restatable}

\iftoggle{full}{%
  \begin{proof}
    \inApp We use Corollary~\ref{cor:liveness}.
  \end{proof}
}{%
}%

\begin{restatable}[Cancellation Termination of Affine and Cancel Processes]{corollary}{corCancellationTermination}
\label{cor:ctermination}%
Suppose $\stJudge{}{}{\mpQ}$ is an initial program.
\begin{enumerate}
   \item If $\mpQ\!\mpMoveStar\!
          \mpCtx{}{[{\mpCancel{\mpChanRole{\mpS}{\roleP}}{\mpQ'}}]}=\mpP'$,
         then we have $\mpP'\mpMoveStar\mpNil$.
   \item
         If $\mpQ\!\mpMoveStar\!
           \mpCtx{}{[\mpEtx{}{[
           \mpQuestionBranch{\mpChanRole{\mpS}{\roleP}}{\roleQ}{i \in I}{%
           \mpLab[i]}{x_i}{\mpP[i]}]}]}=\mpP'$ or
          $\mpQ\!\mpMoveStar\! \mpCtx{}{[\mpEtx{}{[
                  \mpQuestionSel{\mpChanRole{\mpS}{\roleP}}{\roleQ}{\mpLab}{
                    \mpChanRole{\mpS'}{\roleR}}{\mpP}]}]}=\mpP'$,
          then we have $\mpP'\mpMoveStar\mpNil$.
\end{enumerate}
\end{restatable}

\iftoggle{full}{%
  \begin{proof}
    By $\inferrule{\iruleMPCQSel}$,
    $\inferrule{\iruleMPTQSel}$,
    $\inferrule{\iruleMPCQBra}$,
    $\inferrule{\iruleMPTQBra}$,
    and $\inferrule{\iruleMPRedCan}$, we have
    $P'\mpMove \mpCtxi{}{[\kills{s}]}$.
    Then we apply Theorem~\ref{thm:ctermination}.
  \end{proof}
}{%

}%

\begin{restatable}[Termination theorem]{remark}{remarkTerminationTheorem}
The cancellation termination theorem
means that \emph{there always exists a path} which leads to $\mpNil$;
and an initial program might not terminate even if it contains
a process with $\kills{s}$.
This differs from the total termination, \ie all paths are finite --
a program will definitely stop as $\mpNil$. 
However, if we apply \emph{fair traversal sets}, \ie fair scheduling,
from \cite[Definition 5.5]{scalasLess2019}, applying to processes 
in $\mpCtx{}{[\kills{s}]}$, we can prove the total termination.  
Since these extensions require an introduction of labelled transition 
systems for processes, we leave it as future work. 
\end{restatable}

\iftoggle{full}{%
}{%
}%
\section{Design and implementation of~\mpstrust}
\label{sec:implementation}
\iftoggle{full}{%
}{%
}%
\subsection{Challenges for the implementation of~\mpstrust}
\label{sec:challenges}
\iftoggle{full}{%
}{%
}%
The three main challenges underpinning the implementation of
\AMPST in Rust are related to multiparty
communications and ensuring correctness for affine channels.

{\bf (Challenge 1) Realising a multiparty channel by binary channels.} \
\AMPST relies on a \emph{multiparty channel} -- a channel that can communicate with
several roles.
In Rust, communication channels are peer-to-peer,
\eg they are \emph{binary}~\cite{kokkeRusty2019}.
To overcome this limitation, we extend an encoding of \MPST
into binary channels~\cite{scalasLinear2017}.
In this encoding, a multiparty channel
can be represented as an \emph{indexed tuple} of \emph{one-shot binary channels}
used in a sequence depending on the ordering specified by the type.
This design ensures reception error safety by construction.
Since each pair of binary channels is dual, then no communication mismatch can occur.
We piggyback on this result by introducing meshed channels,
which reuse an existing library of binary session types in
Rust~\cite{kokkeRusty2019} with built-in duality guarantees.
We explain the implementation of meshed channels 
in \S~\ref{sec:api}.
\iftoggle{full}{%
      notably \emph{distributed logging} (\Cref{fig:GCLMLogging}),
      and \emph{circuit breaker} (\Cref{fig:GCLMApi}).
}{%
      See~\cite{fullVersion} for usecases that demonstrate
      how to use~\mpstrust for programming
      distributed protocols.
}%

{\bf (Challenge 2) Deadlock-freedom, liveness and termination.}
Duality is unfortunately insufficient to guarantee deadlock-freedom.
The naive decomposition of binary channels leads to
\emph{hard to} detect deadlock errors~\cite{scalasLinear2017}.
To ensure liveness properties and correct termination
of cancellation behaviour, we integrate~\mpstrust with
two state-of-the-art
verification toolchains -- Scribble~\cite{jespersenSession2015}
and \kmc~\cite{langeVerifying2019},
that ensures meshed channel types are \emph{correct}.
The former generates correct meshed channel types in Rust,
while the latter verifies a set of existing meshed channel types.
In both cases, well-typed processes implemented using well-typed meshed channels
are free from deadlocks, orphan messages and reception errors.
We display the Rust types for our running example
in \S~\ref{sec:typing}.

{\bf (Challenge 3) Affinity with \CODE{try-catch} and optional types.} \
Rust does not have a native \CODE{try-catch} construct,
but macros and optional types.
We use them to design and implement
a \CODE{try-catch} block and affine selection and branching.
Channels can be implicitly or explicitly cancelled,
and all processes are
guaranteed to terminate gracefully
in the event of a cancellation,
avoiding endless cascading errors.
We discuss our design choices in~\S~\ref{sec:exceptions}.
\iftoggle{full}{%
}{%
}%
\subsection{Meshed Channels in~\mpstrust}
\label{sec:api}
\iftoggle{full}{%
}{%
}%
\begin{figure}[t]
\begin{minipage}[l]{1\textwidth}
\begin{rustlisting}[basicstyle=\scriptsize\ttfamily]
pub struct MeshedChannels< S1: Session, S2: Session, R: Role, N: Role> {*@ \label{line:mpst:beg} @*
pub session1: S1, pub session2: S, pub stack: R, pub name: N } *@ \label{line:mpst:end} @*
\end{rustlisting}
\iftoggle{full}{%
}{%
}%
\end{minipage}
\caption{Generated MeshedChannles structure.}
\label{fig:MPSTMeshedChannelsPrimitivesShort}
\iftoggle{full}{%
}{%
}%
\end{figure}
A multiparty channel in~\mpstrust is realised as an \emph{affine meshed channel} (hereafter meshed channel),
which has three ingredients:
(1) a list of separate binary channels (one binary channel for each pair of
participants); (2) a stack that imposes the ordering between the binary channels;
and (3) the name of the role, whose behaviour is implemented by the meshed channel.
For example, a meshed channel for a 3-party protocol can be generated using the macro
\CODE{gen_mpst!(MeshedChannels, A, C, S)}:
the name of the structure (\CODE{MeshedChannels}),
and the three roles involved in
the communication (\CODE{A, C, S}).
~\Cref{fig:MPSTMeshedChannelsPrimitivesShort} 
shows the generated structure.

The generated structure, \CODE{MeshedChannels}, holds four fields.
The first two fields, \CODE{session1} and \CODE{session2}, are of type \CODE{Session} which
is a binary session type. Therefore, these fields store binary channels.
\CODE{Session} in Rust is a \CODE{trait}
and a \CODE{trait} is similar to an interface.
The \CODE{Session} trait can be instantiated to three generic (binary session) types:
an \CODE{End} type; a \CODE{Recv<T, S>} or a \CODE{Send<T, S>} type,
with their respective payload of type \CODE{T} and their continuation of a binary session type \CODE{S}.
This has important implications for the design and safety of our system.
Since all pairs of binary channels
are created and distributed across meshed channels at the start of the protocol,
the binary type \CODE{Session} enforces that each pair of binary channels are dual.
For example, the binary channel for \role{S} inside the meshed channels for \role{A};
and the binary channel for \role{A} inside
the meshed channels for \role{S} are dual.
This design ensures that, without using any external tools, our system is communication safe,
no reception error can occur. 
This is insufficient to guarantee deadlock-freedom,
which is why we utilise Scribble or bounded model checking, \ie \kmc, as an additional verification step.

The rest of the fields of the \CODE{struct MeshedChannels} are stack-like structures, \CODE{stack} and \CODE{name}, which represent
respectively the order of the interactions (in what order the binary channels should be used)
and the associated role.
For instance, the behaviour where \role{A} has to communicate first with \role{S},
then with \role{C} and then the session ends, can be specified using a
stack of type \CODE{RoleS<RoleC<RoleEnd>>}. Note that all stack
types such as \CODE{RoleS} and \CODE{RoleC} are generated singleton types.
Role names are codified as \CODE{RoleX<RoleEnd>} where \CODE{X} is the actual
name of the participant. For instance, \role{A} is realised as the singleton type \CODE{RoleA<RoleEnd>}.
We chose this design for its readability and its ease of implementation:
one can guess at a glance the current state of a participant.

The code generation macro \CODE{gen_mpst!}
produces \emph{meshed channels for any finite number of communicating processes}.
For example, in the case of a protocol with four roles, the macro
\CODE{gen_mpst!} will generate a meshed channel with five
fields -- one field for the binary session between each pair of
participants (which is 3 fields in total), one field for the stack and one field for the name of the role that is being implemented.

\iftoggle{full}{%

\subsection{Types for affine meshed channels}
\label{sec:typing}

Meshed channel types -- \CODE{MeshedChannels} -- correspond
to local session types.
They describe the behaviour of each meshed channel
and specify which communication primitives 
are permitted on a meshed channel. 
To better illustrate meshed channel types, we explain the type \CODE{RecA<N>} for 
\role{A} (Authenticator) from~\Cref{fig:implementation:role:a}.
The types are displayed in~\Cref{fig:UsecaseDeclarationTypes}.
The types of the meshed channels for the other roles,
\ie \norole{C} and \norole{S} are available
\iftoggle{full}{%
	in~\Cref{fig:implementation:s,fig:implementation:c} of the Appendix.
}{%
	in~\cite{fullVersion}.
}%

\begin{figure}[t!]
	\begin{rustlisting}[basicstyle=\scriptsize\ttfamily}
	// Declare the name of the role
	type NameA = RoleA<RoleEnd>; *@\label{line:roleA}@*
	
	// Binary session types for A and C
	type AtoCVideo<N> = Recv<N, Send<N, Recv<ChoiceA<N>, End>> *@\label{line:AtoCVideo}@*

	// Binary session types for A and S
	type AtoSVideo<N> = Send<N, Recv<N, End>>; *@\label{line:AtoSVideo}@*
	
	// Declare usage order of binary channels inside a meshed channel
	type StackAInit = RoleC<RoleEnd>; // for the initial meshed channel
	type StackAVideo = RoleC<RoleS<RoleS<RoleC<RoleEnd>>>>; // for branch Video *@\label{line:stack}@*
	
	// Declare the type of the meshed channel
	type RecA<N> = MeshedChannels<Recv<ChoiceA<N>, End>, End, StackAInit, NameA>; *@\label{line:RecA}@*
	
	// Declare an enum with variants corresponding to the different branches, \ie Video and End
	enum ChoiceA<N> { *@\label{line:ChoiceA}@*
		Video(MeshedChannels<AtoCVideo<N>, AtoSVideo<N>, StackAVideo, NameA>), *@\label{line:Video} @*
		Close(MeshedChannels<End, End, RoleEnd, NameA>)
	} *@\label{line:role:endpoint:full:end}@*
	\end{rustlisting}
	\caption{Local Rust types for \role{A} (Authenticator) from \Cref{fig:implementation:role:a}}
	%\Description[Local Rust types for \role{A} (Authenticator) from \Cref{fig:implementation:role:a}]{Local Rust types for \role{A} (Authenticator) from \Cref{fig:implementation:role:a}}
	\label{fig:UsecaseDeclarationTypes}
\end{figure}

Meshed channels consist of
(1) binary channels, (2) a stack that stipulates the order of interactions and
(3) the name of the related role. We explain how to declare each of these below.

\myparagraph{Declaring meshed channel types} 
Following the protocol, the first action on \norole{A}
is an external choice. Role \norole{A} should receive
a choice from \role{C} of either \CODE{Video} or \CODE{Close}.  
External choice is realised in~\mpstrust 
as an \CODE{enum} with a variant for each branch, 
where each variant is parameterised on the meshed channel that will be used for that branch. 
% To create a meshed channel that amalgamates the behaviours of \CODE{AtoCVideo} and \CODE{AtoSVideo}
% we need one more ingredient -- a stack that specifies in what order we are going to use the binary channels. 
Hence, the type for the meshed channel for \norole{A}
must specify that \norole{A} must first receive one of the enum variants 
and its corresponding meshed channel. Then, 
it should use the received meshed channel to realise the interactions for each branch. 
This precise behaviour is captured by the meshed channel type
\CODE{RecA<N>} declared at line~\ref{line:RecA},
see~\Cref{fig:UsecaseDeclarationTypes}.
Note that, in Rust, \CODE{type} \CODE{RecA<N>}
declares a type alias,
not a new type.
Here, \CODE{RecA<N>} for a meshed channel
whose binary channel for \CODE{S} (the type \CODE{Recv<ChoiceA<N>, End>} 
stored at the first position in the meshed channel structure) 
is expected to receive the enum \CODE{ChoiceA<N>}. 
% \rumi{mention StackaCInit}. 

The enum type \CODE{ChoiceA<N>} in line~\ref{line:ChoiceA} 
declares two variants with their respective meshed channels. 
The branch \CODE{Close} is trivial since
no communication apart from closing all channels is expected in this branch. 
Hence, the binary channels for \norole{S} and \norole{A}, and 
\norole{C} and \norole{A} are all \CODE{End}. 
The type of the meshed channel for the branch \CODE{Video} in line~\ref{line:Video} is more elaborate. 
\CODE{MeshedChannels<AtoCVideo<N>, AtoSVideo<N>, StackAVideo, NameA>} 
specifies that the type of the binary channel for \norole{C} and \norole{A} is 
\CODE{AtoCVideo<N>}, the type of the binary channel for \role{S} and \role{A}
is \CODE{AtoSVideo<N>}, the stack of the meshed channel is  \CODE{StackAVideo}. 
The last argument specifies that this is a meshed channel for \role{A}. 

\myparagraph{Declaring meshed channel stacks} 
First, let us recap the behaviour for \role{A} in the \CODE{Video} branch. 
Role \norole{A} first receives a request from \norole{C}, 
then it forwards it to \norole{S}, it waits to receive the response from \norole{S}, 
and finally sends it back to \norole{C}. 
This sequence of interactions is captured in the declaration of the stack \CODE{StackAVideo} 
in line~\ref{line:stack}. 
The declaration \CODE{RoleC<RoleS<RoleS<RoleC<RoleEnd>>>>} specifies 
the order in which binary channels must the used -- first the binary channel with \norole{C}, 
then with \role{S}, then with \norole{S} again, and finally with \norole{C}.    

% For now, we will only focus on the behaviour in the Video branch, later we will explain how choice is handled.
\myparagraph{Declaring binary session types} 
% The \CODE{type} keyword declares an alias of another type.
Finally, we must declare the types for the binary channels that \role{A} must utilise for 
communication with \role{A}  and \role{C} in the \CODE{Video} branch.
The binary type for \role{A} and \role{C} is type \CODE{AtoCVideo<N>}
in line~\ref{line:AtoCVideo}. 
Note that the keyword \CODE{type} declares an alias,
hence the \CODE{AtoCVideo<N>} is an alias 
for the binary channel type \CODE{Recv<N, Send<N, RecvChoiceA<N>>>} 
which specifies that the channel between \norole{S} and \norole{C}
should be used for receiving a payload \CODE{N}, 
then should return a continuation \CODE{Send<N, Recv<ChoiceA<N>, End>>}. 
The continuation is another binary session type with capabilities for sending a payload \CODE{N}, 
and returning a continuation \CODE{Recv<ChoiceA<N>, End>}. 
The type for the binary channel between \role{A}  and \role{S} in the \CODE{Video} 
branch is similar. 
We remind the reader that \role{A} must send a request and then wait to receive a response from \role{S}. 
This is codified in the type \CODE{AtoSVideo<N>}, declared in line~\ref{line:AtoSVideo}. 
The type \CODE{Send<N, Recv<N, End>>} says the channel will first be used to send and then to receive. 

With the above declarations, we are done specifying all behaviours in the branch \CODE{Video}
and the type declaration for the meshed channel for \role{A} is complete.

\myparagraph{Recursive types}
Note that none of the defined stacks is recursive.
Recursion is implicitly specified on binary types, and each stack is only related to
a \CODE{MeshedChannels} up to either a choice or the end of the protocol.
When a choice occurs, each participant adopts a new \CODE{MeshedChannels}.
To \emph{manually} detect the recursion in the current protocol,
one can see that the type \CODE{AtoCVideo},
inside the enum \CODE{ChoiceA},
is sent through the variant \CODE{Video} of this enum.
The type \CODE{AtoCVideo} ends with a \CODE{Recv<ChoiceA<N>, End>},
meaning that, in this branch,
\role{A} will receive this enum again,
making this branch recursive.

% \reviewadd{[$\rightarrow$}
Note also that we assume fair scheduling as in the theory.
In the case of fair scheduling, an infinite reduction
(after cancellation) cannot happen.
At every remaining step for each role,
either it communicates with a non cancelled role
and continues with the following steps or communicates
with a cancelled role and cancels itself.
At the very end of the protocol,
the \CODE{close()} method sends and receives a unit type
from and to every participant,
meaning that every participant will know, at one point,
whether a participant has failed or not.
% \reviewadd{$\leftarrow$]}
%\paragraph{\bf Distributed Execution Environment.} The default transport of~\mpstrust
%is the built-in Rust communication channels (\textbf{crossbeam_channel}).
% To test our example in a more realistic distributed environment,
%we have also connected each process through MQTT (MQ Telemetry Transport)~\cite{hunkelermqtt-spublishsubscribe2008}.
% MQTT is a
%messaging middleware for exchanging messages between devices, predominantly
%used in IoT networks.
% At the start of the protocol, each process connects to a
%public MQTT channel, and a session is established.
% Therefore, we have mapped
%binary channels to MQTT sockets, in addition to the built-in Rust channels.

% If one wants to use the code generation from Scribble,
% they need to download the tool and its Rust API.
% From a checked global protocol written in Scribble,
% this API will generate the types: the users need to implement
% the functions related to those types by themselves as,
% among other limits, the code generator does not know
% the conditions to choose one branch or another in a choice.

\myparagraph{Top-down meets bottom-up} 
The types described so far can be written either by the developers and verified using an external tool, \kmc, 
or generated from a global protocol written in Scribble.
\emph{Reception error safety} is ensured since the underlying~\mpstrust library checks \emph{statically} that
all pairs of binary types are dual to each
other. \emph{Deadlock-freedom} is ensured if meshed channel types are
generated from Scribble or verified using \kmc. Finally, if errors occur,
our implementation ensures all processes in the same session
are safely terminated.

To use the Scribble tool, 
developers have to specify the global protocol in the protocol
description language Scribble.
Examples of the global protocols written in Scribble are available
\iftoggle{full}{%
	in~\Cref{fig:GCLMLogging:scribble,fig:GCLMApi:scribble}.
}{%
	in~\cite{fullVersion}.
}%
Then, by running the Scribble toolchain on those protocols,
they can obtain a Rust file containing 
all meshed channel types needed to implement the processes.
An example of such generated types is provided
in~\Cref{fig:UsecaseDeclarationTypes}.
With this Rust file,
the developers can run the protocol
by implementing the processes,
such as the ones shown
in~\Cref{fig:implementation}.

To use the \kmc~tool, developers \emph{have to write the types} and use 
an~\mpstrust macro that rewrites all Rust types to CFSMs and
invokes the \kmc~tool at compile time.
However, they do not have to write the implementation
of the processes for the meshed channels,
such as the ones shown in~\Cref{fig:implementation}.
The \kmc~tool checks if the system of CFSMs are $k$-multiparty compatible. 
In particular, \kmc~performs a $k$-bounded execution of the set of CFSM and checks
if the system 
% two properties -- safety and exhaustiveness. 
has the progress property 
(no role gets permanently stuck in a receiving state) 
and the eventual reception property 
(all sent messages are eventually received).
It checks that in all system executions where 
each channel contains at most $k$ messages, these properties are respected. 
Note that well-formed global protocols 
correspond to the class of 1-MC safe systems.

% If a system of session types is k-MC, 
% then it is safe [9, Theorem 1], 
% \ie it has the progress property 
% (no role gets permanently stuck in a receiving state) 
% and the eventual reception property 
% (all sent messages are eventually received).

Hence, the two approaches (top-down and bottom-up)
are complementary and can be
used interchangeably for 1-MC systems. 
Note, however, that \kmc~(and hence~\mpstrust) potentially allows 
implementing systems that are safe but cannot be 
expressed as a global protocol.
In this sense, using 
the bottom-up approach inherently permits
more behaviours. 
To stay close to the theory, we have
restricted the execution of \kmc~to only 1-MC 
systems, but the developer can specify the bound as an argument if they wish to 
explicitly diverge from the class of 1-MC systems.
% \reviewadd{[$\rightarrow$}
The bottom-up approach can also be used
with other implementations.
The two approaches can also be combined:
the types can be generated from Scribble
and then checked by \kmc.
However, if the types are generated from Scribble,
they are guaranteed to be compatible,
hence no \kmc-check is required.
In that sense,
the user can choose if they wish to start from
the local types (the types for each communication channel)
and check if they are compatible,
or start from the global type
(and generate the types for each communication channel).
% \reviewadd{$\leftarrow$]}
}{%

\iftoggle{full}{%
}{%
% \vspace{-4mm}
}%
\subsection{Types for affine meshed channels}
\label{sec:typing}
\iftoggle{full}{%
}{%
% \vspace{-2mm}
}%

% \myparagraph{\bf Generated state channel}
% Assume a role $\roleQ$ communicating with role $\roleP$;
% the stack $\roleP.w$ of the order of operations
% where $\roleP$ is the head of the stack
% and $w$, the tail;
% a message labeled $m$ containing
% a payload of type $\stS[]$;
% the continuation of the operations $\stSi[]$;
% the subset $I$ of all branches at a given choice;
% the stack $\perp$ for broadcasting a choice;
% the binary channel between $\roleP$ and $\roleQ$
% in $\roleQ$'s endpoint is the second one;
% then~\Cref{tab:generatedTypes}
% details the generated types from the operations
% provided in our theory.
Meshed channel types -- \CODE{MeshedChannels} -- correspond
to local session types.
They describe the behaviour of each meshed channel
and specify which communication primitives 
are permitted on a meshed channel. 
% The correspondents of local session types and a meshed channel type in Rust is given 
% in Table X.  
% First, we explain table 1 and then we explain the meshed channel types using our previous example.
To better illustrate meshed channel types, we explain the type \CODE{RecA<N>} for 
\role{A} (Authenticator) from~\Cref{fig:implementation:role:a}.
The types are displayed in~\Cref{fig:UsecaseDeclarationTypes}.
% To make the use case more interesting,
% we added an extra interaction between the authenticator and the client:
% at the beginning of the protocol the client (hereafter \role{C})
% sends a request for authentication (hereafter \role{A}) and
% then \role{A} replies with an \CODE{id} token.
% Hereafter, we explain the type declaration for \role{A} in~\Cref{fig:UsecaseDeclarationTypes},
The types of the meshed channels for the other roles,
\ie \norole{C} and \norole{S} are available
\iftoggle{full}{%
	in~\Cref{fig:implementation:s,fig:implementation:c} of the Appendix.
}{%
	in~\cite{fullVersion}.
}%

\begin{figure}[t!]
	% \begin{adjustbox}{padding=20pt 0pt 0pt 0pt, fbox}
	\begin{rustlisting}[basicstyle=\scriptsize\ttfamily}
	// Declare the name of the role
	type NameA = RoleA<RoleEnd>; *@\label{line:roleA}@*
	
	// Binary session types for A and C
	type AtoCVideo<N> = Recv<N, Send<N, Recv<ChoiceA<N>, End>> *@\label{line:AtoCVideo}@*

	// Binary session types for A and S
	type AtoSVideo<N> = Send<N, Recv<N, End>>; *@\label{line:AtoSVideo}@*
	
	// Declare usage order of binary channels inside a meshed channel
	type StackAInit = RoleC<RoleEnd>; // for the initial meshed channel
	type StackAVideo = RoleC<RoleS<RoleS<RoleC<RoleEnd>>>>; // for branch Video *@\label{line:stack}@*
	
	// Declare the type of the meshed channel
	type RecA<N> = MeshedChannels<Recv<ChoiceA<N>, End>, End, StackAInit, NameA>; *@\label{line:RecA}@*
	
	// Declare an enum with variants corresponding to the different branches, \ie Video and End
	enum ChoiceA<N> { *@\label{line:ChoiceA}@*
		Video(MeshedChannels<AtoCVideo<N>, AtoSVideo<N>, StackAVideo, NameA>), *@\label{line:Video} @*
		Close(MeshedChannels<End, End, RoleEnd, NameA>)
	} *@\label{line:role:endpoint:full:end}@*
	\end{rustlisting}
	\iftoggle{full}{%
}{%
}%
	\caption{Local Rust types for \role{A} (Authenticator) from \Cref{fig:implementation:role:a}}
	\label{fig:UsecaseDeclarationTypes}
	\iftoggle{full}{%
}{%
}%
	\end{figure}

Following the protocol, the first action on \norole{A}
is an external choice. Role \norole{A} should receive
a choice from \role{C} of either \CODE{Video} or \CODE{Close}.  
External choice is realised in~\mpstrust 
as an \CODE{enum} with a variant for each branch, 
where each variant is parameterised on the meshed channel that will be used for that branch. 
The enum type \CODE{ChoiceA<N>} in line~\ref{line:ChoiceA} 
precisely specifies this behaviour -- two variants with their respective meshed channels. 
The branch \CODE{Close} is trivial since
no communication apart from closing all channels is expected in this branch. 
Hence, the binary channels for \norole{S} and \norole{A}, and 
\norole{C} and \norole{A} are all \CODE{End}. 
The type of the meshed channel for the branch \CODE{Video} in line~\ref{line:Video} is more elaborate. 
\CODE{MeshedChannels<AtoCVideo<N>, AtoSVideo<N>, StackAVideo, NameA>} 
specifies that the type of the binary channel for \norole{C} and \norole{A} is 
\CODE{AtoCVideo<N>}, the type of the binary channel for \role{S} and \role{A}
is \CODE{AtoSVideo<N>}, the stack of the meshed channel is  \CODE{StackAVideo}. 
The declaration \CODE{RoleC<RoleS<RoleS<RoleC<RoleEnd>>>>} specifies 
the order in which binary channels must the used -- first the binary channel with \norole{C}, 
then with \role{S}, then with \norole{S} again, and finally with \norole{C}. 
The last argument specifies that this is a meshed channel for \role{A}. 

The meshed channel types can be written either by the developers and verified using an external tool, \kmc, 
or generated from a global protocol written in Scribble.

}%

\iftoggle{full}{%
}{%
}%
\subsection{Exception and cancellation}
\label{sec:exceptions}
\iftoggle{full}{%
}{%
}%
\myparagraph{Exception handling}
Rust does not have exceptions.
Instead, it has the type \CODE{Result<T, E>} for
recoverable errors and
the \CODE{panic!} macro that stops execution when
the program encounters an unrecoverable error.
\CODE{Result<T, E>} is a variant type
with two constructors: \CODE{Ok(T)} and \CODE{Err(E)}
where \CODE{T} and \CODE{E} are generic type parameters.

We leverage two mechanisms to implement the semantics presented in \S~\ref{sec:MPST}, both of which rely on
the \CODE{Result} variant type: (1) the \CODE{?} operator and (2) the \CODE{attempt!-catch} macro.
The \CODE{?} is syntactic sugar for error message propagation. More specifically, each communication primitive is wrapped
inside a \CODE{Result} type.
For example, the return type of \CODE{recv()} is \CODE{Result<(T, S), Box<dyn Error>>}.
The call \CODE{recv()} on the multiparty channel \CODE{s}
triggers the attempt of the reception of
a tuple containing a payload of type \CODE{T}
and a continuation of type \CODE{S}.

\iftoggle{full}{%

\begin{figure}[t!]
\begin{subfigure}{0.3\textwidth}
\begin{rustlisting}
match s.recv() {
Ok((v, s)) => M
Err(e) => {cancel(s);
panic!("Error: {:?}", e)
}
}
\end{rustlisting}
\caption{Explicit cancel by the library}
\label{fig:cancellation:explicit:cancel:library}
\end{subfigure}
\begin{subfigure}{0.35\textwidth}
\begin{rustlisting}
attempt! {
s.recv();
get_video(); // Error
}} catch (e) {
    
    ... }
\end{rustlisting}
\caption{Implicit cancel during runtime}
\label{fig:cancellation:implicit:cancel}
\end{subfigure}
\begin{subfigure}{0.3\textwidth}
\begin{rustlisting}
s.recv()?;
match get_video() {
Ok(v) => M;
Err(e) => {cancel(s);
panic!("Error: {:?}", e)
} }
\end{rustlisting}
\caption{Explicit cancel by the user}
\label{fig:cancellation:explicit:cancel:user}
\end{subfigure}
\caption{Examples for implicit and explicit cancellation of meshed channels}
\label{fig:cancellation}
\end{figure}
}{%
}%

If a peer tries to read a cancelled endpoint then an error message is returned.
Therefore, if an error occurs during receive due to,
for example,
the cancellation of the other end of the channel,
the \CODE{?} operator stops the \CODE{recv()}
function and returns an \CODE{Err} value to
the calling code.
Then, the user can decide to handle the error or
\CODE{panic!} and terminate the program.

Similarly, the \CODE{attempt!-catch} block is syntactic sugar that
allows exception handling over multiple communication actions.
For instance, the \CODE{attempt! M catch N} reduces to its failing clause
\CODE{N} if an error occurs in any of the statements in \CODE{M}.
The interested users can try the online Rust playground that demonstrates
the implementation of \CODE{attempt!-catch} using the \CODE{and_then}
combinator~\cite{web:rust:attemptCatch}.
The \CODE{attempt! M catch N} corresponds to the
\CODE{try-catch} in~\S~\ref{sec:MPST}.


The implementation follows  
the behaviour formalised by the reduction rules in~\S~\ref{sec:MPST}. 
In particular, 
it ensures that 
whenever an error happens, 
a session is cancelled ($\kills{s}$). 
We utilise Rust drop mechanism. 
When a value in Rust goes out of scope,
Rust automatically drops it by calling its destructor: the \CODE{Drop} method.
A variable that cannot be cloned, such as a session \CODE{s},
is out scope when used in a function
and not returned,
such as when used in the
\CODE{close()} and \CODE{cancel()} functions.
We have customised this method by implementing the \CODE{Drop} trait,
which explicitly calls \CODE{cancel()}. 
If an error occurs,
and the meshed channel is not explicitly cancelled,
the meshed channel is \emph{implicitly cancelled} from its destructor. 
In the case of a \CODE{panic!}, 
the session \CODE{s} will be dropped, alongside all variables within
the same function, when \CODE{panic!} is called.
Similarly to the theory, \CODE{cancel(s)} is not mandatory and can be placed
arbitrarily within the process. 
Calling \CODE{cancel(s)} is mostly used for expressiveness and mock tests purposes,
when a failure, without \CODE{panic!}, needs to be simulated.

\myparagraph{Session Cancellation} We discuss all cases involving session cancellation below:
\begin{enumerate}
    \item \textbf{Implicit vs explicit cancellation}
          Receiving on or closing disconnected sessions returns
          an error. As a result of the error,
          the multiparty channel \CODE{s} is cancelled by our underlying library,
          and all binary channels associated with
          \CODE{s} are disconnected. We call this an implicit cancellation. 
          This behaviour implements rules $\inferrule{\iruleMPCQSel}$ and
          $\inferrule{\iruleMPCQBra}$. 
          Alternatively, the user can also cancel the session explicitly.
          
    \item \textbf{Raising an exception} An error occurs
          (1) as a result of a communication over a closed/cancelled channel,
          (2) as a result of a timeout on a channel, or
          (3) in case of an error in the user code.
          For example the function \CODE{get_video()} can return an error.
          Then the user can decide to (1) \CODE{cancel(s)} the session, (2) silently drop the session,
          or (3) proceed with the protocol.
          Even if the user does not explicitly call the \CODE{cancel(s)} primitive,
          Rust runtime ensures that the meshed channel is always cancelled in the end.
    \item \textbf{Double cancellation} If a peer tries to cancel a session \CODE{s} that is already cancelled from another endpoint, then
          the cancellation is ignored. Note that in our semantics this behaviour is modelled using the structural congruence
          rules, namely $\kills{\mpS} \mpPar \kills{\mpS} \;\equiv\; \kills{\mpS}$.
    \item \textbf{Cancel propagation} When a session is cancelled, no communication
          action can be used subsequently on that channel. The action \CODE{cancel(s)} cancels all binary channels
          that are a part of the meshed channel, which precisely simulates the kill process \kills{\mpS}.
          When a peer attempts to receive on a channel, if either side of the channel is cancelled, the operation
          returns an error, and the session in scope is dropped.
          This is exactly the behaviour for the channels from the \texttt{crossbeam-channel}
          library, and we inherit and extend this behaviour to our library.
          Since our \emph{receive} happens on a binary channel, our extension
          ensures that all other binary channels that are in scope,
          and the ones that are in the stack, are also closed.
          Since these channels are closed,
          when other peers try to read from them,
          they will also encounter an error,
          and will subsequently close their channels.
\end{enumerate}
%
%
\iftoggle{full}{%
}{%
}%
\section{Evaluations: benchmarks, expressiveness and case studies}
\label{sec:benchmarks}
\iftoggle{full}{%
}{%
}%
We evaluate~\mpstrust in terms of run-time performance
(\textbf{\S~\ref{subsec:benchmarks:performance}}),
compilation time (\textbf{\S~\ref{subsec:benchmarks:compilation}}) and applications
\iftoggle{full}{%
	(\textbf{\S~\ref{subsec:benchmarks:expressiveness}},
	\textbf{\S~\ref{app:subsec:benchmarks:genericComponentLifecycleManagement}}).
}{%
	(\textbf{\S~\ref{subsec:benchmarks:expressiveness}},
	see~\cite{fullVersion}).
}%
Through this section, we demonstrate the applicability of~\mpstrust and
compare its performance with programs written in binary sessions
and untyped implementations (\textsf{Bare}) using
\texttt{crossbeam-channel}.
The purpose of the microbenchmarks is to demonstrate
the best and worst-case scenarios for the implementation:
we have not considered performance as a primary consideration
in the current implementation.
The results show that rewriting multirole
protocols from binary channels to affine meshed channels can have a performance
gain in addition to the safety guarantees provided by MPST.

In summary,~\mpstrust has only a negligible overhead when compared
to the built-in unsafe Rust channels,
provided by \texttt{crossbeam-channel},
and up to two-fold runtime improvement
to binary sessions in protocols with high-degree of synchronisation.
The source files of the benchmarks and a script to reproduce
the results are included in the artifact. 

\iftoggle{full}{%
}{%
}%
\subsection{Performance}
\label{subsec:benchmarks:performance}
\iftoggle{full}{%
}{%
}%
The goal of the microbenchmarks is two-fold.
On one hand, it provides assurance that~\mpstrust does not incur
significant
overhead when compared to alternative libraries. The source of the runtime overhead of~\mpstrust can be attributed
to: (1) the additional data structures that are generated (see
\S~\ref{sec:api}); and
(2) checks for cancellation (as outlined in \S~\ref{sec:exceptions}).
We also evaluate the efficiency of~\mpstrust when implementing multiparty (as opposed to binary) protocols.
Multiparty protocols specify interaction dependencies between multiple threads.
It is well-understood that a naive decomposition
of multiparty protocol to a binary one (without preserving interaction dependencies) not only causes race conditions
and wrong results but also deadlocks~\cite{scalasLinear2017}.
One may mitigate this problem by utilising a
synchronisation mechanism, which is an
off-the-shelf alternative to meshed channels.
We compare the performance of~\mpstrust and meshed channels to a binary-channels-only
implementation that
uses thread-synchronisation.

We compare implementations, written using (1)~\mpstrust API (\MPST) without cancellation; (2)~\mpstrust API with
cancellation (\AMPST); (3) binary channels, following~\cite{kokkeRusty2019} (\textsf{BC}); and (4) a Bare-Rust
implementation (\textsf{Bare}) using untyped channels as provided by the corresponding transport
library \texttt{crossbeam-channel}.
As a reminder,~\mpstrust uses~\cite{kokkeRusty2019}'s channels
(which are binary only and technically non-meshed),
and~\cite{kokkeRusty2019}'s channels use
\texttt{crossbeam-channel} for actually
sending and receiving payloads:
the scaffolding of all programs differs only
in the final communication primitives used.
In addition, the \textsf{BC} implementations synchronise between 
threads when messages must be received in order.

~\Cref{fig:visualise} shows simple visualisation,
displayed for illustrative purpose,
of the three examples that we
benchmark.~\Cref{fig:benchmarks} reports the results on runtime performance,
\ie the time to complete a
protocol by the implemented endpoints in Rust,
and compilation time, \ie
the time to compile the implementations for all roles.
We stress tested the library up to
20 participants but only show
the results up to 10 participants
for readability.


\emph{\textbf{Setup:} }Our machine configurations are AMD Opteron\textsuperscript{TM} Processor 6282 SE @ 1.30 GHz
with 32 cores/64 threads, 128 GB of RAM and 100 GB of HDD
with Ubuntu 20.04, and with the latest version available
for Rustup (1.24.3) and the Rust cargo compiler (1.56.0).
We use \emph{criterion}~\cite{web:rust:criterion}, a popular benchmark framework in Rust.
We repeat each benchmark 10000 times and report the average execution time with a fairly narrow confidence interval of 95\%.

\iftoggleverb{full}
\begin{figure}[t!]
	\scriptsize
	\begin{tabular}{|C{0.6cm}|C{7cm}|C{5cm}|}
		\hline
		name & Scribble representation & diagram \\ \hline
		ping-pong
		&
		\begin{minipage}{\textwidth}
 			\begin{SCRIBBLELISTING}
rec loop {
	ping() from A$_1$ to A$_2$;
	pong() from A$_2$ to A$_1$;
loop }
 			\end{SCRIBBLELISTING}
		\end{minipage}
		&
		\begin{center}


\begin{tikzpicture}[
    role/.style={circle,text width=2em,text centered,draw,inner sep=0,font=\small},
    arrow/.style={>=latex,thick,draw},
    label/.style={inner sep=3,font=\sf\small},
    transform shape,
    scale = 0.7
]
\node (p0) [role] {$\text{\normalfont A}_\text{\normalfont 1}$};
\node (p1) [role,right=of p0] {$\text{\normalfont A}_\text{\normalfont 2}$}
edge [bend left=30,arrow,->] node [midway,below,label]{pong} (p0)
edge [bend right=30,arrow,<-] node [midway,above,label]{ping}(p0);

\end{tikzpicture}
\end{center}
		\\ \hline
		ring
		&
		\begin{minipage}{\textwidth}
			\begin{SCRIBBLELISTING}
rec loop {
	ping() from A$_1$ to A$_2$;
	ping() from A$_2$ to A$_3$;...
	ping() from A$_1$ to A$_n$;
	pong() from A$_n$; to A$_{n-1}$;
	pong() from A$_2$ to A$_1$;...
loop }\end{SCRIBBLELISTING}
\end{minipage}%
		&
		\begin{center}


\begin{tikzpicture}[
    role/.style={circle,text width=2em,text centered,draw,inner sep=0,font=\small},
    arrow/.style={>=latex,thick,draw},
    label/.style={inner sep=3,font=\sf\small},
    transform shape,
    scale = 0.7
]
\node (p0) [role] {$\text{\normalfont A}_\text{\normalfont 1}$};
\node (p1) [role,right=of p0] {$\text{\normalfont A}_\text{\normalfont 2}$}
edge [bend left=30,arrow,->] node [midway,below,label]{pong} (p0)
edge [bend right=30,arrow,<-] node [midway,above,label]{ping}(p0);
\node (p2) [role,right=of p1] {$\text{\normalfont A}_\text{\normalfont 3}$}
edge [bend left=30,arrow,->] node [midway,below,label]{pong} (p1)
edge [bend right=30,arrow,<-] node [midway,above,label]{ping}(p1);
\node (p3) [role,right=of p2] {$\text{\normalfont A}_\text{\normalfont n}$}
edge [densely dashed,bend left=30,arrow,->] node [midway,below,label]{pong} (p2)
edge [densely dashed,bend right=30,arrow,<-] node [midway,above,label]{ping}(p2);
\end{tikzpicture}
\end{center} \\ \hline
		full-mesh &
		\begin{minipage}{\textwidth}
			\begin{SCRIBBLELISTING}
rec loop {
	ping() from A to B; pong() from B to A;
	ping() from B to C; pong() from C to B;
	ping() from A to C; pong() from C to A;
loop }\end{SCRIBBLELISTING}
\end{minipage}
		&
		\begin{center}


\begin{tikzpicture}[
    role/.style={circle,text width=2em,text centered,draw,inner sep=0,font=\small},
    arrow/.style={>=latex,thick,draw},
    label/.style={inner sep=2,font=\sf\small},
    transform shape,
    scale = 0.7
  ]
  \node (p0)[role] {A};
  \node (p1) [role,below=of p0] {B}
    edge [arrow,<-] node [label,midway,fill=white] {ping} (p0)
    edge [arrow,->,bend left=45] node [label,midway,left] {pong} (p0);
  \node (c) [role,right=2 of p1] {C}
    edge [bend right=20,arrow,->] node [label,midway,fill=white] {ping} (p0)
    edge [arrow,<-] node [label,midway,fill=white] {pong} (p0)
    edge [bend left=10,arrow,->] node [label,midway,fill=white] {ping} (p1)
    edge [bend right=10,arrow,<-] node [label,midway,fill=white] {pong} (p1);
\end{tikzpicture}\end{center} \\ \hline
	\end{tabular}
	\caption{Protocols for Microbenchmarks}
	\label{fig:visualise}
	
\end{figure}
\else
\begin{figure}[t!]
	\scriptsize
	\begin{tabular}{|C{1cm}||C{2cm}|C{4cm}|C{3.5cm}|}
		\hline
		name & ping-pong & ring & full-mesh \\ \hline
		diagram
		&
		\begin{center}


\begin{tikzpicture}[
    role/.style={circle,text width=2em,text centered,draw,inner sep=0,font=\small},
    arrow/.style={>=latex,thick,draw},
    label/.style={inner sep=3,font=\sf\small},
    transform shape,
    scale = 0.7
]
\node (p0) [role] {$\text{\normalfont A}_\text{\normalfont 1}$};
\node (p1) [role,right=of p0] {$\text{\normalfont A}_\text{\normalfont 2}$}
edge [bend left=30,arrow,->] node [midway,below,label]{pong} (p0)
edge [bend right=30,arrow,<-] node [midway,above,label]{ping}(p0);

\end{tikzpicture}
\end{center}
		&
		\begin{center}


\begin{tikzpicture}[
    role/.style={circle,text width=2em,text centered,draw,inner sep=0,font=\small},
    arrow/.style={>=latex,thick,draw},
    label/.style={inner sep=3,font=\sf\small},
    transform shape,
    scale = 0.7
]
\node (p0) [role] {$\text{\normalfont A}_\text{\normalfont 1}$};
\node (p1) [role,right=of p0] {$\text{\normalfont A}_\text{\normalfont 2}$}
edge [bend left=30,arrow,->] node [midway,below,label]{pong} (p0)
edge [bend right=30,arrow,<-] node [midway,above,label]{ping}(p0);
\node (p2) [role,right=of p1] {$\text{\normalfont A}_\text{\normalfont 3}$}
edge [bend left=30,arrow,->] node [midway,below,label]{pong} (p1)
edge [bend right=30,arrow,<-] node [midway,above,label]{ping}(p1);
\node (p3) [role,right=of p2] {$\text{\normalfont A}_\text{\normalfont n}$}
edge [densely dashed,bend left=30,arrow,->] node [midway,below,label]{pong} (p2)
edge [densely dashed,bend right=30,arrow,<-] node [midway,above,label]{ping}(p2);
\end{tikzpicture}
\end{center}
		&
		\begin{center}


\begin{tikzpicture}[
    role/.style={circle,text width=2em,text centered,draw,inner sep=0,font=\small},
    arrow/.style={>=latex,thick,draw},
    label/.style={inner sep=2,font=\sf\small},
    transform shape,
    scale = 0.7
  ]
  \node (p0)[role] {A};
  \node (p1) [role,below=of p0] {B}
    edge [arrow,<-] node [label,midway,fill=white] {ping} (p0)
    edge [arrow,->,bend left=45] node [label,midway,left] {pong} (p0);
  \node (c) [role,right=2 of p1] {C}
    edge [bend right=20,arrow,->] node [label,midway,fill=white] {ping} (p0)
    edge [arrow,<-] node [label,midway,fill=white] {pong} (p0)
    edge [bend left=10,arrow,->] node [label,midway,fill=white] {ping} (p1)
    edge [bend right=10,arrow,<-] node [label,midway,fill=white] {pong} (p1);
\end{tikzpicture}\end{center} \\ \hline
	\end{tabular}
	\iftoggle{full}{%
}{%
}%
	\caption{Protocols for Microbenchmarks}
	\label{fig:visualise}
	\iftoggle{full}{%
}{%
}%
\end{figure}
\fi

\myparagraph{Ping-pong} benchmark measures the execution time for completing a recursive protocol between
two roles repeatedly increasing the number of executions for request-response unit messages.
~\Cref{fig:benchmarksPingPong} displays the running time \wrt the number of iterations.
This protocol is binary, and this benchmark measures the pure overhead of \MPST implementation.
\MPST directly reuses the \textsf{BC} library, adding the structure \CODE{MeshedChannels} on top of it.
Since both implementations need the
same number of threads, the benchmark compares only the overhead of \CODE{MeshedChannels}.
Both \MPST and \AMPST have a linear performance increase compared to \textsf{BC} and \textsf{Bare}.
\MPST is about 2.5 times slower than \textsf{BC} and about 6.5
times slower than \textsf{Bare} for 500 iterations.





\myparagraph{Ring} protocol, as seen in~\Cref{fig:visualise}, specifies \CODE{N} roles, connected in a ring,
sending one message in a sequence.
This example is sequential and stress tests the usage of numerous binary channels in an
\mpstrust implementation.
~\Cref{fig:benchmarksRing100} displays the running time \wrt the number of participants.
We measure the time to complete 100 rounds of a message for an increasing number of roles.
This benchmark demonstrates a worst-case scenario for~\mpstrust since
the \MPST implementation requires \CODE{N*N} binary channels, hence \CODE{N*N} interactions at most,
meanwhile the other implementations only need \CODE{2*N} binary channels.
\mpstrust is increasingly slower than the other implementations following a quadratic curve.
All the implementations are running at the same speed for 2 participants;
\MPST becomes almost 2 times slower than \textsf{BC} for 10 participants
and almost 3.25 times slower for 20 participants.
\AMPST implementation has a negligible overhead compared to \MPST.


\myparagraph{Full-mesh} benchmark measures the execution time for completing a
recursive protocol between \CODE{N} roles mutually exchanging the same message together:
for every iteration,
each participant sends and receives once with every other participant.
For simplicity, we show the pattern in~\Cref{fig:visualise} for three roles only.
\Cref{fig:benchmarksMesh100} displays the running time \wrt the number of participants.
This is a best-case scenario protocol for~\mpstrust since the protocol requires a
lot of explicit synchronisation if implemented as a composition of binary protocols.
The slowdown of \textsf{BC} is explained by the difference of implementation and the management of threads:
the~\mpstrust needs only one thread for each participant,
meanwhile for the binary case,
two threads per pair of interactions are
required to ensure that the message causalities are preserved.
All implementations have similar running time for 2 participants
but \MPST is about 2.3 times faster than \textsf{BC},
and about 11 times slower than \textsf{Bare} for 10 participants.
The figure only displays the results for up to 10 participants,
since this is sufficient to show the overhead trend.
In practice, we measured for up to 20 participants.
For reference, 
at 20 participants,
\MPST is about 12 times slower than \textsf{Bare}
and about 3.75 times faster than \textsf{BC}.
As expected, \AMPST has almost the same running time as \MPST.

\begin{figure}
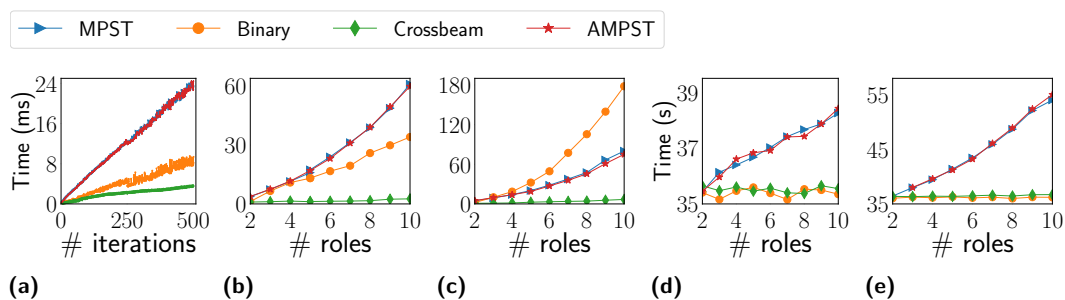

	\includegraphics[scale=0.04]{pdf/Legend.pdf}

	\begin{subfigure}[b]{0.195\textwidth}
		\includegraphics[width=\textwidth]{pdf/graphPingPong.pdf}
		\caption{}
		\label{fig:benchmarksPingPong}
	\end{subfigure}
	\begin{subfigure}[b]{0.195\textwidth}
		\centering
		\includegraphics[width=\textwidth]{pdf/graphRing100.pdf}
		\caption{}
		\label{fig:benchmarksRing100}
	\end{subfigure}
	\begin{subfigure}[b]{0.195\textwidth}
		\centering
		\includegraphics[width=\textwidth]{pdf/graphmesh100.pdf}
		\caption{}
		\label{fig:benchmarksMesh100}
	\end{subfigure}
	\begin{subfigure}[b]{0.195\textwidth}
		\centering
		\includegraphics[width=\textwidth]{pdf/graphAverageCompileRing.pdf}
		\caption{}
		\label{fig:benchmarksRingCompilation}
	\end{subfigure}
	\begin{subfigure}[b]{0.195\textwidth}
		\centering
		\includegraphics[width=\textwidth]{pdf/graphAverageCompileMesh.pdf}
		\caption{}
		\label{fig:benchmarksMeshCompilation}
	\end{subfigure}
	\iftoggle{full}{%
}{%
}%
	\caption{Execution time (ms) for Ping-pong (a), Ring (b), Mesh (c) and compile time (s) for Ring (d), Mesh (e)}
	\label{fig:benchmarks}
	\iftoggle{full}{%
}{%
}%
\end{figure}

\myparagraph{Results summary on execution time} Overall,
\mpstrust is faster than the \textsf{BC} implementation when there
are numerous interactions and participants,
thanks to the encapsulation of each participant as a thread;
the worst-case scenario for~\mpstrust
is for protocols with many participants but no causalities
between them which results in a
slowdown when compared with \textsf{BC}.
\AMPST adds a negligible running time due to the simple checking of the status of the binary channels.


\myparagraph{Results summary on compilation time}
\label{subsec:benchmarks:compilation}
We also compare the compilation time of the three protocols using \texttt{cargo build}.
The results are presented in~\Cref{fig:benchmarksRingCompilation,fig:benchmarksMeshCompilation}.
As expected, the more participants there are, the higher is the compilation time for \MPST,
with up to 40\% increase for the full-mesh protocol and only 11\% for the ring protocol.
We omit the graph for the ping-pong protocol since the number of iterations does not
affect compilation time and the number of generated types,
hence the compilation stays constant at 36.4s (\MPST),
36.6s (\AMPST),
36.1s (\textsf{BC}) and 36.3s (\textsf{Bare}).

The compilation time of \textsf{BC} and \textsf{Bare} are very close thanks to
Rust's features, a mechanism to express conditional compilation and optional dependencies.
This allows compiling only specific parts of libraries, instead of the whole libraries,
depending on the needs of each file.
For \textsf{BC} and \textsf{Bare}, we only compile~\mpstrust's \emph{default} features,
meanwhile for \MPST and \AMPST, we also compile the \emph{macros} features,
which include heavy blocks of code and new dependencies for the creation of the new roles,
meshed channels and associated functions.

\iftoggle{full}{%
}{%
}%
\subsection{Expressiveness}
\label{subsec:benchmarks:expressiveness}
\iftoggle{full}{%
}{%
}%
We demonstrate the expressiveness and applicability of~\mpstrust by
implementing protocols for a range of applications.
We also draw the examples from
the session types literature,
well-established \emph{application protocols} (OAuth, SMTP),
and distributed protocols (logging, circuit breaker).
Protocols with more
than 5 participants are not considered
since having one global protocol with more
participants can quickly become intractable in terms of
protocol logic and is considered bad practice.
The global protocols and patterns in the literature
that have many participants are parameterised~\cite{castroDistributed2019},
participants can be grouped in kinds having the same type.
Thereby, this will avoid a combinatorial explosion.

~\Cref{tab:examples} displays the examples and related metrics.
In particular, we report compilation time (Check./Comp./Rel.),
execution time (Exec. Time),
the number of lines of code (LoC) for implementing all roles in~\mpstrust,
the lines of code generated from Scribble (Gen Types) and
the total lines of code (All);
the two following columns indicate whether the protocol involves three participants or more (MP),
and if the protocol is recursive (Rec). 

We report three compilation times corresponding to the different compilation options in
Rust -- \texttt{cargo check} which only type checks the code without producing binaries,
\texttt{cargo build} which compiles the code with binaries and \texttt{cargo build --release}
which, in addition, optimises the compiled artifact.
Each recursive protocol is built/checked 100 times,
and we display the average in the table.
All protocols are type-checked within 27 seconds,
while the basic compilations range between 36s and 41s and the
optimised compilations vary between 80s and 97s.
Those results represent the longest time we can expect for the respective build/check:
Rust compilation is iterative, therefore, the usual compilation time should be shorter.
A 30 seconds pause is short enough to not break the
\emph{flow}~\cite{web:flowPsychology}
of the mental headspace focused on the current task.
Building the binaries takes longer, because of two heavy libraries used by~\mpstrust
(\texttt{tokio}~\cite{web:rust:tokio} and \texttt{hyper}~\cite{web:rust:hyper}).
The execution time of the protocols is measured by implementing only the communication aspects of the protocol,
and orthogonal computation-related aspects are omitted.
The execution time is the time to complete all protocol interactions,
and even for larger protocols, it is negligible.

~\Cref{tab:examples} does not contain protocols with more than 5 distinct participants
because, in our experience, whenever more participants are needed,
the protocol is parameterised~\cite{castroDistributed2019}.
We leave such extension for future investigation. 

\begin{table}[t!]
	\centering
	\caption{Selected examples from the literature}
	\iftoggle{full}{%
}{%
}%
	\resizebox{\textwidth}{!}{
		\begin{tabular}{|l|r|c|c|c|c|c|}
			\hline
			Example (Endpoint)                                                                            & Check./Comp./Rel./Exec. Time       & LoC Impl. & Gen Types/All & MP     & Rec    \\ 
			\hline
			\hline
			Three buyers~\cite{jiaMonitors2016}                                                           & 26.7s / 37.1s / 81.3s / 568 $\mu$s & 143       & 37 / 180      & \cmark & \cmark \\ 
			\hline
			Calculator~\cite{huHybrid2016}                                                                & 26.5s / 36.9s / 81.3s / 467 $\mu$s & 136       & 32 / 168      & \xmark & \xmark \\ 
			\hline
			Travel agency~\cite{huSessionBased2008}                                                       & 26.5s / 37.6s / 84.8s / 8 ms       & 200       & 47 / 247      & \xmark & \cmark \\ 
			\hline
			Simple voting~\cite{huHybrid2016}                                                             & 26.3s / 36.7s / 82.4s / 396 $\mu$s & 207       & 61 / 268      & \xmark & \xmark \\ 
			\hline
			Online wallet~\cite{neykovaSpy2013}                                                           & 26.4s / 37.8s / 84.4s / 759 $\mu$s & 231       & 76 / 307      & \xmark & \cmark \\ 
			\hline
			Fibonacci~\cite{huHybrid2016}                                                                 & 26.6s / 36.7s / 80.9s / 9 ms       & 141       & 23 / 164      & \xmark & \cmark \\ 
			\hline
			Video Streaming service (\textbf{\S~\ref{sec:overview}})                                      & 26.3s / 37.4s / 83.0s / 11 ms      & 104       & 39 / 143      & \cmark & \cmark \\ 
			\hline
			oAuth2~\cite{neykovaSpy2013}                                                                  & 26.4s / 37.5s / 83.2s / 12 ms      & 215       & 61 / 276      & \cmark & \cmark \\ 
			\hline
			\iftoggle{full}{%
				Distributed logging (\textbf{\S~\ref{app:subsec:benchmarks:genericComponentLifecycleManagement}})
			}{%
				Distributed logging (\cite{fullVersion})
			}%
			& 26.5s / 36.8s / 82.6s / 5 ms       & 252       & 59 / 311      & \xmark & \cmark \\ 
			\hline
			\iftoggle{full}{%
				Circuit breaker (\textbf{\S~\ref{app:subsec:benchmarks:genericComponentLifecycleManagement}})
			}{%
				Circuit breaker (\cite{fullVersion})
			}%
			& 26.5s / 38.5s / 87.0s / 18 ms      & 375       & 142 / 517     & \cmark & \cmark \\ 
			\hline
			SMTP~\cite{fieldingHypertext2014}                                                             & 26.4s / 41.1s / 97.3s / 5 ms       & 571       & 143 / 714     & \xmark & \cmark \\ 
			\hline
		\end{tabular}
		\footnotesize
	}
	
	\label{tab:examples}
	\iftoggle{full}{%
}{%
}%
\end{table}

\iftoggle{full}{%
}{%
}%
\section{Related work}
\label{sec:related-works}
\iftoggle{full}{%
}{%
}%
A vast amount of session types
implementations based on theories exist, as detailed in the recent surveys on
language implementations~\cite{anconaBehavioral2016} and tools~\cite{gayBehavioural2017}.
We discuss closely related works,
focusing on (1) session types implementations in Rust
(\S~\ref{subsec:bst-rust});
(2) \MPST top-down implementations (including
\iftoggle{full}{
	other programming languages) (\S~\ref{subsec:mpst-in-other});
	and (3) Affine types and exceptions/error handling in session types
	(\S~\ref{app:subsec:affine-types}).
}{%
	other programming languages) (\S~\ref{subsec:mpst-in-other}).
	For related work about Affine types and
	exceptions/error handling in session types, see~\cite{fullVersion}.
}%


\iftoggle{full}{%
}{%
}%
\subsection{Session types implementations in Rust}
\label{subsec:bst-rust}
\iftoggle{full}{%
}{%
}%
Binary session types (\BST) have been implemented in Rust by
\cite{jespersenSession2015},
\cite{kokkeRusty2019} and
\cite{chenFerrite2021}, whereas,
to our best knowledge,
~\cite{noauthorDeadlockFree2022} is the only
implementation of
multiparty session types in Rust.

~\cite{jespersenSession2015} implemented binary session types,
following~\cite{hondaLanguage1998},
while~\cite{kokkeRusty2019} based their library on the EGV calculus by
\cite{fowlerExceptional2019}
\iftoggle{full}{%
	(see \S~\ref{app:subsec:affine-types}).
}{%
	(See~\cite{fullVersion}).
}%
Both verify at compile-time that the behaviours of two endpoint
processes are \emph{dual}, \ie ~the processes are compatible.
The latter library allows to write and check session typed communications,
and supports exception handling constructs.
Rust originally did not support \emph{recursive types}
so~\cite{jespersenSession2015} had to use \emph{de Bruijn}
indices to encode recursive session types, while
\cite{kokkeRusty2019} uses Rust's native recursive types
but only handles failure for \CODE{recv()} actions:
according to~\cite{kokkeRusty2019},
this is generally the case with asynchronous implementations.
This is because once an endpoint has received several messages,
it makes sense to cancel them at the receiver
rather than the sender.
In fact, raising an exception on a send operation
in an asynchronous calculus actually breaks confluence.

The library by~\cite{jespersenSession2015}
relies on an older version of Rust, hence we build
\mpstrust on top of~\cite{kokkeRusty2019}.
Notice that we formalised \AMPST
guaranteeing the \MPST properties of~\mpstrust (such as
deadlock-freedom, liveness and cancellation termination), which are
not present in~\cite{kokkeRusty2019}.
In addition, our benchmarks
confirmed that, in protocols where most of the participants
mutually communicate,~\mpstrust is up to two times
faster than~\cite{kokkeRusty2019}.

~\cite{chenFerrite2021} introduces their library,
Ferrite, that implements \BST in Rust,
adopting \emph{intuitionistic logic-based typing}~\cite{cairesIntuitionistic2010}.
The library ensures \emph{linear} typing of channels,
and includes a recently shared name extension by~\cite{BalzerP17},
but cannot statically handle prematurely dropped channel endpoints.
Since Ferrite lacks an additional causal analysis for ensuring
deadlock-freedom by~\cite{BalzerTP19},
deadlock-freedom and liveness among more than two
participants are not guaranteed,
unlike~\mpstrust.
Ferrite also lacks documentation and tests,
making it hard to use.
 
\cite{DuarteRetrofitting2021} presents an implementation of 
a library for programming typestates in Rust.
The library ensures that Rust programs
follow a typestate specification. 
The tool, however, has several limitations.
Differently than other works on typestates (\eg typestates in Java~\cite{kouzapasTypechecking2016}), 
\cite{DuarteRetrofitting2021} implements and verifies only binary non-recursive protocols,
without a static guarantee that all branches are exhaustively implemented. 

Note that all the above implementations are
limited to \emph{binary} and
no formalism is proposed in their papers
(see~\Cref{tab:compareMPST}).

\cite{noauthorDeadlockFree2022} implements
\MPST using \CODE{async} and \CODE{await} primitives.
Their main focus is a performance analysis
of asynchronous message reordering and comparisons of
their asynchronous subtyping algorithm with existing
tools, including the \kmc~tool~\cite{langeVerifying2019}.
Their algorithm is a sound approximation
of the (undecidable) asynchronous subtyping relation~\cite{ghilezanPrecise2021},
by which  their tool enables to check whether an unoptimised (projected from a
global type) CFSM
and its optimised CFSM are under the subtyping relation or not.
The main disadvantage of~\cite{noauthorDeadlockFree2022}
is that their library depends on external tools for checking not only
deadlock-freedom, but also communication-safety.
Differently,~\mpstrust can guarantee \emph{dual compatibility} (inherited from
\cite{kokkeRusty2019}) in a multiparty protocol, 
based on our meshed channels implementation.

Unlike~\mpstrust, 
neither failure handling nor cancellation
termination is implemented or formalised in any of the above-mentioned works.  



\iftoggle{full}{%
}{%
}%
\subsection{Multiparty session types implementations in other languages}
\label{subsec:mpst-in-other}
\iftoggle{full}{%
}{%
}%


We compare implementations of (top-down) \MPST, ordered by date of publication,
in~\Cref{tab:compareMPST},
focusing on statically typed languages:
we exclude \MPST implementations by runtime monitoring
such as Erlang~\cite{neykovaLet2017} and Python
\cite{demangeonPractical2015}.

The table is composed as follows, row by row:

\begin{description}[noitemsep,topsep=0pt]
	\item[Languages] lists the programming languages introduced or used.

	\item[Mainstream language] states if the language is broadly used among developers or not.

	\item[MPST top-down] characterises the framework: Multiparty session types (\MPST) or
		binary session types (\BST).
		If the implementation allows the user to write \MPST global types, it is
		called a top-down approach.

	\item[Linearity checking] describes whether the linear usage of channels is not checked,
		checked at compile-time (\emph{static}) or checked at runtime (\emph{dynamic}).

	\item[Exhaustive choices check] indicates whether the implementation
		can \emph{statically} enforce
		the correct handling of potential input 
		types.~\xmark${}$ denotes implementations that do not support
		pattern-matching to carry out choices (branching)
		using switch statements on \CODE{enum} types.

	\item[Formalism] defines the theoretical foundations
		of the implementations, such as (1) the end point
		calculus (the $\pi$-calculus (noted as $\pi$-cal.) or FJ~\cite{igarashiFeatherweight2001});
		(2) the (global) types formalism without any endpoint
		calculi (no typing system is given, and no subject
		reduction theorem is proved); or (3)
		no formalism is given (no theory is developed).


	\item[Communication safety]
		outlines the presence or the absence of session type-soundness demonstration.
		Four languages, marked as $\triangle$, provide the type safety only at type level.
		\xmark${}^\bullet$ means that the theoretical formalism does not provide
		linear types, therefore only type safety of base values is proved.

	\item[Deadlock-freedom] is a property guaranteeing that all components
		are progressing or ultimately terminate
		(which correspond to deadlock-freedom in \MPST).
		Four languages marked by $\triangle$ proved deadlock-freedom only at the type level.
		\cmark${}^\bullet$ implies the absence of a formal link with
		the local configurations reduced from the projection
		of a global type.
		\footnote{
			\cite{harveyMultiparty2021} did
			not prove that any typing context
			reduced from
			a projection of a well-formed global type satisfies
			a safety property (a statement corresponding to~\Cref{lem:globalsafety}).
			Hence, deadlock-freedom is not provided
			for processes initially typed by a given global type.
			Note that their typing contexts contain
			new elements not found in those defined
			in~\cite{huismanExplicit2017},
			which weakens the link
			with the top-down approach.
		}
	\item[Liveness]
		is a property which ensure that all actions
		are eventually communicated with other parties (unless killed by
		an exception in the case of \AMPST).

	\item[Cancellation termination:]
		once a cancellation happens at one of the participants in a multiparty
		protocol, the cancellation is propagated correctly, and
		all processes can terminate.
	
\end{description}

The Rust implementations in the first column of~\Cref{tab:compareMPST}
are included for reference.

\begin{table}[t!]
	\begin{center}
		\caption{\MPST top-down implementations}
		\iftoggle{full}{%
}{%
}%
		\resizebox{\textwidth}{!}{%
			\begin{tabular}{| c || c | c | c | c | c | c | c | c | c | c | c | c | c | c ||}
				\hline
				 & \thead{~\cite{kokkeRusty2019,jespersenSession2015,chenFerrite2021}}
				 & \thead{~\cite{ngProtocols2015}}
				 & \thead{~\cite{huHybrid2016,huismanExplicit2017}}
				 & \thead{~\cite{kouzapasTypechecking2018}}
				 & \thead{~\cite{scalasLinear2017}}
				 & \thead{~\cite{neykovaSession2018}}
				 & \thead{~\cite{castroDistributed2019}}
				 & \thead{~\cite{imaiMultiparty2020}}
				 & \thead{~\cite{miuGenerating2020}}
				 & \thead{~\cite{zhouStatically2020}}
				 & \thead{~\cite{harveyMultiparty2021}}
				 & \thead{~\cite{vieringMultiparty2021}}
				 & \thead{~\cite{noauthorDeadlockFree2022}}
				 & \thead{~\mpstrust}                                                                         \\
				\hline\hline
				\makecell{Language\\}
				 & Rust                                                                & MPI-C
				 & Java                                                                & Java
				 & Scala                                                               & F\#
				 & Go                                                                  & OCaml
				 & Typescript                                                          & F*
				 & EnsembleS                                                           & Scala
				 & Rust                                                                & Rust                 \\
				\hline
				\makecell{Mainstream                                                                          \\language}
				 & \cmark                                                              & \cmark
				 & \cmark                                                              & \cmark
				 & \cmark                                                              & \cmark
				 & \cmark                                                              & \cmark
				 & \cmark                                                              & \cmark
				 & \xmark                                                              & \cmark
				 & \cmark                                                              & \cmark               \\
				\hline
				\makecell{MPST                                                                                \\Top-Down}
				 & \xmark                                                              & \cmark
				 & \cmark                                                              & \cmark
				 & \cmark                                                              & \cmark
				 & \cmark                                                              & \cmark
				 & \cmark                                                              & \cmark
				 & \cmark                                                              & \cmark
				 & \cmark                                                              & \cmark               \\
				\hline
				\makecell{Linearity                                                                           \\  check }
				 & static                                                              & \xmark
				 & dynamic                                                             & \xmark
				 & dynamic                                                             & dynamic
				 & dynamic                                                             & static
				 & static                                                              & static
				 & static                                                              & dynamic
				 & static                                                              & static               \\
				\hline
				\makecell{Exhaustive                                                                          \\ choices check}
				 & \cmark                                                              & \xmark
				 & \xmark                                                              & \xmark
				 & \cmark                                                              & \xmark
				 & \xmark                                                              & \cmark
				 & \cmark                                                              & \cmark
				 & \cmark                                                              & \cmark
				 & \cmark                                                              & \cmark               \\
				\hline
				\hline
				\makecell{  Formalism                                                                         \\ }
				 & \xmark                                                              & \xmark
				 & types                                                               & FJ
				 & $\pi$-cal.                                                          & \xmark
				 & types                                                               & $\pi$-cal.
				 & types                                                               & types
				 & $\pi$-cal.                                                          & $\pi$-cal.
				 & types                                                               & $\pi$-cal.           \\
				\hline
				\makecell{Communication                                                                       \\safety}
				 & \xmark                                                              & \xmark
				 & $\triangle$                                                         & \cmark
				 & \cmark                                                              & \xmark
				 & $\triangle$                                                         & \xmark${}^{\bullet}$
				 & $\triangle$                                                         & $\triangle$
				 & \cmark                                                              & \cmark
				 & $\triangle$                                                         & \cmark               \\
				\hline
				\makecell{Deadlock                                                                            \\freedom}
				 & \xmark                                                              & \xmark
				 & $\triangle$                                                         & \xmark
				 & \cmark                                                              & \xmark
				 & $\triangle$                                                         & \xmark
				 & $\triangle$                                                         & $\triangle$
				 & \cmark${}^{\bullet}$                                                & \cmark
				 & $\triangle$                                                         & \cmark               \\
				\hline
				\makecell{Liveness}
				 & \xmark                                                              & \xmark
				 & \xmark                                                              & \xmark
				 & \xmark                                                              & \xmark
				 & \xmark                                                              & \xmark
				 & \xmark                                                              & \xmark
				 & \xmark                                                              & \xmark
				 & \xmark                                                              & \cmark               \\
				\hline
				\makecell{Cancellation                                                                        \\termination}
				 & \xmark                                                              & \xmark
				 & \xmark                                                              & \xmark
				 & \xmark                                                              & \xmark
				 & \xmark                                                              & \xmark
				 & \xmark                                                              & \xmark
				 & \xmark                                                              & \xmark
				 & \xmark                                                              & \cmark               \\
				\hline
			\end{tabular}
		}
		\iftoggle{full}{%
}{%
}%
		\label{tab:compareMPST}
	\end{center}
\end{table}

Most of the \MPST implementations
\cite{huismanExplicit2017,scalasLinear2017,neykovaSession2018,castroDistributed2019,miuGenerating2020,zhouStatically2020}
follow the methodology given
by~\cite{huHybrid2016}, which generates Java communicating APIs from
Scribble
\cite{yoshidaScribble2013,imperialCollegeLondonScribble2021}.
They exploit the equivalence between local session types and
finite state machines to generate session types APIs for mainstream
programming languages.
\cite{huHybrid2016,huismanExplicit2017,scalasLinear2017,neykovaSession2018,castroDistributed2019}
are not completely static: they check linearity dynamically.
\mpstrust can check linearity using the built-in affinity type checking from Rust.
\cite{ngProtocols2015,neykovaSession2018,castroDistributed2019}
do not enforce exhaustive handling of input types;
and~\cite{huHybrid2016,huismanExplicit2017,kouzapasTypechecking2018}
rely on runtime checks to correctly handle branching.

~\cite{miuGenerating2020,zhouStatically2020} provide static
checking using the call-back style API generation.
\mpstrust uses a decomposition of \AMPST to \BST;
in~\cite{scalasLinear2017},
\MPST in Scala is implemented combining binary channels
on the top of the existing \BST library
from~\cite{scalasLightweight2016}.
Unlike~\mpstrust,~\cite{scalasLinear2017} lacks static linearity check
and uses a continuation-passing style translation from
\MPST into linear types.~\cite{kouzapasTypechecking2018} implements
static type-checking of
communication protocols by linking
Java classes and their respective
typestate definitions generated from Scribble.
Objects declaring a typestate should be used linearly,
but a linear usage of channels is not statically enforced.

All above implementations generate multiparty APIs from protocols.
To our knowledge,~\cite{imaiMultiparty2020}
is the only type-level embedding of classic multiparty
channels in a mainstream language, OCaml.
However, the library heavily relies on OCaml-specific parametric polymorphism
for variant types to ensure type-safety.
Their formalism lacks linear types and deadlock-freedom is not formalised
nor proved.
In addition, this implementation uses a non-trivial,
complicated encoding of polymorphic variant types and lenses,
while~\mpstrust uses the built-in affine type system in Rust.

The work most closely related to ours
is~\cite{harveyMultiparty2021} which implements
handling of dynamic environments by \MPST with explicit
connections from~\cite{huismanExplicit2017},
where actors can dynamically connect and disconnect.
It relies on the actor-like research language, Ensemble;
and generates endpoint code from Scribble.
Their core calculus includes a syntax of the \textbf{try $L$ catch $M$} construction
where $M$ is evaluated if $L$ raises an exception.
The type system
follows~\cite{vasconcelosTypechecking2006},
and is not as expressive as the previous paper on binary exception handling
~\cite{fowlerExceptional2019} that extends the richer type system of GV
~\cite{lindleyBananas2016,lindleySemantics2015}.
Due to this limitation of their base typing system,
and since their main focus is \emph{adaptation},
there are several differences from \AMPST, listed below:
(1) they do not model general failure of multiple
(interleaved) session endpoints (such as
failures of \emph{selection} and \emph{branching} constructs as
shown in rules $\inferrule{\iruleMPCSel}$, $\inferrule{\iruleMPCBra}$);
(2) their \CODE{try-catch} scope (handler) is limited to
a single action unlike \AMPST and~\cite{fowlerExceptional2019} where
its scope can be an arbitrary process $P$, participants and session
endpoints (\inferrule{\iruleMPRedCat}); (3) they do not model any
Rust specific $\mathbf{?}$-options where an arbitrary process $P$
can self-fail ($\inferrule{\iruleMPTryCatch}$, $\inferrule{\iruleMPCQSel}$); and
(4) their kill process is weaker than ours
(it is point-to-point, it does not broadcast the failure notification to the same session).

As a consequence,
their progress result (\cite[Theorem 18]{fowlerExceptional2019})
is weaker than our theorems since their configuration can be stuck
with an exception process that contains $\mathbf{raise}$,
while our termination theorem (Theorem~\ref{thm:ctermination})
guarantees that there always exists a path 
such that the process will move or terminate as $\mathbf{0}$,
\emph{cleaning up} all intermediate processes
which interact non-deterministically.
More precisely, in \cite[Theorem 18]{fowlerExceptional2019}, 
a cancellation in a session is propagated, but 
$\mathbf{raise}$ blocks a reduction when the actor is not
involved in a session, and its behaviour is also $\mathbf{stop}$, 
meaning it is terminated. 
Otherwise, the actor will leave the session and restart.
In contrast,~\mpstrust ensures
the strong progress properties by construction (see
\S~\ref{sec:mpst-rust}).
We also implemented interleaved sessions
\iftoggle{full}{%
	(as shown in~\Cref{subsec:app:implementation:rust:interleaved}),
}{%
	(as shown in~\cite{fullVersion}),
}%
where one participant is involved in two different protocols
at the same time.

\iftoggle{full}{%
}{%
}%
\section{Conclusion and future work}
\label{sec:conclusion}
\iftoggle{full}{%
}{%
}%
%
%

Rust's pledge to guarantee memory safety does not extend to communication safety.
Rust's built-in binary channels and affine type
system are insufficient to ensure correct interaction and termination of multiple communicating processes.
This paper overcomes this limitation by providing two main contributions.
We proposed a new typing discipline, \emph{affine multiparty session
types}, which captures implicit/explicit cancellation mechanisms in
Rust, and proved its cancellation termination theorem.
In addition to progress and liveness properties,
our end-point processes can guarantee that
all processes terminate safely and
cancellation is correctly propagated across all channels in a session,
whenever and wherever a failure happens.
We embedded the theory in Rust and
developed a practical library for safe multiparty communication,~\mpstrust,
which ensures deadlock-freedom and liveness in the presence of cancellations of
arbitrary processes.
Evaluation of~\mpstrust shows that it has only a negligible overhead when compared with
the built-in unsafe Rust channels.
We demonstrated the use of~\mpstrust for programming distributed application
protocols with exception handling patterns.


As part of future work, we would like to develop recovery strategies
based on causal analysis, along the lines of~\cite{neykovaLet2017}.
In addition,
it would be interesting to
verify role-parametric session types following~\cite{castroDistributed2019}
in an affine setting.
Finally, we plan to study polymorphic meshed channels 
with different delivery guarantees
such as TCP and UDP.

\bibliography{main}

\iftoggle{full}{
	\appendix
	\section{Appendix for \S~\ref{sec:MPST}}
\label{app:AMPST}

\thmMPSTSubectCongruence*

\begin{proof}
  \textbf{Case} $\kills{\mpS} \mpPar
    \kills{\mpS}\;\equiv\;\kills{\mpS}$. \\[1mm]
  ($\Rightarrow$)
  $\stJudge{\mpEnv}{\stEnv_1}{\kills{\mpS}}$ with
  $\stEnv_i =
    \stEnv_i' \stEnvComp
    \stEnvMap{\mpChanRole{\mpS}{\roleP_{1i}}}{\stS[1i]}\stEnvComp
    \stEnvMap{\mpChanRole{\mpS}{\roleP_{2i}}}{\stS[2i]}\stEnvComp
    \dots
    \stEnvComp
    \stEnvMap{\mpChanRole{\mpS}{\roleP_{ni}}}{\stS[ni]}$ with
  $\stEnvEndP{\stEnv_i'}$
  ($i=1,2$).
  Let $\stEnv_1\cup \stEnv_2=\stEnv''$. Then we have:
  $\stJudge{\mpEnv}{\stEnv''}{\kills{\mpS}}$ with
  $\stEnvEndP{\stEnv_1'\cup\stEnv_2'}$, as desired. \\[1mm]
  ($\Leftarrow$) Similar with the case of ($\Rightarrow$). \\[1mm]
  \textbf{Case} $\mpRes{\mpS}{\kills{\mpS}}%
    \;\equiv\; \mpNil$\\[1mm]
  By~\Cref{lem:globalsafety},
  we only have to assume we use Rule [\iruleMPSafeRes].\\[1mm]
  ($\Rightarrow$) Suppose
  $\stJudge{\mpEnv}{\stEnv}{\mpRes{\mpS:\stEnv'}{\kills{\mpS}}}$ by [\iruleMPSafeRes].
  Then $\stEnvEndP{\stEnv}$. Hence
  $\stJudge{\mpEnv}{\stEnv}{\mpNil}$, as desired. \\[1mm]
  ($\Leftarrow$) Similar with the case of ($\Rightarrow$) applying
    [\iruleMPKills] (taking the session environment $\stEnv'$ safe)
  and [\iruleMPSafeRes]. 
\end{proof}

\thmMPSTSubectReduction*

\begin{proof}
The case analysis of the reduction rules.\\[2mm] 
  \textbf{Case \inferrule{\iruleMPRedComm}} \quad
  $\mpEtx_{1}[\mpDaggerBranch{\mpChanRole{\mpS}{\roleP}}{\roleQ}{i \in I}{%
        \mpLab[i]}{x_i}{\mpP[i]}]%
    \,\mpPar\,%
    \mpEtx_{2}[\mpDaggerSel{\mpChanRole{\mpS}{\roleQ}}{\roleP}{\mpLab[k]}{%
        \mpChanRole{\mpSi}{\roleR}%
      }{\mpQ}]%
    \;\;\mpMove\;\;%
    \mpP[k]\subst{\mpFmt{x_k}}{\mpChanRole{\mpSi}{\roleR}}%
    \,\mpPar\,%
    \mpQ$ with
  $k \!\in\! I$
  \\[1mm]%
In the case both contexts are empty coincides with the proof in~\cite[Theorem 4.8]{scalasLess2019}. Hence
 we assume
 $\mpEtx_{i} \,\coloncolonequals\, \trycatch{[\ ]}{\mpR_i}$ for some
 $\mpR_i$ ($i=1,2$).
Recall the typing rule: 
\[
         \inference[{\iruleMPTryCatch}]{%
         {\stJudge{\mpEnv}{%
                \stEnv%
                }{%
                  \mpP%
                }}%
            \qquad%
            {\sbj{P}=c}
            \qquad%
            {\stJudge{\mpEnv}{%
                  \stEnv%
                }{%
                  {\mpQ}%
                }}%
          }{%
            {\stJudge{\mpEnv}{\stEnv}
                {%
                  \trycatch{P}{Q}
                }%
             }} %
\]
\ie for any $\mpP$, $\mpEtx{}[\mpP]$ is typed by the same environments 
$\mpEnv$, $\stEnv$ as $\mpP$;
hence the proof is the same as the case of $\mpEtx{}=[\ ]$ 
(\cite[Theorem 4.8]{scalasLess2019}).\\[1mm]
\textbf{Case \inferrule{\iruleMPCQSel}:}
$\mpQuestionSel{\mpChanRole{\mpS}{\roleP}}{\roleQ}{\mpLab}{
            \mpChanRole{\mpS'}{\roleR}}{\mpP}
          \mpMove
          \mpSel{\mpChanRole{\mpS}{\roleP}}{\roleQ}{\mpLab}{
            \mpChanRole{\mpS'}{\roleR}}{\mpP}
          \mpPar \kills{\mpS}$\\[1mm]
Assume, by \inferrule{\iruleMPSel}, 
\[
\stJudge{\mpEnv}{%
                  \stEnv \stEnvComp \stEnv[1] \stEnvComp \stEnv[2]%
                }{%
\mpQuestionSel{\mpChanRole{\mpS}{\roleP}}{\roleQ}{\mpLab}{
            \mpChanRole{\mpS'}{\roleR}}{\mpP}
               }
\]
where 
\begin{equation}\label{eq:proof:one}
            \stEnvEntails{\stEnv[1]}{\mpChanRole{\mpS}{\roleP}}{%
              \stIntSum{\roleQ}{}{\stChoice{\stLab}{\stS} \stSeq \stSi}%
            }%
           \quad                             %
            \stEnvEntails{\stEnv[2]}{\mpChanRole{\mpS'}{\roleR}}{\stS}%
           \quad                             %
            \stJudge{\mpEnv}{%
              \stEnv \stEnvComp \stEnvMap{\mpChanRole{\mpS}{\roleP}}{\stSi}%
            }{%
              \mpP%
            }%
\end{equation}
Then by $\inferrule{\iruleMPKills}$,
we have
\[
\stJudge{\mpEnv}{\emptyset 
                }
                {
                  \kills{s}
                } 
\]
From (\ref{eq:proof:one}), we have:
\[
\stJudge{\mpEnv}{%
                  \stEnv \stEnvComp \stEnv[1] \stEnvComp \stEnv[2]%
                }{%
\mpSel{\mpChanRole{\mpS}{\roleP}}{\roleQ}{\mpLab}{
            \mpChanRole{\mpS'}{\roleR}}{\mpP}
               }
\]
By \inferrule{\iruleMPPar}, we have:
\[
\stJudge{\mpEnv}{%
                 \stEnv \stEnvComp \stEnv[1] \stEnvComp \stEnv[2]%
                }{%
\mpSel{\mpChanRole{\mpS}{\roleP}}{\roleQ}{\mpLab}{
            \mpChanRole{\mpS'}{\roleR}}{\mpP}\mpPar \kills{s}
               }
\]
as required. \\[1mm]
\textbf{Case \inferrule{\iruleMPCSel}}
$\mpSel{\mpChanRole{\mpS}{\roleP}}{\roleQ}{\mpLab}{
      \mpChanRole{\mpS'}{\roleR}}{\mpP}
    \mpPar \kills{\mpS}
    \mpMove
    \mpP \mpPar \kills{\mpS} \mpPar \kills{\mpS'}$\\[1mm]
Suppose: 
\[
\stJudge{\mpEnv}{%
              \stEnv \stEnvComp \stEnv[1] \stEnvComp \stEnv[2] \stEnvComp \stEnv[3]%
                }{%
\mpSel{\mpChanRole{\mpS}{\roleP}}{\roleQ}{\mpLab}{
            \mpChanRole{\mpS'}{\roleR}}{\mpP}
\mpPar \kills{\mpS}
               }
\]
with $\stEnv \stEnvComp \stEnv[1] \stEnvComp \stEnv[2] \stEnvComp
\stEnv[3]$ safe 
and assume, by \inferrule{\iruleMPSel}: 
\[
\stJudge{\mpEnv}{%
                  \stEnv \stEnvComp \stEnv[1] \stEnvComp \stEnv[2]%
                }{%
\mpSel{\mpChanRole{\mpS}{\roleP}}{\roleQ}{\mpLab}{
            \mpChanRole{\mpS'}{\roleR}}{\mpP}}
\]
where 
\begin{equation}
\label{eq:proof:two}
            \stEnvEntails{\stEnv[1]}{\mpChanRole{\mpS}{\roleP}}{%
              \stIntSum{\roleQ}{}{\stChoice{\stLab}{\stS} \stSeq \stSi}%
            }%
           \quad                             %
            \stEnvEntails{\stEnv[2]}{\mpChanRole{\mpS'}{\roleR}}{\stS}%
           \quad                             %
            \stJudge{\mpEnv}{%
            \stEnv \stEnvComp \stEnvMap{\mpChanRole{\mpS}{\roleP}}{\stSi}%
            }{%
              \mpP%
            }%
\end{equation}
with $\stEnv[2]=\stEnv[5]\stEnvComp 
\stEnvMap{\mpChanRole{\mpS'}{\roleR}}{\stS_0}$
with 
$\stS_0\stSub\stS$ 
and 
$\stEnvEndP{\stEnv[5]}$.
By \inferrule{\iruleMPKills}: 
\begin{equation}
\label{eq:proof:three}
\stJudge{\mpEnv}{\stEnv[3]}
                {
                  \kills{s}
                } 
\end{equation}
where, by safety of the session environments in (\ref{eq:proof:two}),
we have:
\[
\stEnvEntails{\stEnv[3]}{\mpChanRole{\mpS}{\roleQ}}{
\stExtSum{\roleP}{i \in I}{\stChoice{\stLab[i]}{\stS[i]} \stSeq \stSii[i]}}
\]
By $\inferrule{\iruleMPKills}$, we have: 
\begin{equation}
\label{eq:proof:four}
\stJudge{\mpEnv}{\stEnv[2]}
                {
                  \kills{s'}
                } 
\end{equation}
By (\ref{eq:proof:two}), (\ref{eq:proof:three}) and 
(\ref{eq:proof:four})
we have: 
\[
\stJudge{\mpEnv}{%
              \stEnv \stEnvComp \stEnvi[1]
\stEnvComp \stEnv[2] \stEnvComp \stEnvi[3]%
                }{%
\mpP \mpPar \kills{\mpS}\mpPar \kills{\mpSi}
               }
\quad\text{and} \quad
\stJudge{\mpEnv}{\stEnvi[3]}
                {
                  \kills{s}
                } 
\]
where 
\[
\stEnvEntails{\stEnvi[1]}{\mpChanRole{\mpS}{\roleP}}{
              \stSi}
\quad\text{and} \quad 
\stEnvEntails{\stEnvi[3]}{\mpChanRole{\mpS}{\roleQ}}{
              \stSii[i]}
\quad\text{and} \quad
\stEnv[1],\stEnv[3] 
\stEnvMove
\stEnvi[1],
\stEnvi[3]
\quad 
\quad\text{and} \quad 
\stEnv \stEnvComp \stEnvi[1]
\stEnvComp \stEnv[2] \stEnvComp \stEnvi[3]\quad\mbox{safe}
\]
as desired.\\[1mm] 
\textbf{Case \inferrule{\iruleMPCQBra}:}
Similar with \inferrule{\iruleMPCQSel} above.
  \\[1mm]
\textbf{Case \inferrule{\iruleMPTQBra}:} 
Similar with \inferrule{\iruleMPTQSel} above.
  \\[1mm]
\textbf{Case \inferrule{\iruleMPCBra}:}
  $\mpBranch{\mpChanRole{\mpS}{\roleP}}{\roleQ}{i \in
      I}{\mpLab[i]}{x_i}{\mpP[i]} \mpPar \kills{\mpS}
    \mpMove
    \mpRes{\mpS'}(\mpP[k]\subst{\mpFmt{x_k}}{\mpChanRole{\mpS'}{\roleR}}
    \mpPar \kills{\mpS'})
    \mpPar \kills{\mpS}$ with
  $\mpS'\not\in \fc{P_k}, k\in I$.
  \\[2mm]%
  Assume by $\inferrule{\iruleMPBranch}$,
  we have
  $\stJudgeTry{\mpEnv}{%
      \stEnv \stEnvComp \stEnv[1]%
    }{%
      \mpBranch{\mpChanRole{\mpS}{\roleP}}{\roleQ}{i \in I}{\mpLab[i]}{y_i}{\mpP[i]}%
    }$
  with
  $\stEnvEntails{\stEnv[1]}{\mpChanRole{\mpS}{\roleP}}{%
      \stExtSum{\roleQ}{i \in I}{\stChoice{\stLab[i]}{\stS[i]}
        \stSeq \stSi[i]}}$ for
  all $i \!\in\! I$
  and $\stJudge{\mpEnv}{%
      \stEnv \stEnvComp%
      \stEnvMap{y_i}{\stS[i]} \stEnvComp%
      \stEnvMap{\mpChanRole{\mpS}{\roleP}}{\stSi[i]}%
    }{%
      \mpP[i]%
    }$.
  Then
  by the substitution lemma,
  $\stJudge{\mpEnv}{%
      \stEnv \stEnvComp%
      \stEnvMap{\mpChanRole{\mpS'}{\roleR}}{\stS[k]} \stEnvComp%
      \stEnvMap{\mpChanRole{\mpS}{\roleP}}{\stSi[k]}%
    }{%
      \mpP[k]\subst{\mpFmt{x_k}}{\mpChanRole{\mpS'}{\roleR}}%
    }$.
  By typing $\kills{s'}$ with an appropriate session typing context $\stEnv'$
  such that $\stEnvSafeP{\stEnv'\stEnvComp
      \stEnvMap{\mpChanRole{\mpS'}{\roleR}}{\stS[k]}}$, hence by 
  applying $\inferrule{\iruleMPSafeRes}$, we obtain the result.
  \\[1mm]
\textbf{Case \inferrule{\iruleMPRedCan}:}
  $\mpEtx{}[\mpCancel{\mpChanRole{\mpS}{\roleP}}{Q}]
    \mpMove \kills{s}\mpPar Q$\\[1mm]
By the same reason as the case 
$\inferrule{\iruleMPRedComm}$, we can set $\mpEtx{}= [\ ]$.
  Assume
$\stJudge{\mpEnv}{\stEnv\stEnvComp \stEnvMap{\mpChanRole{\mpS}{\roleP}}{S}}
                {
                  \mpCancel{\mpChanRole{\mpS}{\roleP}}{Q}
                }$
  with $\stJudge{\mpEnv}{\stEnv}{\mpQ}$.
  Then by $\inferrule{\iruleMPKills}$, 
  $\stJudge{\mpEnv}{\stEnvMap{\mpChanRole{\mpS}{\roleP}}{S}}
    {
      \kills{s}
    }$. Applying $\inferrule{\iruleMPPar}$, we obtain
  $\stJudge{\mpEnv}{\stEnv \stEnvComp \stEnvMap{\mpChanRole{\mpS}{\roleP}}{S}}
    {
      \mpQ\mpPar \kills{s}
    }$, as desired. \\[1mm]
  \textbf{Case \inferrule{\iruleMPCCat}:}
  $\trycatch{P}{Q} \mpPar \kills{\mpS}
    \mpMove Q \mpPar \kills{\mpS}$ with
  $\mpChanRole{\mpS}{\roleR}=\sbj{P}$ for some $\roleR$.
  \\[1mm]%
Similar to the case of \inferrule{\iruleMPCQSel} 
noting that $Q$ is typed by the same typing contexts as 
$\trycatch{P}{Q}$ (see the case of $\inferrule{\iruleMPRedComm}$).
\\[1mm]
  \textbf{Cases \inferrule{\iruleMPRedCall} and \inferrule{\iruleMPRedCtx}:} By
  ~\Cref{lem:globalsafety} and Subject Reduction Theorem
  in~\cite[Theorem 4.8]{scalasLess2019}. \\[1mm]
  \textbf{Case \inferrule{\iruleMPRedSt}:} By
  Theorem~\ref{thm:subjectcong}.
\end{proof}

\lemSessionFidelity*

\begin{proof} The proof is analogous with~\cite{scalasLess2019} except
that we have the additional cases to consider:
\begin{enumerate}
\item \label{caseOne:SF} $\kills{\mpS}$ as a unique role $\roleP$ in $s$.
\item \label{caseTwo:SF} $\kills{\mpS'}$ for some $s'$ is newly generated.
\end{enumerate}
\textbf{Case (\ref{caseOne:SF}):} Suppose $\mpP[\roleP]=\kills{s}$ for
some $\roleP$. The only interesting cases are: 
$\inferrule{\iruleMPTQSel}$, $\inferrule{\iruleMPCBra}$ and 
$\inferrule{\iruleMPCCat}$. These cases are trivial since if $P\mpMove
P'$ then $P'=R \mpPar \kills{s}$ for some $R$ thus $\kills{s}$ stays as
a unique role $\roleP$ in $s$ of $\mpPi$. \\[1mm]
\textbf{Case (\ref{caseTwo:SF}):} There are two main cases:
\begin{enumerate}
\item By $\inferrule{\iruleMPCQSel}$, $\inferrule{\iruleMPCBra}$, 
$\inferrule{\iruleMPCQBra}$, and $\inferrule{\iruleMPRedCan}$, 
$\kills{s}$ is newly generated. 
\begin{enumerate}
\item In the cases
of $\inferrule{\iruleMPCQSel}$ and $\inferrule{\iruleMPCQBra}$, 
we have $\stJudge{\emptyset}{\stEnvMap{\mpChanRole{\mpS}{\roleQ}}{\stEnd}}
{\kills{s}}$ where $\roleQ\not\in I$.
Hence, $\kills{s}$ is included in $\kills{\mpQ'}$. 
\item 
In the cases $\inferrule{\iruleMPCSel}$,
$\inferrule{\iruleMPCBra}$ and  
$\inferrule{\iruleMPRedCan}$, 
$\kills{s}$ will take a place as role $\roleP$ in $s$ of 
the selection and branching, and typed by $\stEnv[\roleP]$.
\end{enumerate}
\item $\kills{s'}$ for some $s'\not = s$ is newly generated by
  $\inferrule{\iruleMPCSel}$. In this case, by the definition of 
a unique role, we have 
$\stJudge{\emptyset}{\stEnvMap{\mpChanRole{\mpSi}{\roleP}}{\stEnd}}{\kills{s'}}$.
Hence, $\kills{s'}$ is included in $\kills{\mpQ'}$.
\end{enumerate}
This concludes the proof.
\end{proof}

\thmProcessDeadfree*

\begin{proof}
Recall that $\stEnvDFP{\stEnv}$ means  
$\stEnv \stEnvMove^\ast \stEnvi \not\stEnvMove$ %
    \;\;implies\;\; %
$\stEnvEndP{\stEnvi}$.
Assume 
$\stJudge{\mpEnvEmpty\!}{\!\stEnv}{\!\mpP}$ 
with 
$\mpP \equiv \mpBigPar{\roleP \in I}{\mpP[\roleP]}\mpPar \kills{Q}$, %
 \,and\, $\stEnv = \bigcup_{\roleP \in I}\stEnv[\roleP]\cup
 \stEnv[0]$
 such that, for each $\mpP[\roleP]$, %
 we have\, $\stJudge{\mpEnvEmpty\!}{\stEnv[\roleP]}{\!\mpP[\roleP]}$;
 and $\stJudge{\mpEnvEmpty\!}{\stEnv_0}{\kills{Q}}$.
By Theorem~\ref{lem:session-fidelity}, 
we have $\mpP \mpMove^\ast \mpPi$ and 
  Then, %
  $\stEnvMoveP{\stEnv}$ %
  \,implies\; %
  $\exists \stEnvi,\mpPi$ %
  such that\, %
  $\stEnv \!\stEnvMove^\ast\! \stEnvi$, %
  $\mpP \!\mpMoveStar\! \mpPi$ %
  \,and\, %
  $\stJudge{\mpEnvEmpty\!}{\!\stEnvi}{\mpPi}$, %
  \;with\; %
  $\stEnvi$ safe, %
  \,$\mpPi \equiv \mpBigPar{\roleP \in I}{\mpPi[\roleP]}\mpPar \kills{\mpQi}$, %
  \,and\, $\stEnvi = \bigcup_{\roleP \in I}\stEnvi[\roleP]\cup
  \stEnvi[0]$ 
such that, for each $\mpPi[\roleP]$, %
we have\, $\stJudge{\mpEnvEmpty\!}{\stEnvi[\roleP]}{\!\mpPi[\roleP]}$,
\,and each $\mpPi[\roleP]$
is either\, $\mpNil$, %
or only plays $\roleP$ in $\mpS$, by $\stEnvi[\roleP]$; 
and $\stJudge{\mpEnvEmpty\!}{\stEnvi_0}{\kills{\mpQi}}$.
By the definition of $\stEnvDFP{\stEnv}$, 
$\stEnvEndP{\stEnvi[\roleP]}$ and 
$\stEnvEndP{\stEnvi_0}$. By applying \inferrule{\iruleMPSafeRes} to $\kills{Q}$, 
we have $\mpRes{\stEnvMap{\vec{s}}{{\stEnvi_0}}}{\kills{Q}}\equiv
\mpNil$.
Then by $\stEnvEndP{\stEnvi[\roleP]}$, 
we have $\mpRes{\stEnvMap{\mpS}{\stEnvi}}{
\mpBigPar{\roleP \in I}{\mpPi[\roleP]}}\equiv 
\mpNil$.
\end{proof}


\thmProcessLiveness*

\begin{proof}
Assume 
$\stJudge{\mpEnvEmpty\!}{\!\stEnv}{\!\mpP}$ 
with 
$\mpP \equiv \mpBigPar{\roleP \in I}{\mpP[\roleP]}\mpPar \kills{Q}$, %
 \,and\, $\stEnv = \bigcup_{\roleP \in I}\stEnv[\roleP]\cup
 \stEnv[0]$
 such that, for each $\mpP[\roleP]$, %
 we have\, $\stJudge{\mpEnvEmpty\!}{\stEnv[\roleP]}{\!\mpP[\roleP]}$;
 and $\stJudge{\mpEnvEmpty\!}{\stEnv_0}{\kills{Q}}$.
By Theorem~\ref{lem:session-fidelity}, 
we have $\mpP \mpMove^\ast \mpPi$ and 
  Then, %
  $\stEnvMoveP{\stEnv}$ %
  \,implies\; %
  $\exists \stEnvi,\mpPi$ %
  such that\, %
  $\stEnv \!\stEnvMove^\ast\! \stEnvi$, %
  $\mpP \!\mpMoveStar\! \mpPi$ %
  \,and\, %
  $\stJudge{\mpEnvEmpty\!}{\!\stEnvi}{\mpPi}$, %
  \;with\; %
  $\stEnvi$ safe, %
  \,$\mpPi \equiv \mpBigPar{\roleP \in I}{\mpPi[\roleP]}\mpPar \kills{\mpQi}$, %
  \,and\, $\stEnvi = \bigcup_{\roleP \in I}\stEnvi[\roleP]\cup
  \stEnvi[0]$ 
such that, for each $\mpPi[\roleP]$, %
we have\, $\stJudge{\mpEnvEmpty\!}{\stEnvi[\roleP]}{\!\mpPi[\roleP]}$,
\,and each $\mpPi[\roleP]$
is either\, $\mpNil$, %
or only plays $\roleP$ in $\mpS$, by $\stEnvi[\roleP]$; 
and $\stJudge{\mpEnvEmpty\!}{\stEnvi_0}{\kills{\mpQi}}$.
The only important difference from 
\cite[Theorem 5.15]{scalasLess2019} is that 
$\mpPi[\roleP]$ and $\kills{\mpQi}$ interact
together by $\inferrule{\iruleMPCSel}$ and $\inferrule{\iruleMPCBra}$.
In the case of $\inferrule{\iruleMPCSel}$, it matches with 
in~\Cref{def:process-properties}(\ref{item:process-liveness:sel}).
In the case of $\inferrule{\iruleMPCBra}$, it matches with 
in~\Cref{def:process-properties}(\ref{item:process-liveness:branch}).
\end{proof}

\thmCancellationTermination*

\begin{description}
\item[Case 1:]
Suppose $\mpQ$ is finite, \ie it does not contain any (free or bound)
process calls. 
Then by Corollary \ref{cor:liveness}, for all $\mpQi$ such that 
$\mpQ\!\mpMoveStar\! \mpQi$, there exits a reduction 
sequence such that $\mpQi \!\mpMoveStar\! \mpQii\equiv 
\mpNil$.
\item[Case 2:]
Assume $\mpQ\!\mpMoveStar\!
\mpCtx{}{[\kills{s}]}$ and $\mpQ$ contains 
the process definitions and calls, \ie ~$\mpQ$ can produce 
the infinite sequence of reductions. In this case, 
the only interesting case is
$\mpCtx{}{[\kills{s}]} \mpMoveStar
{\mpCtx{}'}[\mpDef{\mpX}{\widetilde{x}}{\mpQii}{\mpR}\mpPar \kills{s}]$
where, by Definition~\ref{def:imps}(3), 
$\mpQii=\trycatch{\mpPi}{\mpR}$ where $\mpR$ is finite. 
By Definition~\ref{def:unique-role-proc}, 
$\sbj{\mpPi}=\mpChanRole{s}{\roleR}$ for some $\roleR$.
Hence, using $\inferrule{\iruleMPRedCall}$ and 
$\inferrule{\iruleMPRedCat}$, we have 
${\mpCtx{}'}[\mpDef{\mpX}{\widetilde{x}}{\mpQii}{\mpPi}\mpPar
\kills{s}]\mpMove\mpMove {\mpCtx{}'}[\mpR\mpPar \kills{s}]$ 
where $\mpR$ is finite. The rest is the same as 
\textbf{Case (1)}. 
\end{description}

\subsection{Additional examples and definitions for our theory}
\label{subsec:app:additional:examples}

\begin{figure*}[t]
    \small
    \[
        \begin{array}{c}
            \colorbox{ColourShade}{$\kills{\mpS} \mpPar
                    \kills{\mpS}
                    \;\equiv\;%
                    \kills{\mpS}$}
            \quad%
            \colorbox{ColourShade}{$\mpRes{\mpS}{\kills{\mpS}}%
                    \;\equiv\;%
                    \mpNil$}
            \\[1mm]%
            \mpP \mpPar \mpQ%
            \;\equiv\;%
            \mpQ \mpPar \mpP%
            \quad%
            \mpFmt{(\mpP \mpPar \mpQ) \mpPar \mpR}%
            \;\equiv\;%
            \mpP \mpPar \mpFmt{(\mpQ \mpPar \mpR)}%
            \quad%
            \mpP \mpPar \mpNil%
            \;\equiv\;%
            \mpP%
            \quad%
            \mpRes{\mpS}{\mpNil}%
            \;\equiv\;%
            \mpNil%
            \quad%
            \mpRes{\mpS}{%
                \mpRes{\mpSi}{%
                    \mpP%
                }%
            }%
            \;\equiv\;%
            \mpRes{\mpSi}{%
                \mpRes{\mpS}{%
                    \mpP%
                }%
            }%
            \\[1mm]%
            \mpRes{\mpS}{(\mpP \mpPar \mpQ)}%
            \;\equiv\;%
            \mpP \mpPar \mpRes{\mpS}{\mpQ}%
            \quad%
            \text{\footnotesize{}if\, $\mpS \not\in \fc{\mpP}$}%
            \quad%
            \mpDefAbbrev{\mpDefD}{\mpNil}%
            \;\equiv\;%
            \mpNil%
            \\[1mm]%
            \mpDefAbbrev{\mpDefD}{\mpRes{\mpS}{\mpP}}%
            \;\equiv\;%
            \mpRes{\mpS}{(
                \mpDefAbbrev{\mpDefD}{\mpP}%
                )}%
            \quad%
            \text{\footnotesize{}if\, $\mpS \not\in \fc{\mpP}$}%
            \\[1mm]%
            \mpDefAbbrev{\mpDefD}{(\mpP \mpPar \mpQ)}%
            \;\equiv\;%
            \mpFmt{(\mpDefAbbrev{\mpDefD}{\mpP})} \mpPar \mpQ%
            \quad%
            \text{\footnotesize{}if\, $\dpv{\mpDefD} \cap \fpv{\mpQ} = \emptyset$}%
            \\[1mm]%
            \mpDefAbbrev{\mpDefD}{%
                (\mpDefAbbrev{\mpDefDi}{\mpP})%
            }%
            \;\equiv\;%
            \mpDefAbbrev{\mpDefDi}{%
                (\mpDefAbbrev{\mpDefD}{\mpP})%
            }%
            \\
            \text{\footnotesize%
                if\,%
                $(\dpv{\mpDefD} \cup \fpv{\mpDefD}) \cap \dpv{\mpDefDi}%
                    \,=\,%
                    (\dpv{\mpDefDi} \cup \fpv{\mpDefDi}) \cap \dpv{\mpDefD}%
                    \,=\,%
                    \emptyset$%
            }%
        \end{array}
    \]
    
    \caption{%
    Structural congruence rules, extended version. 
    }%
    \label{app:fig:mpst-pi-semantics}%
    \label{app:fig:structure}%
    
\end{figure*}

\begin{restatable}[Nested \CODE{try-catch} processes]{example}{exampleNestedTryCatchProcesses}
  \label{app:ex:nestedtrycatch}
  The following shows the examples of nested \CODE{try-catch} processes 
  which are typable by our typing system introduced later. 
  Below $R_1$ has a simple nested \CODE{try-catch} process,
  $R_2$ has a session delegation,
  $R_3$ has a parallel composition (spawning after the first try),
  and $R_4$ has a scope opening after the first try.  
  Notice that our Rust implementation can write the interleaved 
  sessions and delegations,
  some related case studies can be found
  in~\Cref{subsec:app:implementation:rust:interleaved}.

  \begin{align}
     & \trycatchbreakequal{R_1}{
        \mpSel{
            \mpChanRole{\mpS}{\roleP}
        }{
            \roleR
        }{\stLab[1]}{}{
            (
            \trycatch{
                \mpSel{
                    \mpChanRole{\mpSi}{\roleQ}
                }{\roleR}{\stLab[2]}{}{P}
            }{
                \mpCancel{\mpChanRole{\mpSi}{\roleQ}}{
                    \mpNil
                }
            }
            )
        }
    }{
        \mpCancel{\mpChanRole{\mpS}{\roleP}}{
            \mpCancel{\mpChanRole{\mpSi}{\roleQ}}{
                \mpNil
            }
        }
    }
    \label{example:r:one}        \\
     & \trycatchbreakequal{R_2}{
        \mpSel{
            \mpChanRole{\mpS}{\roleP}
        }{\roleR}{\stLab[1]}{
            \mpChanRole{\mpSii}{\roleR}
        }{
            (
            \trycatch{
                \mpSel{
                    \mpChanRole{\mpSi}{\roleQ}
                }{\roleR}{\stLab[2]}{}{\mpNil}
            }{
                \mpCancel{\mpChanRole{\mpSi}{\roleQ}}{\mpNil}
            }
            )
        }
    }{
        \mpCancel{\mpChanRole{\mpS}{\roleP}}{
            \mpCancel{\mpChanRole{\mpSi}{\roleQ}}{
                \mpCancel{\mpChanRole{\mpSii}{\roleR}}{\mpNil}}
        }
    } \label{example:r:two}      \\
     & \trycatchbreakequal{R_3}{
        \mpSel{
            \mpChanRole{\mpS}{\roleP}
        }{\roleR}{\stLab[1]}{
        }{
            (
            \mpSel{
                \mpChanRole{\mpS}{\roleP}
            }{\roleQ}{\stLab[3]}{}{
                \mpNil
            }
            \mpPar
            \trycatch{
                \mpSel{
                    \mpChanRole{\mpSi}{\roleQ}
                }{\roleR}{\stLab[2]}{}{
                    \mpNil
                }
            }{
                \mpCancel{\mpChanRole{\mpSi}{\roleQ}}{
                    \mpNil
                }
            }
            )
        }
    }{
        \mpCancel{\mpChanRole{\mpS}{\roleP}}{
            \mpCancel{\mpChanRole{\mpSi}{\roleQ}}{
                \mpNil
            }
        }
    } \label{example:r:three}    \\
     & \trycatchbreakequal{R_4}{
        \mpSel{
            \mpChanRole{\mpS}{\roleP}
        }{\roleR}{\stLab[1]}{}{
            \mpRes{\mpSi}{
                (
                \mpSel{
                    \mpChanRole{\mpS}{\roleP}
                }{\roleR}{\stLab[3]}{
                    \mpChanRole{\mpSi}{\roleR}
                }{
                    \mpNil
                }
                \mpPar
                \trycatch{
                    \mpSel{
                        \mpChanRole{\mpSi}{\roleQ}
                    }{\roleR}{\stLab[2]}{}{
                        \mpNil
                    }
                }{
                    \mpCancel{\mpChanRole{\mpSi}{\roleQ}}{
                        \mpNil
                    }
                }
                )
            }
        }
    }{
        \mpCancel{\mpChanRole{\mpS}{\roleP}}{
            \mpNil
        }
    } \label{example:r:four}
\end{align}
  \label{app:example:r:two:to:four}
\end{restatable}

\begin{restatable}[Typing nested Try-catch]{example}{exampleTypingNestedTryCatchProcesses}
  \label{ex:typing:nested}
  
  Typing derivation for~\Cref{example:r:one}.
  Assume
  $
  \mpP
  =
  \mpNil
  $
  in~\Cref{example:r:one}
  and
  $
  R_{1.1}
  =
  \trycatch{
    \mpSel{
      \mpChanRole{\mpSi}{\roleQ}
    }{\roleR}{\stLab[2]}{}{\mpNil}
  }{
    \mpCancel{\mpChanRole{\mpSi}{\roleQ}}{
      \mpNil
    }
  }
  $
  .

  {
  \footnotesize
  \[
    \begin{array}{l}
      \inference[\iruleMPTryCatch]{
        \inference[\iruleMPSel]{
          \text{See~\Cref{ex:typing:nested:aux}}
        }{
          \stJudge{}{
            \stEnv
          }{
            \mpSel{
              \mpChanRole{\mpS}{\roleP}
            }{
              \roleR
            }{
              \stLab[1]
            }{}{
              R_{1.1}
            }
          }   
        }{}
        &
        \sbj{R_1}=
        \{
          \mpChanRole{\mpS}{\roleP}
        \}
        &
        \inference[\iruleMPCancel]{
          \inference[\iruleMPCancel]{
            \inference[\iruleMPNil]{
              \stEnvEndP{
                \stEnvEmpty
              }
            }{
            \stJudge{}{\stEnvEmpty}{\mpNil}
            }
          }{
            \stJudge{}{\stEnv[1]}
            {
              \mpCancel{\mpChanRole{\mpSi}{\roleQ}}{\mpNil}
            }
          }
        }{
          \stJudge{}{\stEnv}{
          \mpCancel{\mpChanRole{\mpS}{\roleP}}{
            \mpCancel{\mpChanRole{\mpSi}{\roleQ}}{\mpNil}
          }
          }
        }     
      }
      {
        \stJudge{}{
          \stEnv
        }{
          {R_1}
        }
      }
    \end{array}
  \]
  }

  where
  $
  \stEnv[1]
  =
  \stEnvMap{
    \mpChanRole{\mpSi}{\roleQ}
  }{
    \stIntSum{
      \roleR
    }{}{
      \stLab[2]
    }
  }
  $
  and
  $
  \stEnv
  =
  \stEnv[1]
  \stEnvComp
  \stEnvMap{
    \mpChanRole{\mpS}{\roleP}
  }{
    \stIntSum{\roleR}{}{\stChoice{\stLab[1]}{}}
  }
  $
  .
\end{restatable}

\begin{restatable}[Typing nested Try-catch aux]{example}{exampleTypingNestedTryCatchProcessesAux}
  \label{ex:typing:nested:aux}
  
  {
    \footnotesize
    \[
      \begin{array}{lll}
        \inference[\iruleMPSel]{
          \inference[\iruleMPSub]{
            \dots
          }{
            \stEnvEntails{\stEnv[2]}{
              \mpChanRole{\mpS}{\roleP}
            }{
              \stIntSum{\roleR}{}{\stChoice{\stLab[1]}{}}
            }
          }
          &
          \inference[\iruleMPTryCatch]{
            \inference[\iruleMPSel]{
              \dots
            }{
              \stJudge{}{
                +
                \stEnv[3]
              }{
                \mpSel{
                  \mpChanRole{\mpSi}{\roleQ}
                }{
                  \roleR
                }{
                  \stLab[2]
                }{}{
                  \mpNil
                }
              }   
            }{}
            &
            \sbj{
              \mpSel{
                \mpChanRole{\mpSi}{\roleQ}
              }{\roleR}{\stLab[2]}{}{\mpNil}
            }
            =
            \{
              \mpChanRole{\mpSi}{\roleQ}
            \}
            &
            \inference[\iruleMPCancel]{
              \inference[\iruleMPNil]{
                \stEnvEndP{
                  \stEnvMap{
                  \mpChanRole{\mpS}{\roleP}
                  }{
                  \stEnd
                  }
                }
              }{
              \stJudge{}{
                \stEnvMap{
                  \mpChanRole{\mpS}{\roleP}
                }{
                  \stEnd
                }
              }{\mpNil}
              }
            }{
              \stJudge{}{\stEnv[3]}{
                \mpCancel{\mpChanRole{\mpSi}{\roleQ}}{\mpNil}
              }
            }      
          }{
            \stJudge{}{
              \stEnv[3]
            }{
              \trycatch{
                \mpSel{
                  \mpChanRole{\mpSi}{\roleQ}
                }{\roleR}{\stLab[2]}{}{\mpNil}
              }{
                \mpCancel{\mpChanRole{\mpSi}{\roleQ}}{
                  \mpNil
                }
              }
            }
          }{}
        }{
          \stJudge{}{
            \stEnv
          }{
            \mpSel{
              \mpChanRole{\mpS}{\roleP}
            }{
              \roleR
            }{
              \stLab[1]
            }{}{
              R_{1.1}
            }
          }   
        }{}
      \end{array}
    \]
    }
  
    because $\stLab[1]$ does not have any payload
    and
    where
    $
    \stEnv[3]
    =
    \stEnv[1]
    \stEnvComp
    \stEnvMap{\mpChanRole{\mpS}{\roleP}}{\stEnd}
    .
    $
    .
\end{restatable}

\begin{restatable}[Typing nested Try-catch with rec]{example}{exampleTypingNestedTryCatchProcessesRec}
  \label{ex:typing:nested:rec}

Typing derivation for~\Cref{example:r:four}.
Assume
$
    R_{4.1}
    =
    \mpRes{\mpSi}{
        (
        \mpSel{
            \mpChanRole{\mpS}{\roleP}
        }{\roleR}{\stLab[3]}{
            \mpChanRole{\mpSi}{\roleR}
        }{
            \mpNil
        }
        \mpPar
        \trycatch{
            \mpSel{
                \mpChanRole{\mpSi}{\roleQ}
            }{\roleR}{\stLab[2]}{}{
                \mpNil
            }
        }{
            \mpCancel{\mpChanRole{\mpSi}{\roleQ}}{
                \mpNil
            }
        }
        )
    }
$
.

    {
        \footnotesize
        \[
            \begin{array}{l}
                \inference[\iruleMPTryCatch]{
                    \inference[\iruleMPSel]{
                        \text{See~\Cref{ex:typing:nested:rec:aux}}
                    }{
                        \stJudge{}{
                            \stEnv
                        }{
                            \mpSel{
                                \mpChanRole{\mpS}{\roleP}
                            }{
                                \roleR
                            }{
                                \stLab[1]
                            }{}{
                                R_{4.1}
                            }
                        }
                    }{}
                 &
                    \sbj{R_4}=
                    \{
                    \mpChanRole{\mpS}{\roleP}
                    \}
                 &
                    \inference[\iruleMPCancel]{
                        \inference[\iruleMPNil]{
                            \stEnvEndP{
                                \stEnvEmpty
                            }
                        }{
                            \stJudge{}{\stEnvEmpty}{\mpNil}
                        }
                    }{
                        \stJudge{}{
                            \stEnv
                        }{
                            \mpCancel{\mpChanRole{\mpS}{\roleP}}{\mpNil}
                        }
                    }
                }
                {
                    \stJudge{}{
                        \stEnv
                    }{
                        {R_4}
                    }
                }
            \end{array}
        \]
    }

where
$
    \stEnv
    =
    \stEnvMap{
        \mpChanRole{\mpS}{\roleP}
    }{
        \stIntSum{\roleR}{}{\stChoice{\stLab[1]}{}}
    }
$
.
\end{restatable}
  
\begin{restatable}[Typing nested Try-catch with rec aux]{example}{exampleTypingNestedTryCatchProcessesRecAux}
\label{ex:typing:nested:rec:aux}
Assume
$
    R_{4.2}
    =
    \mpSel{
        \mpChanRole{\mpS}{\roleP}
    }{\roleR}{\stLab[3]}{
        \mpChanRole{\mpSi}{\roleR}
    }{
        \mpNil
    }
$
and
$
    R_{4.3}
    =
    \trycatch{
        \mpSel{
            \mpChanRole{\mpSi}{\roleQ}
        }{\roleR}{\stLab[2]}{}{
            \mpNil
        }
    }{
        \mpCancel{\mpChanRole{\mpSi}{\roleQ}}{
            \mpNil
        }
    }
$
.

    {
        \footnotesize
        \[
            \begin{array}{lll}
                \inference[\iruleMPSel]{
                    \inference[\iruleMPSub]{
                        \dots
                    }{
                        \stEnvEntails{\stEnv[2]}{
                            \mpChanRole{\mpS}{\roleP}
                        }{
                            \stIntSum{\roleR}{}{\stChoice{\stLab[1]}{}}
                        }
                    }
                 &
                    \inference[\iruleMPSafeRes]{
                        \stEnv[4] = \setenum{
                            \stEnvMap{
                                \mpChanRole{\mpS}{\roleP}
                            }{\stS[\roleP]}
                        }_{\roleP \in I}
                 &
                        \mpS \not\in \stEnv[3]
                 &
                        \stEnvSafeP{\stEnv[4]}
                 &
                        \inference[\iruleMPPar]{
                            \inference[\iruleMPSel]{
                                \dots
                            }{
                                \stJudge{}{
                                    \stEnv[3]
                                }{
                                    R_{4.2}
                                }
                            }{}
                 &
                            \inference[\iruleMPTryCatch]{
                                \dots
                            }{
                                \stJudge{}{
                                    \stEnv[4]
                                }{
                                    R_{4.3}
                                }
                            }{}
                        }{
                            \stJudge{}{
                                \stEnv[3] \stEnvComp \stEnv[4]
                            }{
                                R_{4.2}
                                \mpPar
                                R_{4.3}
                            }
                        }{}
                    }{
                        \stJudge{}{
                            \stEnv[3]
                        }{
                            \mpRes{
                                \stEnvMap{\mpSi}{\stEnv[4]}
                            }{
                                (
                                R_{4.2}
                                \mpPar
                                R_{4.3}
                                )
                            }
                        }
                    }{}
                }{
                    \stJudge{}{
                        \stEnv
                    }{
                        \mpSel{
                            \mpChanRole{\mpS}{\roleP}
                        }{
                            \roleR
                        }{
                            \stLab[1]
                        }{}{
                            R_{4.1}
                        }
                    }
                }{}
            \end{array}
        \]
    }

because $\stLab[1]$ does not have any payload
and
where
$
    \stEnv[3]
    =
    \stEnvMap{
        \mpChanRole{\mpS}{\roleP}
    }{
        \stIntSum{
            \roleR
        }{}{
            \stChoice{\stLab[3]}{
                \stS[
                    \mpChanRole{\mpSi}{\roleR}
                ]}
        }
    }
$
and
$
    \stEnv[4]
    =
    \stEnvMap{
        \mpChanRole{\mpSi}{\roleQ}
    }{
        \stIntSum{
            \roleR
        }{}{
            \stLab[2]
        }
    }
    \stEnvComp
    \stEnvMap{\mpChanRole{\mpS}{\roleP}}{\stEnd}
$
.
$
    \stEnvSafeP{\stEnv[4]}
$
is trivial because it only contains one endpoint.

\end{restatable}


\section{Additional implementation details}
\label{app:implementation}

\subsection{Interleaved sessions in~\mpstrust}
\label{subsec:app:implementation:rust:interleaved}


\begin{figure}[!t]
\begin{subfigure}{0.48\textwidth}
\begin{rustlisting}
fn endpoint_controller(
 s_circuit_breaker: CircuitBreakerController,
 s_logging: LogController,
) -> Result<(), Box<dyn Error>> {
 let s_circuit_breaker = *@\label{line:implementation:interleaved:usage:start}@*
  s_circuit_breaker.send(0)?;
 let s_logging = *@\label{line:implementation:interleaved:usage:middle}@*
  s_logging.send(0)?;
 let s_circuit_breaker = 
  s_circuit_breaker.send(0)?;
 let (_, s_circuit_breaker) = *@\label{line:implementation:interleaved:usage:end}@*
  s_circuit_breaker.recv()?;
 ...
}
\end{rustlisting}
\caption{Interleaved session example}
\label{fig:implementation:interleaved:usage}
\end{subfigure}
\quad
\begin{subfigure}{0.48\textwidth}
\begin{rustlisting}[basicstyle=\scriptsize\ttfamily]
create_fork_interleaved!( ... );*@\label{line:implementation:interleaved:fork:macro}@*

 let (*@\label{line:implementation:interleaved:fork:function}@*
  thread_api,
  thread_storage,
  thread_user,
  thread_logs,
  thread_controller
 ) = fork_interleaved(
  endpoint_api,
  endpoint_storage,
  endpoint_user,
  endpoint_logs,
  endpoint_controller
 ); // The function for the interleaved session is always at the end
\end{rustlisting}
\caption{Interleaved session forking}
\label{fig:implementation:interleaved:fork}
\end{subfigure}
\caption{Interleaved sessions}
\label{fig:implementation:interleaved}

\end{figure}


In some cases, participants are involved in multiple protocols at the same time,
but the different sets of participants are completely disconnected.
Instead of having one thread for each participant of each protocol,
we may be tempted to run each shared participant
on one thread.
This way, the shared participants will give the rhythm between
the different protocols because
all theirs operations will run sequentially
For instance, in \S~\ref{app:subsec:benchmarks:genericComponentLifecycleManagement},
\role{Controller} can be shared between the two
protocols \emph{circuit breaker} and \emph{distributed logging}:
its related session will be qualified as \emph{interleaved}.
We implemented a way to work with such interleaved sessions:
the main idea is that the function which represents
the shared function needs two \CODE{MeshedChannels}
as inputs, instead of one for the others.
~\Cref{fig:implementation:interleaved:usage} provides an
overview of what the implementation of \role{Controller}
would look like.
From line~\ref{line:implementation:interleaved:usage:start}
to line~\ref{line:implementation:interleaved:usage:end},
we can see that the two endpoints that represent
the participant work sequentially:
first we \CODE{send()} and \CODE{recv()} to the participants
of the \emph{circuit breaker} protocol,
then we \CODE{send()} on the \emph{distributed logging} protocol.
~\Cref{fig:implementation:interleaved:fork} shows
how to \emph{fork} the different endpoints:
the \CODE{fork_interleaved()} function
(line~\ref{line:implementation:interleaved:fork:function}),
created through the macro \CODE{create_fork_interleave!},
(line~\ref{line:implementation:interleaved:fork:macro}),
will accept 5 endpoints in our case:
one for each participant, grouped by protocol, and
the last argument of \CODE{fork_interleaved()}
is always the function representing the interleaved session.
The function returns 5 threads, one for each participant,
in the same order as the arguments.


\begin{figure}[t!]
\begin{minipage}[l]{0.48\textwidth}
\begin{rustlisting}[basicstyle=\scriptsize\ttfamily]
type StoAClose = <AtoSClose as Session>::Dual;
type StoCClose = End;
type StoAVideo<N> = <StoBVideo<N> as Session>::Dual;

type RecursStoC<N> = Recv<Branches0StoC<N>, End>;


// Declare the name of the role
type NameS = RoleS<RoleEnd>;


// Declare meshed channel
type EndpointSRecurs<N> = MeshedChannels<End, RecursStoC<N>, StackSRecurs, RoleS<RoleEnd>>;
\end{rustlisting}
\end{minipage}
\quad
\begin{minipage}[r]{0.48\textwidth}
{\lstset{firstnumber=14,}
 \begin{rustlisting}[basicstyle=\scriptsize\ttfamily,]
// Declare usage order of channels for choice
type StackSEnd = RoleEnd;
type StackSVideo = RoleA<RoleA<RoleC<RoleEnd>>>;
type StackSRecurs = RoleC<RoleEnd>;

enum ChoiceS<N: marker::Send> {
 End(MeshedChannels<StoAClose, StoCClose, StackSEnd, RoleS<RoleEnd>>),
 Video(MeshedChannels<StoAVideo<N>, RecursStoC<N>, StackSVideo, RoleS<RoleEnd>>),
}

type RecS<N> = MeshedChannels<End, Recv<ChoiceS<N>, End>, RoleC<RoleEnd>, RoleS<RoleEnd>>;
\end{rustlisting}}
\end{minipage}
\caption{Local Rust types for \role{S} (Server) from \Cref{fig:implementation:role:s}}
\label{fig:implementation:s}
\end{figure}


\begin{figure}[t!]
\begin{minipage}[l]{0.48\textwidth}
\begin{rustlisting}[basicstyle=\scriptsize\ttfamily]
// The types to send the new MeshedChannels
type Choose0fromCtoA<N> = Send<ChoiceA<N>, End>;
type Choose0fromCtoS<N> = Send<ChoiceS<N>, End>;

type InitC<N> = Send<N, Recv<N, Choose0fromCtoA<N>>>;


// Declare the name of the role
type NameC = RoleC<RoleEnd>;
\end{rustlisting}
\end{minipage}
\quad
\begin{minipage}[r]{0.48\textwidth}
{\lstset{firstnumber=10,}
\begin{rustlisting}[basicstyle=\scriptsize\ttfamily,]
// Declare usage order of channels for choice
type StackCRecurs = RoleBroadcast;
type StackCFull = RoleA<RoleA<StackCRecurs>>;

// Declare meshed channels
type EndpointCRecurs<N> = MeshedChannels<Choose0fromCtoA<N>, Choose0fromCtoS<N>, StackCRecurs, RoleC<RoleEnd>>;
type RecC<N> = MeshedChannels<InitC<N>, Choose0fromCtoS<N>, StackCFull, RoleC<RoleEnd>>;
\end{rustlisting}}
\end{minipage}
\caption{Local Rust types for \role{C} (Client) from \Cref{fig:implementation:role:c}}
\label{fig:implementation:c}
\end{figure}

\iftoggle{full}{%
}{%
  Details of the pseudocode
  provided in~\Cref{fig:MPSTMeshedChannelsPrimitives}.
}%


\subsection{Communication primitives}
\label{sec:primitives}
To enable polymorphic behaviour for meshed channels, we leverage Rust
generic \CODE{impl} blocks.
In particular,
consider the code snippet
\CODE{impl<T> Bar<T>\{ fn foo()\{...\}\}}.
It specifies the implementation of the function
\CODE{foo()} for the structure \CODE{Bar},
parameterised by a generic type \CODE{T}.
Similarly,
the \CODE{gen_mpst!} macro generates functions
for sending and receiving on meshed channels.

\myparagraph{Generating send/recv communication primitives}
\listing{app:fig:MPSTMeshedChannelsPrimitives} left
(lines~\ref{app:line:mpst:send} --~\ref{app:line:mpst:close})
shows an example of a \CODE{send()} and \CODE{recv()} functions that are
generated from \CODE{gen_mpst!}.
We explain the implementation of \CODE{send()}; \CODE{recv()} is similar.
The \CODE{send()} function requires a caller of a type
\CODE{MeshedChannels<S1, Send<T, S2>, RoleS<R>, RoleA<RoleEnd>>}.
From the types, we can deduce that this is a meshed channel
for the \role{A}, as clear from the last argument
\CODE{RoleA<RoleEnd>}.
The next operation on the meshed channel should be a communication
with the binary channel for \role{S}
as stipulated by the stack type \CODE{RoleS<R>}.
Note that binary channels in a meshed channel are ordered in alphanumerical order.
We also see that the binary channel for \role{S} should be of type \CODE{Send<T, S2>},
which is a channel with capabilities to send payloads of type \CODE{T} and to return a continuation
of type \CODE{S2}. Let us consider that \CODE{s} is a meshed channels of the given \CODE{MeshedChannels} type,
then \CODE{s.send(p)} will send the payload \CODE{p} of type \CODE{T} on the binary
channel between \role{A} and \role{S}.

The return type of the function \CODE{send()} is specified as
\CODE{RT<S1, S2, R>}, which is essentially a meshed channel encapsulated inside an option type.
Remember that all communication primitives of~\mpstrust return option types:
they need to be unwrapped to check first whether they are an error
or an expected result,
before retrieving the said result.
Notice that the returned meshed channel should have as a binary session type for \role{S} the continuation \CODE{S2},
the type for the binary channel for \role{C} stays the same as \CODE{S1}
and the stack should be \CODE{R}.

Let us now look at the body of the function, which essentially sends the message of type \CODE{T}
on the binary channel stored in the second field,
\CODE{session2} (corresponding to the binary session with \role{S}),
of the meshed channel \CODE{s}.
At line~\ref{app:line:mpst:next}, the head of the \textbf{stack} field is consumed
with \CODE{s.stack.continuation()} and its continuation is returned.
Since the communication is on a binary channel,
we reuse the binary \CODE{send()} primitive (line~\ref{app:line:mpst:send:binary})
from~\cite{kokkeRusty2019}.

\myparagraph{Communication Transport} To effectively send a message,
the binary \CODE{send()} primitive from~\cite{kokkeRusty2019}
uses \texttt{crossbeam-channel's} primitives,
a community-driven alternative library
to the standard Rust library \CODE{std::sync::mpsc}
with more features and better performance.
As a consequence,~\mpstrust also relies on the
\texttt{crossbeam-channel} message-passing library as a default transport,
included in the standard Rust distribution.
However, other transports are possible.
We have chosen
the \texttt{crossbeam-channel} library because
it provides the best trade-off.
In comparison,
\CODE{std::sync::mpsc} (a standard library of Rust)
has fewer features and worse performance~\cite{web:rust:crossbeamChannel}.

\myparagraph{External and internal choices} are implemented as separate macros that
require an argument of type \CODE{MeshedChannels}.
The implementation of the external choice, \CODE{offer_mpst!}, is given in
lines~\ref{app:line:offer:start} --~\ref{app:line:offer:end}.
In essence, a choice is implemented as a broadcast.
In our usecase, the active role that makes the choice is \role{C}.
Hence, the macro \CODE{offer_mpst!} explicitly performs a
receive (\CODE{s.recv()}) on the meshed channel \CODE{s}.
The received value is pattern matched and passed to any of the
functions given as arguments to \CODE{offer_mpst!}.
This macro expects only  those two types of arguments:
the receiving channel does not need to be provided
as the type of the \CODE{MeshedChannels} is following
a strict pattern with only one \CODE{Recv<enum, End>} type
and \CODE{End} types for the rest of the binary types.
Similarly,
\CODE{choose_c!} in lines~\ref{app:line:choice:start} --~\ref{app:line:choice:end}
is a macro that performs a select operation.
This macro expects the running session \CODE{s} and two variants of \CODE{enum},
each one being sent to the corresponding passive role.
\CODE{choose_c!} broadcasts the new choice
to every other
at lines~\ref{app:line:broadcast:one} and~\ref{app:line:broadcast:two}.
The name of \CODE{choose_c!}
as well as the number of expected variants of
\CODE{enum} corresponding to the number of passive roles
is decided by the macro \CODE{gen_mpst!}. 
In our specific example,
\role{C} sends the selected labels \CODE{l1} and \CODE{l2} alongside the new meshed channels
\CODE{s1} and \CODE{s2} that \role{A} and \role{S} should
use to communicate with \role{C}.

Note that all \CODE{send()}, \CODE{recv()}, \CODE{offer_mpst!} and
\CODE{choose_c!} primitives that fit the given \CODE{MeshedChannels}
and roles and all the possible interactions between them
are generated through the macro \CODE{gen_mpst!}.
\begin{figure}[t]
\begin{minipage}[l]{1\textwidth}
\begin{rustlisting}[basicstyle=\scriptsize\ttfamily]
pub struct MeshedChannels< S1: Session, S2: Session, R: Role, N: Role> {*@ \label{app:line:mpst:beg} @*
pub session1: S1, pub session2: S, pub stack: R, pub name: N } *@ \label{app:line:mpst:end} @*
\end{rustlisting}
\hrule
\end{minipage}
\begin{minipage}[l]{0.515\textwidth}
{\lstset{firstnumber=4}
\begin{rustlisting}[basicstyle=\scriptsize\ttfamily]
impl<S1: Session, S2: Session, R: Role, T: marker::Send> *@ \label{app:line:mpst:send:start} @*
    MeshedChannels<S1, Send<T, S2>, RoleS<R>, RoleA<RoleEnd>> {
    pub fn send(self, x: T) -> RT<S1, S2, R> { *@ \label{app:line:mpst:send} @*
        let new_binary_channel = send(x, self.session2)?; *@ \label{app:line:mpst:send:binary} @*
		let new_stack = s.stack.continuation(); *@ \label{app:line:mpst:next} @*
		// construct and return a new meshed channel
    }  *@ \label{app:line:mpst:close} @*
} *@ \label{app:line:mpst:send:end} @*
\end{rustlisting}
}
\end{minipage}
\hspace{2mm}
\begin{minipage}[l]{0.515\textwidth}
{\lstset{firstnumber=4}
\begin{rustlisting}[basicstyle=\scriptsize\ttfamily]
impl<S1: Session, S2: Session, R: Role, T: marker::Send>
    MeshedChannels<S1, Recv<T, S2>, RoleS<R>, RoleA<RoleEnd>> {
    /// Receive a payload from role S
    pub fn recv(self) -> RST<T, S1, S2, R> { *@ \label{app:line:mpst:recv} @*
		let (x, new_binary_channel) = recv(x, self.session2)?;
		let new_stack = s.stack.continuation();
		// return x and a newly constructed meshed channel
    } }
\end{rustlisting}}
\end{minipage}
\hrule
\begin{minipage}{0.515\textwidth}
{\lstset{firstnumber=14}
\begin{rustlisting}[basicstyle=\scriptsize\ttfamily]
#[macro_export]
macro_rules! offer_mpst {
    ($session: expr, { $( $pat: pat => $result: expr, )+ }) => { *@ \label{app:line:offer:start} @*
        (move || -> Result<_, _> {
            let (l, s) = $session.recv()?; // receive a label l and a new meshed channel s
            cancel(s);
			// Invoke the function matching the label l
			// and pass the new meshed channel s
            match l { $( $pat => $result, )* } *@ \label{app:line:offer:end} @*
        })() }; }
\end{rustlisting}}
\end{minipage}
\hspace{2mm}
\begin{minipage}{0.515\textwidth}
{\lstset{firstnumber=14}
\begin{rustlisting}[basicstyle=\scriptsize\ttfamily]
#[macro_export]
macro_rules! choose_c($s, $l1, $l2) { *@ \label{app:line:choice:start} @*
    let (s1, s1_dual) = MeshedChannels::new();
    let (s2, s2_dual) = MeshedChannels::new();
    // create new meshed channels, send to the passive roles the new labels l1 and l2 and the new meshed channels s1 and s2
    crossbeam::send($l1::(s1), $s.session1);  *@ \label{app:line:broadcast:one} @*
    crossbeam::send($l2::(s2), $s.session2);  *@ \label{app:line:broadcast:two} @*
    // return the last newly created meshed channels
    MeshedChannels { ... } } *@ \label{app:line:choice:end} @*
\end{rustlisting}}
\end{minipage}
\hrule

\caption{\AMPST communication primitives in~\mpstrust}
\label{app:fig:MPSTMeshedChannelsPrimitives}

\end{figure}

\section{Usecases: distributed logging and circuit breaker}
\label{app:subsec:benchmarks:genericComponentLifecycleManagement}

In this section, we present the implementations of two commonly occurring patterns related to
failure -- a logging-management protocol and
a circuit breaker~\cite{Falahah2021}.
Those two usecases were implemented with~\mpstrust; relevant metrics are available in~\Cref{tab:examples}.
Both protocols mix complex interactions. 

The first usecase is a protocol for distributed logging management.
The Scribble protocol for this pattern is given in~\Cref{fig:GCLMLogging:scribble}, 
while the~\mpstrust implementation is in~\Cref{fig:GCLMLogging:rust}.

We have chosen this protocol because it allows us to exemplify all basic constructs in Scribble and~\mpstrust.
In essence, a logging \emph{supervisor} ensures that whenever the logging process fails,
it is restarted if requested by the system \emph{controller}.
Note that the protocol contains two roles only -- a system controller (\emph{Controller})
that deals with the management of all systems components,
and a logger supervisor (\emph{Supervisor}), a wrapper that controls the logging service(s).
The logging process is not part of the protocol since the supervision mechanism between
a logger and its supervisor is an orthogonal matter;
it can be specified as another protocol, but we omit it here for simplicity.
Initially, the \emph{Controller} starts the \emph{Supervisor} (line~\ref{line:log:start}).
The \emph{Supervisor} monitors the logger process for their status and reports the
status back to the \emph{Controller} -- \CODE{Success}
(line~\ref{line:log:success}),
or \CODE{Failure} (line~\ref{line:log:failure}). If the status is a \CODE{Success},
the \emph{Supervisor} keeps monitoring,
otherwise, the \emph{Supervisor} receives a request to either \CODE{Restart}
the failing logger process and continue monitoring the new process
(line~\ref{line:log:restart}) or to \CODE{Stop} the protocol (line~\ref{line:log:stop}).

\begin{figure}[!t]
\begin{subfigure}{0.5\textwidth}
\begin{SCRIBBLELISTING}
global protocol Logging(role Controller, role Supervisor)
{
Start(int) from Controller to Supervisor; *@ \label{line:log:start} @*

rec Loop {

 choice at Supervisor
 {
  Success(int) from Supervisor to Controller;*@ \label{line:log:success} @*
  continue Loop; *@ \label{line:log:loop} @*
 } or {
  Failure(int) from Supervisor to Controller; *@ \label{line:log:failure} @*
  choice at Controller
  {
   Restart(int) from Controller to Supervisor;*@ \label{line:log:restart} @*
   continue Loop;

  } or {
   Stop(int) from Controller to Supervisor; *@ \label{line:log:stop} @*
  }
 }
}}
\end{SCRIBBLELISTING}
\caption{Scribble protocol}
\label{fig:GCLMLogging:scribble}
\end{subfigure}
\begin{subfigure}{0.5\textwidth}
\begin{rustlisting}[basicstyle=\scriptsize\ttfamily]
fn start_supervisor(s: SInit<i32>) -> R *@ \label{line:gclm:logging:init} @*
{
 let (_start, s) = s.recv()?; *@ \label{line:gclm:logging:start} @*

 recurs_supervisor(s) } *@ \label{line:gclm:logging:req} @*

 fn recurs_supervisor(s: SRec<i32>) -> R {
  let status = ... ;
  // ping logger to get logger status
  if status > 0 *@ \label{line:gclm:logging:choice} @* {
   let s = choose_s!(s, ChoiceC::Success); *@ \label{line:gclm:logging:choice:up} @*
   let s = s.send(1)?; *@ \label{line:gclm:logging:signal:success} @*
   recurs_supervisor(s) } *@ \label{line:gclm:logging:up:loop} @*
  else { failure_supervisor(s) } }

 fn failure_supervisor(s: SFail<i32>) -> R {
  let s = choose_s!(s, ChoiceC::Failure); *@ \label{line:gclm:logging:signal:failure} @*
  let s = s.send(-1)?; *@ \label{line:gclm:logging:signal:payload} @*
  offer_mpst!(s, {
	ChoiceS::Restart(s) => {
    let (_restart, s) = s.recv()?; *@ \label{line:gclm:logging:signal:restart} @*
    recurs_supervisor(s) }, *@ \label{line:gclm:logging:down:loop} @*
	ChoiceS::Stop(s) => {
    let (_stop, s) = s.recv()?; *@ \label{line:gclm:logging:signal:stop} @*
    s.close() } *@ \label{line:gclm:logging:signal:close} @*
} ) }
\end{rustlisting}
\caption{Rust protocol}
\label{fig:GCLMLogging:rust}
\end{subfigure}

\caption{Distributed logging protocol}
\label{fig:GCLMLogging}

\end{figure}

The Rust implementation of the \emph{supervisor} is given in~\Cref{fig:GCLMLogging:rust}.
It is split into three methods -- \CODE{start_supervisor()}, \CODE{recurs_supervisor()} and \CODE{failure_supervisor()};
each method takes a \emph{meshed channel} \CODE{s} and its respective type.
The implementation starts with a \emph{start} message received from the
\emph{Controller} via the~\mpstrust API call, \CODE{s.recv()?}
(line~\ref{line:gclm:logging:start}).
As explained in \S~\ref{sec:api}, the \CODE{?}-operator at the end of each method is
syntactic sugar for error message propagation.
Hence, if the \CODE{s.recv()} at line~\ref{line:gclm:logging:start} fails,
the whole session will be safely cancelled, otherwise, the execution continues to
line~\ref{line:gclm:logging:req}
which calls the recursive function
\CODE{recurs_supervisor()}.
Line~\ref{line:gclm:logging:choice:up} uses the
\CODE{choose_s!} macro to perform an internal choice.
\CODE{ChoiceC} is an enum type,
while, \CODE{Success} and \CODE{Failure} are variants of the enum.
We remind the reader that all \CODE{enum}
types, and their variants 
(\eg \CODE{ChoiceC::Failure, ChoiceS::Restart}, etc) 
are either generated from Scribble, or written by the user. 
Their respective name does not matter,
as long as the enum and its branches exist.

In particular, \CODE{choose_s!(s, ChoiceC::Success)} sends
the label \CODE{Success} to the \emph{Controller}.
This line corresponds to line~\ref{line:log:success} from the Scribble protocol in~\Cref{fig:GCLMLogging:scribble}.
The \CODE{Failure} branch (line~\ref{line:log:failure} --~\ref{line:log:stop}
in~\Cref{fig:GCLMLogging:scribble}) is implemented in the function \CODE{failure_supervisor()}.
First, we send to the \emph{Controller} the choice label \CODE{Failure},
\CODE{choose_s!(s, ChoiceC::Failure)}
at line~\ref{line:gclm:logging:signal:failure},
then the payload -1 (a failure code), \CODE{s.send(-1)}, at line~\ref{line:gclm:logging:signal:payload}.
Afterwards, we use another~\mpstrust primitive, \CODE{offer_mpst!(s, ...)} which is a macro for external choice.
The \CODE{offer_mpst!} macro requires an expression corresponding to each of the enum
types -- \CODE{ChoiceS::Restart} and \CODE{ChoiceS::Stop}.
If the supervisor receives a \CODE{Restart} label,
lines~\ref{line:gclm:logging:signal:restart} --~\ref{line:gclm:logging:down:loop} are executed,
otherwise, the \CODE{Stop} message is received, and the session is correctly closed
via \CODE{s.close()} in line~\ref{line:gclm:logging:signal:close}.
Note the difference between \CODE{cancel(s)} and \CODE{close(s)}:
the former drops the channel \CODE{s}, while the latter
cleanly and safely closes the channel \CODE{s}.
Using \CODE{close(s)} after \CODE{cancel(s)}
will result in a compilation error because \CODE{s} will not be accessible.

\begin{figure}[!t]
\begin{subfigure}{0.48\textwidth}
\begin{SCRIBBLELISTING}
global protocol CBreaker(role Api, role Controller,
	role Storage, Role User)
{
Start(int) from Controller to Storage, Api;*@ \label{line:api:start} @*
HardPing(int) from Storage to Controller;
rec Loop {
	Request() from User to Api;
	GetMode() from Api to Controller; *@ \label{line:api:mode} @*

	choice at Controller {
		Up() from Controller to Api; *@ \label{line:api:success} @*
		Request(int) from Api to Storage;
		Response(int) from Storage to Api; *@ \label{line:gclm:api:signal:start:api} @*
		Response(int) from Api to User; *@ \label{line:api:response} @*
		continue Loop; *@ \label{line:api:contres} @*
	} or {
		Failure(int) from Controller to Api; *@ \label{line:api:failure:api} @*
		Restart(int) from Controller to Storage; *@ \label{line:api:restart} @*
		Failure(int) from Api to User; *@ \label{line:api:failure:user} @*
		continue Loop; } } }
\end{SCRIBBLELISTING}
\caption{Scribble protocol}
\label{fig:GCLMApi:scribble}
\end{subfigure}
\begin{subfigure}{0.5\textwidth}
\begin{rustlisting}[basicstyle=\scriptsize\ttfamily]
fn endpoint_api(s: AInit<i32>) -> R*@ \label{line:gclm:api:init} @*
{
	let (_start, s) = s.recv()?; *@ \label{line:gclm:api:start} @*
	recurs_api(s)}


fn recurs_api(s: ARec<i32>) -> R {
	let (req, s) = s.recv()?; *@ \label{line:gclm:api:signal:user} @*
	let s = s.send(1)?; 

	offer_mpst!(s, {
		ChoiceA::Up => { *@ \label{line:gclm:api:signal:up} @*
			let (_, s) = s.recv()?;
			let s = s.send(0)?; *@ \label{line:gclm:api:signal:request} @*
			let (res, s) = s.recv()?;
			let s = s.send(res)?;  
			recurs_api(s)}, *@ \label{line:gclm:api:signal:rec1} @*
		ChoiceA::Failure => { *@ \label{line:gclm:api:signal:down} @*
			let (err, s) = s.recv()?; *@ \label{line:gclm:api:signal:error} @*
			let s = s.send(err);
			recurs_api(s) } *@ \label{line:gclm:api:signal:rec} @*
})}
\end{rustlisting}
\caption{Rust protocol}
\label{fig:GCLMApi:rust}
\end{subfigure}

\caption{Circuit breaker protocol}
\label{fig:GCLMApi}

\end{figure}

The second protocol is presented in~\Cref{fig:GCLMApi}.
This protocol is called a circuit breaker~\cite{Falahah2021}. It can be implemented
to protect any scarce resource in the system. In this case, we showcase the pattern using a \emph{storage} process.
The protocol includes four roles -- a system controller (\emph{Controller}); a proxy (\emph{Api}),
a \emph{Storage} that represents a scarce resource, and a \emph{User}.
As shown in~\Cref{fig:GCLMApi:scribble},
initially, the \emph{Controller} starts both the \emph{Storage} and the \emph{Api} (line~\ref{line:api:start}).
At each iteration, the \emph{Api} checks the status of the \emph{Storage} by querying the \emph{Controller} (line~\ref{line:api:mode}).
If the \emph{Controller} decides that the \emph{Storage} can handle requests,
then the \emph{Api} forwards any incoming \CODE{Request}s from the \emph{User} (lines~\ref{line:api:success} --~\ref{line:api:contres}).
Otherwise, the \emph{Api} immediately sends a default \CODE{Failure} to the \emph{User} (line~\ref{line:api:failure:user}),
while the \emph{Controller} attempts to \CODE{Restart} the \emph{Storage} process (line~\ref{line:api:restart}).

The implementation of the \emph{Api} role is given in~\Cref{fig:GCLMApi:rust}.
It uses the same primitives as in the previous example,
and precisely follows the protocol. Similarly to the previous example
to check for cancellation we use the \CODE{?}--operator after each \CODE{send()} and \CODE{recv()}.
If any of the communicating actions fail,
the notification of the failure will propagate to the relevant processes.

\section{Related work: extension}
\label{app:related:work}

\subsection{Affine typing systems, exceptions and error-handling in session types}
\label{app:subsec:affine-types}
We discuss related works which treat affinity,
exceptions and error-handling in session types
(except~\cite{harveyMultiparty2021} discussed in the previous paragraph),
focusing on recent work. 
\cite{carboneStructured2008} first proposes  
asynchronous exceptions in binary session types; and 
\cite{capecchiGlobal2010} extends it to multiparty session types. 
Both works are theoretical only.   
\cite{adameitSession2017} introduces session types for links failures,
and extend session types with an optional block
surrounding a process and containing default values used
when a failure occurs.
This default value, returned in case of a failure, has to be
declared by developers.
\cite{vieringTyping2018} provide a formalism and
typing discipline for handling crash failure
in asynchronous distributed systems,
alongside a domain-specific language
based on ZooKeeper,
a niche language unlike Rust.
Besides explicitly raised application-level failures,
it also handles participant crashes, but not concurrent
failures nor the spreading of failure messages to all participants
involved in the protocol.
~\cite{adameitSession2017,vieringTyping2018}
is limited to theory.
~\cite{demangeonPractical2015} model
an interruptible session
where exceptions are explicitly handled by declared
participants. They syntactically
extended Scribble with special interruptible blocks, and
implement runtime monitoring in Python.
Their models are limited to global types (no endpoint processes),
and their implementations
require runtime checking, while
\mpstrust ensures fully static safety and liveness.


\cite{mostrousAffine2018}
proposed
affine \emph{binary} session types
with explicit cancellation, which
\cite{fowlerExceptional2019} extend
to define Exceptional GV
for binary asynchronous communication.
Exceptions can be nested and handled over multiple communication
actions,
and their implementation is an extension of the research language Links.
~\cite{mostrousAffine2018} is limited to the theory,
and both~\cite{mostrousAffine2018,fowlerExceptional2019}
are limited to binary and
linear logic-based session types without recursive session
types.\footnote{%
The first author's thesis~\cite{fowlerTypedNodate}
discusses how to extend~\cite{fowlerExceptional2019} to recursive
types by adding recursions at the term level.
}
In addition,
the progress proof in~\cite{mostrousAffine2018} requires either
a parallel composition with a characteristic process
at each active prefixed session process~\cite{mostrousAffine2018}.
As a result, a global termination for cancellation
among more than two participants, like
Theorem~\ref{thm:ctermination} and
Corollary~\ref{cor:liveness},
cannot be guaranteed in their systems.

\cite{vieringMultiparty2021} proposed a theory to extend
\MPST with fault-tolerance. 
The implementation is in Scala and is integrated with Apache Spark.
Linearity is checked dynamically.
Unlike our theory, they assume the existence of
resilient components (such as \emph{monitors} and \emph{robust roles})
that cannot fail. Upon failure
a session can continue by skipping the interactions with the faulty components.
This can be problematic in a real case scenario
where a faulty participant handles
critical information required by others to
actually progress.

\subsection{Code generation:~\mpstrust for an arbitrary number of roles}
Currently, we have only demonstrated a communication API for three hard-coded roles, but our code generation facilities
provide utility functions for generating API for an arbitrary number of roles, parametric on the role names.
In essence, each generated function is a
generalisation of the functions presented in~\Cref{fig:CreateMeshedChannels}.
We will show only the most important of the code-generation functions -- the one that generates a new \AMPST structure.
All utility functions are implemented as Rust macros.
Rust supports two types of macros -- regular and procedural, we utilise both.
The former are expanded into a syntactic form during compilation,
before any static checking, while the latter is a function over the Rust AST and creates syntax extensions.

\begin{figure}[!ht]
	\begin{minipage}[l]{0.6\textwidth}
		\begin{rustlisting}[basicstyle=\scriptsize\ttfamily]
// The procedural macro
let sessions: Vec<proc_macro2::TokenStream> = ... // Creates """S#N ,""" for #N from 1 to nsessions (excluded) *@ \label{line:meshedchannels:multiple:procedural:meshedchannels:vec:sessions} @*
... *@ \label{line:meshedchannels:multiple:procedural:meshedchannels:rest:vec} @*
quote! { // Concatenates the previous code with specific code *@ \label{line:meshedchannels:multiple:procedural:meshedchannels:quote:start} @*
	...
	pub struct #meshedchannels_name<
		#( #sessions )* // Include the content of *sessions* *@ \label{line:meshedchannels:multiple:procedural:meshedchannels:quote:vec:sessions} @*
		R, N
	>
	where
		#( #sessions_struct )*
		R: mpstthree::role::Role,
		N: mpstthree::role::Role
	{
		#( #sessions_pub )*
		pub stack: R,
		pub name: N,
	}
	... *@ \label{line:meshedchannels:multiple:procedural:meshedchannels:rest:quote} @*
} *@ \label{line:meshedchannels:multiple:procedural:meshedchannels:quote:end} @*
		\end{rustlisting}
	\end{minipage}
	\begin{minipage}[r]{0.4\textwidth}
		{
            \lstset{firstnumber=22}
		        \begin{rustlisting}
// The regular macro usage
create_meshedchannels!(MeshedChannelsFour, 4); *@ \label{line:meshedchannels:multiple:regular:call} @*


// Structure created:
//
// 	MeshedChannelsFour< *@ \label{line:meshedchannels:multiple:structure:created:start} @*
//		S1: mpstthree::binary::Session,
//		S2: mpstthree::binary::Session,
//		S3: mpstthree::binary::Session,
//		R: mpstthree::role::Role,
//		N: mpstthree::role::Role,
//	> {
//		pub session1: S1,
//		pub session1: S2,
//		pub session1: S3,
//		stack: R,
//		name: N,
//	}
//	... *@ \label{line:meshedchannels:multiple:structure:created:end} @*
		    \end{rustlisting}
        }
	\end{minipage}
	\caption{Code to generate the new \CODE{MeshedChannels} for multiple participants}
	\label{fig:CreateMeshedChannels}
	
\end{figure}

~\Cref{fig:CreateMeshedChannels} (left) shows the macro that generates \MPST structure with an arbitrary number of
roles,~\Cref{fig:CreateMeshedChannels} (right) shows its usage for creating \MPST structure with four participants,
along with the generated structure.
The macro inputs for the regular macro (line~\ref{line:meshedchannels:multiple:regular:call})
are the name of the structure and the number of roles, and are passed to the procedural macro.
\CODE{mpst_seq::create_meshedchannels} calls a method
that creates the code with \CODE{quote!}, using the arguments provided by the user.
After retrieving the arguments,
the macro creates the different groups of \CODE{ident},
such as \emph{sessions} (line~\ref{line:meshedchannels:multiple:procedural:meshedchannels:vec:sessions}).
Those \CODE{Vec} contain the blocks of code that are repeated,
such as \CODE{S1 , S2 , S3 ,}, \dots
for \CODE{sessions}.
The \CODE{...} at line~\ref{line:meshedchannels:multiple:procedural:meshedchannels:rest:vec} materialises the other \CODE{Vec} that are created.
Those groups are concatenated along with complementary code between lines~\ref{line:meshedchannels:multiple:procedural:meshedchannels:quote:start}
and~\ref{line:meshedchannels:multiple:procedural:meshedchannels:quote:end}.
The rest of the code, materialised with \CODE{...}
at line~\ref{line:meshedchannels:multiple:procedural:meshedchannels:rest:quote},
contain the methods for the new structure.
The line~\ref{line:meshedchannels:multiple:regular:call} displays the usage of the regular macro for developers,
and the code created for 4 participants is show between lines~\ref{line:meshedchannels:multiple:structure:created:start}
and~\ref{line:meshedchannels:multiple:structure:created:end}.



\begin{wrapfigure}{r}{0.2\textwidth}
	\begin{tabular}{l " l | l | l | l }
		  & a  & b  & c  & d \\
		\thickhline
		a &    & 1  & 2  & 3 \\
		\hline
		b & -1 &    & 4  & 5 \\
		\hline
		c & -2 & -4 &    & 6 \\
		\hline
		d & -3 & -5 & -6 &
	\end{tabular}
	\caption{Distribution of the channels for \CODE{branching} for 4 participants}
	\label{fig:choose:multiple:distribution}
	
\end{wrapfigure}

The most intricate part when implementing multiparty channel is the mapping between binary
channels and their name in their index in the \MPST structure.
When the session starts, all the newly created binary channels should be correctly distributed among all roles.
We will rely on the kind of structure displayed in~\Cref{fig:choose:multiple:distribution}.
The figure demonstrates the distribution for a protocol involving 4 roles.
The headers, column and row, contain the different roles, and the rest details the index of the channels between two participants.
This is a skew-symmetric matrix, where each number on the top right has an opposite in the bottom left.
A negative number means that the role on the corresponding row will receive the dual
of the binary session type of the role on the corresponding column.

%
}{
}

\end{document}
\endinput